\documentclass[12pt]{article}
\pdfoutput=1
\usepackage{jheppub}
\usepackage{amsmath}
\usepackage{amssymb}
\usepackage{wrapfig}
\usepackage{bbm}
\usepackage{arydshln}
\usepackage{multirow}
\usepackage{mathtools}
\usepackage{blkarray}
\usepackage{setspace}
\usepackage{dsfont}
\usepackage{float}
\usepackage{subcaption}
\usepackage{braket}

\allowdisplaybreaks

\title{$\mathcal{W}$-algebra Modules, Free Fields, and Gukov-Witten Defects}

\abstract{We study the structure of modules of corner vertex operator algebras arrising at junctions of interfaces in $\mathcal{N}=4$ SYM. In most of the paper, we concentrate on truncations of $\mathcal{W}_{1+\infty}$ associated to the simplest trivalent junction. First, we generalize the Miura transformation for $\mathcal{W}_{N_1}$ to a general truncation $Y_{N_1,N_2,N_3}$. Secondly, we propose a simple parametrization of their generic modules, generalizing the Yangian generating function of highest weight charges. Parameters of the generating function can be identified with exponents of vertex operators in the free field realization and parameters associated to Gukov-Witten defects in the gauge theory picture. Finally, we discuss some aspect of degenerate modules. In the last section, we sketch how to glue generic modules to produce modules of more complicated algebras. Many properties of vertex operator algebras and their modules have a simple gauge theoretical interpretation.}

\author[a]{Tom\'{a}\v{s} Proch\'{a}zka,}
\author[b]{Miroslav Rap\v{c}\'{a}k}
\affiliation[a]{Arnold Sommerfeld Center for Theoretical Physics,\\ Ludwig Maximilian University of Munich,\\
Theresienstr. 37, D-80333 M\"unchen, Germany}
\affiliation[b]{Perimeter Institute for Theoretical Physics,\\
31 Caroline St N, Waterloo, ON N2L 2Y5, Canada}
\emailAdd{Tomas.Prochazka@lmu.de}
\emailAdd{miroslav.rapcak@gmail.com}

\begin{document}
\maketitle

\section{Introduction}

The theory of vertex operator algebras (VOA) is an enormously rich subject with a long history. Recently, a new way to study VOAs and to connect them with various physical and mathematical applications was initiated in \cite{Gaiotto:2017euk} based on the previous work of \cite{Nekrasov:2010aa,Gaiotto:2011nm} and it was further explored in \cite{Creutzig:2017uxh,Prochazka:2017qum}. The new perspective is based on a realization of VOAs as algebras of local operators within a topological twist of a particular configuration in the four-dimensional $\mathcal{N}=4$ super Yang-Mills theory \cite{Kapustin:aa,Witten:2010aa,Witten:2011aa,Mikhaylov:2014aa}. Configurations of interest are $(p,q)$ webs \cite{Aharony:aa} of supersymmetric interfaces studied in \cite{Gaiotto:2008aa,Gaiotto:2008ab,Gaiotto:2008ac}. Local operators of the twisted theory turn out to live at the two-dimensional junction and give rise to VOAs \cite{Gaiotto:2017euk}. 

The simplest configuration of the triple junction of D5, NS5 and $(1,1)$ interfaces between $U(N_1)$, $U(N_2)$,  $U(N_3)$ gauge theories leads to the VOA labeled as $Y_{N_1,N_2,N_3}$. These corner algebras were originally identified in terms of a BRST reduction of Kac-Moody super-algebras in \cite{Gaiotto:2017euk}. Later, it was argued in \cite{Prochazka:2017qum} that the algebras can be also viewed as truncations (quotients) of the $\mathcal{W}_{1+\infty}$ algebra. The study of this algebra has a very long history. Originally, a linear version of the algebra was constructed as $N \to \infty$ of $W_N$ algebras \cite{Pope:1989sr,Pope:1989ew,Pope:1990kc,Kac:1995sk}. Later, it was gradually realized that there exists in fact a two-parametric family of non-linear algebras \cite{Yu:1991bk,deBoer:1993gd,Khesin:1994ey,Hornfeck:1994is,Blumenhagen:1994wg,Gaberdiel:2012aa,Prochazka:2014aa,Linshaw:2017tvv}. Recently, this algebra appeared in connection with equivariant cohomology of instanton moduli spaces \cite{Schiffmann:2012gf,Maulik:2012rm,Braverman:2014ys,tsymbaliuk2017affine} and its equivalence to Yangian of affine $\mathfrak{gl}(1)$ was found \cite{Maulik:2012rm,tsymbaliuk2017affine,Prochazka:2015aa,Zhu:2015nha,Gaberdiel:2017dbk}. This makes it possible to use the techniques of integrability to study the properties of vertex operator algebras.

Apart from local operators living at the corner, line operators supported at interfaces \cite{Kapustin:aa,Witten:2011aa,Mikhaylov:2014aa} and surface defects \cite{Gukov:2006jk,Mikhaylov:2014aa} supported in the bulk survive the twisting procedure. If we let line operators to end at the junction, the fusion of the endpoint with local insertions at the junction generates a module for the corresponding VOA. Similarly, Gukov-Witten (GW) surface defects ending at interfaces also play the role of VOA modules. The main objective of this paper is the study of modules associated to such higher dimensional operators together with free field realization of the algebras.

\paragraph{Free field realization}

In this work, we identify $Y_{N_1,N_2,N_3}$ with algebras defined previously in terms of a kernel of screening charges by \cite{bershtein,Litvinov:2016mgi}. It is well known that the kernel of screening charges realizing $\mathcal{W}_N\times \widehat{\mathfrak{gl}}(1)=Y_{0,0,N}$  has an explicit construction in terms of the Miura transformation \cite{Fateev:1987zh}. Generators of the algebra are coefficients of an $N$-th order differential operator which is a product of first order differential operators $R^{(3)}_i$ for $i=1,\dots,N$. We generalize this construction to $Y_{N_1,N_2,N_3}$ by introducing two classes of pseudo-differential operators $R^{(1)}_i$ and $R^{(2)}_i$ and taking the product of $N_1$ operators of the first type, $N_2$ of the second type and $N_3$ of the first order differential operators $R^{(3)}_i$. This provides us with a simple way to determine the free field realization of $Y_{N_1,N_2,N_3}$ generators.

\paragraph{Generic modules}

The representation theory of $W_{N}\times \widehat{\mathfrak{gl}}(1)=Y_{0,0,N}$ algebras is relatively well-understood. Generic modules are parametrized by $N$ complex numbers modulo the action of Weyl group. On the other hand, maximally degenerate modules are known to be parametrized by a pair of Young tableau \cite{Prochazka:2015aa}. Gukov-Witten defects for the corresponding gauge theory configuration are parametrized by a complex $N$-dimensional torus which is a product of $N$ tori with the modular parameter being the canonical parameter of the Kapustin-Witten twist $\Psi$. These lead to generic modules. On the other hand, maximally degenerate modules correspond to a pair of line operators (parametrized by finite dimensional representations of $U(N)$) supported at the two boundaries. Note that line operators can be fused with the end-line of the Gukov-Witten defect. This fusion changes the boundary condition imposed on the GW defect that has been implicit in the discussion above. Such a fusion (or the choice of a boundary condition) lifts the $N$-dimensional torus to the full $\mathbb{C}^N$. We call the corresponding parameters lifted GW parameters.

The situation of a general $Y_{N_1,N_2,N_3}$ seems to be more complicated at first sight. The representation theory of $Y_{0,1,1}$ and $Y_{0,1,2}$ from the point of view of (non-freely generated) extensions of the Virasoro algebra by generators of spin 1,3 in the first case and by generators of spin 1,3,4,5 in the second case was studied in \cite{Wang:1998bt} and \cite{1207.3909}. Generic modules can be parametrized by a two-dimensional subvariety inside $\mathbb{C}^3$ for $Y_{0,1,1}$ and by a three dimensional subvariety inside $\mathbb{C}^5$ for $Y_{0,1,2}$. In general, we argue that generic highest weight modules of the $Y_{N_1,N_2,N_3}$ algebra should be parametrized by $N_1+N_2+N_3$ dimensional subvariety (the number of GW parameters in the setup) inside 
\begin{eqnarray}
\mathbb{C}^{(N_1+1)(N_2+1)(N_3+1)-1}.
\end{eqnarray}
The gauge theory setup suggests that the parametrization of representations should be simpler. Indeed, we find that generic modules can be parametrized by $N_1$ complex parameters $x^{(1)}_i$, $N_2$ parameters $x^{(2)}_i$ and $N_3$ parameters $x^{(3)}_i$. As discussed above, the algebra $\mathcal{W}_{1+\infty}$ is isomorphic as an associative algebra to the well known affine Yangian of $\mathfrak{gl}(1)$ generated by an infinite number of generators $\psi_i,f_i,e_i$. The modules of interest can be defined in terms of an action of the commuting generators $\psi_i$ on the highest weight state. Such an action is encoded in a generating function of $\psi_i$ charges whose poles are parametrized by $x^{(\kappa)}_i$ for $\kappa=1,2,3$.

Moreover, we identify $x^{(\kappa)}_i$ with the three families of lifted GW parameters discussed above. In terms of the free field realization, one can construct modules of $Y_{N_1,N_2,N_3}$ by an action of $\mathcal{W}$-algebra generators on the highest weight vector of a tensor product of $N_1+N_2+N_3$ free-boson Fock modules. Parameters $x^{(\kappa)}_i$ can be identified (up to constant shifts) with such highest weights, giving the third perspective on $x^{(\kappa)}_i$. The change of basis of algebra generators allows us to translate $\psi_i$ charges of the affine Yangian to $(W_i)_0$ charges (eigenvalues of zero modes of $\mathcal{W}$-algebra generators) and recover the Zhu varieties from \cite{Wang:1998bt} and \cite{1207.3909}.

\paragraph{Degenerate modules}

A generic GW defect breaks the gauge group at the defect to the maximal torus $U(1)^{N_1}$. A degeneration of a generic module appears when we specialize GW parameters such that a Levi subgroup is preserved at the defect. In particular, when two of the parameters specifying singularity of the complexified gauge field at the interface are equal, the preserved gauge group is enhanced to the next-to-minimal Levi subgroup $U(2)\times U(1)^{N_i-2}$. This configuration can be further dressed by turning on a Wilson or 't Hooft line of the preserved $SU(2)$ factor. The corresponding degeneration of the module appears when the lifted GW parameters satisfy\footnote{The parameters $h_\kappa$ are related to the canonical parameter $\Psi$ of the Kapustin-Witten twist by
\begin{eqnarray}
h_1=\frac{1}{\sqrt{\Psi}},\qquad h_2=-\sqrt{\Psi},\qquad h_3=\sqrt{\Psi}-\frac{1}{\sqrt{\Psi}}.
\end{eqnarray}}
\begin{eqnarray}
x^{(3)}_i-x^{(3)}_{j}=h_1 n+h_2 m\qquad \mbox{or} \qquad x^{(3)}_j- x^{(3)}_i=h_1 n+h_2 m
\end{eqnarray}
for some $i,j$ and positive integers $n,m$ that parametrize the line operators (finite dimensional $SU(2)$ representations) at the two boundaries of the third\footnote{We call the corner between D5 and NS5 interfaces with the gauge theory of the gauge group $U(N_3)$ the third corner. Similarly, the corner between D5 and $(1,1)$ interfaces is the second and the corner between $(1,1)$ and NS5 interfaces is the first.} corner. 

Further degenerations appear when the following specialization
\begin{eqnarray}
x^{(3)}_i-x^{(2)}_{j}=-h_3+h_1 n
\end{eqnarray}
happens between GW parameters in different corners for any integer $n$ and similarly for the other pairs of parameters. 

When more parameters are specialized, one gets further degenerations associated to more complicated Levi subgroups. If a maximal number of them are specialized, one gets maximally degenerate modules that can be identified with the configuration of line operators with trivial surface defects.

\paragraph{Gluing of generic modules}

In \cite{Prochazka:2017qum}, we proposed a construction that associates a VOA to an arbitrary $(p,q)$ web of interfaces between $U(N_i)$ gauge theories. The corresponding VOA is an extension of the tensor product of $Y$-algebras associated to trivalent juncions of the diagram by a fusion of bi-modules associated to line operators supported at internal edges of the diagram. The free field realization discussed above points towards the completion of the gluing proposal from \cite{Prochazka:2017qum} by finding a way to possibly determine all OPEs of gluing fields by which we extend the product of Y-algebras. In particular, as discussed in the section \ref{maxdeg}, one can realize the fundamental and the anti-fundamental representation associated to each interface as a Fock descendant of a vertex operator of free bosons. There are actually many possible choices for a given free field realization and we conjecture (and test in examples) that the result (if non-vanishing) is independent of the choice of the representant as long as we include contour integrals of screening currents along the lines of \cite{Dotsenko:1984nm,Felder:1988zp}.

The gauge theory picture suggests that generic modules of glued algebras can be obtained as a tensor product of corresponding modules of each vertex with GW parameters correctly identified. The total number of continuous parameters of a generic module is thus a sum of all the numbers of D3-branes at each face. From the VOA point of view, this identification of GW parameters is needed for generic modules to have trivial braiding with bi-modules added in the gluing procedure. As a non-trivial example, we discuss the structure of generic modules of the $\widehat{\mathfrak{gl}}(N)$ Kac-Moody algebra and corresponding $\mathcal{W}$-algebras. We conjecture that a subclass of modules coming from GW defects can be identified with modules induced from generic Gelfand-Tsetlin modules of $\widehat{\mathfrak{gl}}(n)$ from \cite{1409.8413} and their $\mathcal{W}$-algebra analogues.

\section{Gauge theory setup}

In this section, we briefly review the gauge theory setup from \cite{Gaiotto:2017euk} and comment on the main players (line operators and Gukov-Witten defects) in the discussion of modules. Finally, we discuss the simple example of the $Y_{0,0,1}=\widehat{\mathfrak{gl}}(1)$ Kac-Moody algebra that serves as a prototype for the general discussion in later sections.

\subsection{The corner}

\begin{figure}[h]
\centering
\includegraphics[width=0.44\textwidth]{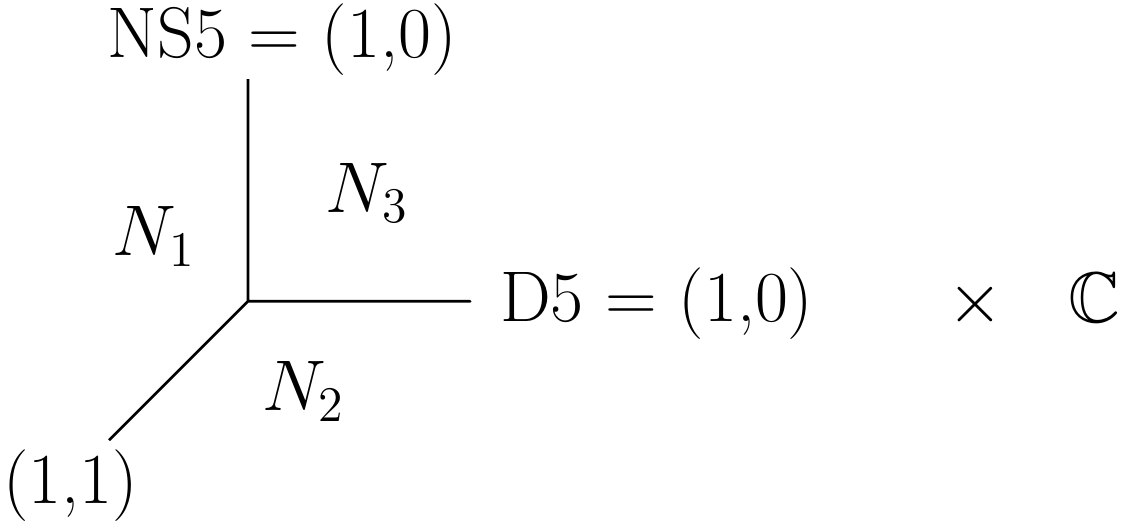}
\caption{The simplest trivalent junction of NS5, D5 and $(1,1)$ interfaces between $U(N_1)$, $U(N_2)$ and $U(N_3)$ gauge theories with a two-dimensional corner. The junction on the left shows the configuration of interfaces in the $x^1,x^2$ plane of the gauge theory with $\mathbb{C}$ factor corresponds to the $x^3,x^4$ directions.}
\label{corner1}
\end{figure}

There exists a class of half-BPS domain walls between four-dimensional $\mathcal{N}=4$ super Yang-Mills theories with gauge groups $U(N_1)$ and $U(N_2)$ associated to co-prime numbers $(p,q)$. The gauge theory setup descends from $N_1$ and $N_2$ D3-branes ending from the left and from the right on a $(p,q)$-brane.\footnote{One identifies the NS5-brane with $(0,1)$ and the D5-brane with $(1,0)$.} The simplest quarter-BPS trivalent junction of NS5, D5 and $(1,1)$ interfaces between $U(N_1)$, $U(N_2)$ and $U(N_3)$ gauge theories as shown in the figure \ref{corner1} was analyzed in \cite{Gaiotto:2017euk}. These triple junctions serve as building blocks of more complicated junctions coming form various $(p,q)$-web configurations studied in \cite{Prochazka:2017qum}. 

Let us restrict to Kapustin-Witten twist \cite{Kapustin:aa} of the configuration with the canonical parameter $\Psi \in \mathbb{CP}^1$ and deformed boundary conditions in such a way that the Kapustin-Witten supercharge is preserved. It was argued in \cite{Gaiotto:2017euk} that local operators in the cohomology of the Kapustin-Witten supercharge are supported at the junction of domain walls and give rise to the vertex operator algebra $Y_{N_1,N_2,N_3}[\Psi]$. For each choice of ranks of gauge groups $N_1,N_2,N_3$, one obtains a one parameter family of VOAs parametrized by the canonical parameter $\Psi$. In the following, we often suppress the dependence on $\Psi$ in $Y_{N_1,N_2,N_3}[\Psi]$.

\subsection{Line operators}

Apart from the local operators living at the two-dimensional corner, line operators supported at each of the three interfaces are part of the twisted theory as well. Consider line operators supported at one of the three interfaces, going from the infinity and ending at the corner at point $z\in \mathbb{C}$. The endpoint $z$ determines the insertion of the corresponding vertex operator from the CFT point of view. The process of fusing local operators living at the corner with the line endpoint generates a module for $Y_{N_1,N_2,N_3}$. 

Line operators supported at the NS5-interface can be identified with the Wilson lines associated to a finite-dimensional representation $\mu$ of the Lie super-group $U(N_1|N_3)$ as discussed in \cite{Mikhaylov:2014aa}. Similarly, line operators at the D5-interface are 't Hooft operators associated to $U(N_3|N_2)$ representations and line operators at the (1,1)-interface are Wilson line operators associated to representations of $U(N_2|N_1)$. These modules play the role of degenerate modules of $Y_{N_1,N_2,N_3}$. The algebra $Y_{N_1,N_2,N_3}$ has a natural grading by spin and degenerate modules are characterized by the fact that they contain less states in some graded component compared to a generic module.

\subsection{Gukov-Witten defects}

Apart from the line operators discussed above, Gukov-Witten (GW) surface defects \cite{Gukov:2006jk} also survive the GL twist. Inserting such a GW defect at a point $z\in \mathbb{C}$ and attaching it to one of the corners of the Y-shaped junction, one gets a new (continuous) family of modules for the corner VOA. 

\begin{figure}[h]
\centering
\includegraphics[width=0.33\textwidth]{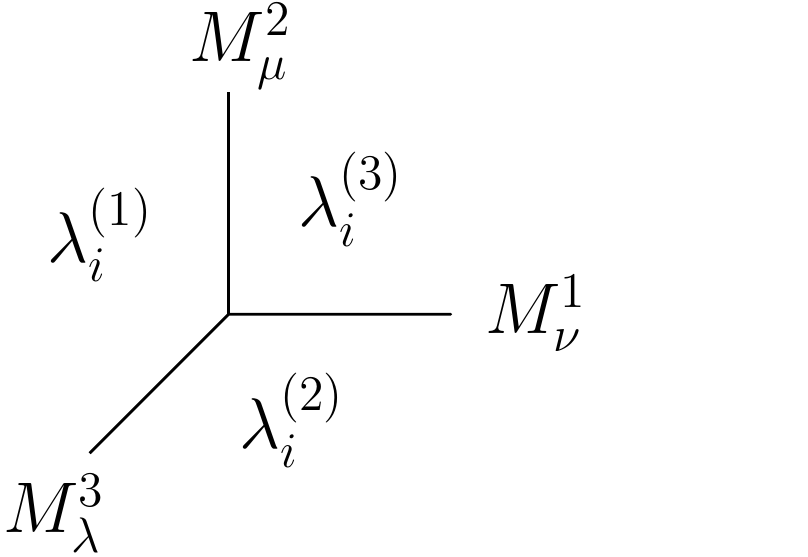}
\caption{Line operators $M^i_\mu$ labeled by finite representations of the gauge groups are supported at interfaces and give rise to degenerate modules of VOA. GW defects attached to the corners of the diagram are labeled by $\lambda^{(1)}_i,$ $\lambda^{(2)}_i,$ $\lambda^{(3)}_i$ in $N_1,N_2$ and $N_3$ complex tori with modular parameters $\Psi$ and give rise to generic modules.}
\label{corner2}
\end{figure}

GW defects in the $U(N)$ gauge theory are labeled according to \cite{Gukov:2006jk,Witten:2011aa,Mikhaylov:2014aa} by four real parameters\footnote{In general, the parameter $\eta$ lives in the Cartan subalgebra of the Langlands dual gauge group $T^{\vee}$. Since $U(N)$ is left invariant under the Langlands duality, we do not distinguish them in this work.} $(\alpha,\beta,\gamma, \eta)\in (T,\mathfrak{t},\mathfrak{t},T)$, where $T$ is the Cartan of the gauge group $U(N)$ and $\mathfrak{t}$ the Cartan subalgebra of the Lie algebra $\mathfrak{u}(N)$. In the GL-twisted theory, parameters $\beta$ and $\gamma$ were argued in \cite{Mikhaylov:2014aa} to deform the integration contour of the complexified Chern-Simons theory. On the other hand, the combination
\begin{eqnarray}
\lambda=\Psi \alpha -\eta
\end{eqnarray}
parametrizes the monodromy of the complexified gauge connection $\mathcal{A}=A+\omega\phi$ around the defect, i.e.
\begin{eqnarray}
\mathcal{A}(z)\sim \frac{\mbox{diag}(\lambda_1,\dots,\lambda_N)}{z}
\end{eqnarray}
near the defect at the origin $z=0$. The parameter $\omega$ is related to $\Psi$ in such a way that $\mathcal{A}$ is a closed combination at the interface (modulo a gauge transformation). Since both $\alpha$ and $\eta$ live in the Cartan subgroup $\alpha, \eta \in \left (S^1\right )^N$ of the gauge group $U(N)$, we see that the corresponding monodromies (and Gukov-Witen defects in the GL-twisted theory) are labeled by points in $N$ complex tori of modular parameter $\Psi$.

Let us discuss S-duality transformation of the GW parameters identified in \cite{Gukov:2006jk}. The pair $(\beta, \gamma)$ transforms as
\begin{eqnarray}
S:(\beta,\gamma)\rightarrow |\tau|(\beta,\gamma)
\end{eqnarray}
under the S-transformation and it is unaffected by the T-transformation. On the other hand, the pair $(\alpha,\eta)$ relevant to us transforms as 
\begin{eqnarray}
S:(\alpha,\eta)\rightarrow (\eta,-\alpha),\qquad T:(\alpha,\eta)\rightarrow (\alpha,\eta-\alpha).
\end{eqnarray}
The complex parameter $\lambda$ of the twisted theory transforms as
\begin{eqnarray}
S:\lambda=\Psi \alpha -\eta \rightarrow \lambda^\prime = \alpha-\frac{1}{\Psi}\eta,\qquad T:\lambda \rightarrow \lambda.
\end{eqnarray}
We see that $\lambda$ is invariant under the T-transformation and the S-transformation simply multiplies the Gukov-Witten parameter by $1/\Psi$ and exchanges the role of $\alpha$ and $\eta$. In later sections, we will see that this transformation is consistent with the triality covariance of $Y_{N_1,N_2,N_3}$.

When a GW defect ends at an interface, one needs to further specify a boundary condition for the defect. We will see later that the choice of the boundary condition lifts $\lambda^{(\kappa)}$ for $\kappa=1,2,3$ living in the $N_\kappa$ complex-dimensional torus of in each corner to $\tilde{\lambda}^{(\kappa)}\in \mathbb{C}^{N_\kappa}$. The boundary line of the surface operator can be fused with line operators discussed above. Such a fusion changes the boundary condition for the GW defect. For example in the $Y_{0,0,1}$ configuration, line operators supported at the NS5 interface produce a defect with charge $n\in \mathbb{Z}$ that lifts the parameter $\eta$ and the line defect supported at the other interface creates a vortex of monodromy $\Psi m\in \Psi \mathbb{Z}$ lifting the parameter $\alpha$. Similarly in the other two corners, the fundamental domain of the torus is lifted to the full $\mathbb{C}$ by modules coming from line operators at the corresponding two boundaries.

For generic values of GW-parameters, the defect breaks the gauge group to the maximal torus at the defect. Corresponding modules are going to be associated to generic modules for the corner VOA. For special values of parameters, a Levi subgroup of the gauge group is preserved and we expect the corresponding representations to be (partially) degenerate, i.e. the associated Verma module contains some null states. For example, if two of the monodromy parameters are specialized, the next-to-minimal Levi subgroup $U(2)\times U(1)^{N_i-2}$ is preserved. One can decorate such a configuration by line operators in some representation of the preserved $SU(2)$ gauge group. In the parameter space of the lifted GW parameters, one gets a discrete set of codimension one walls corresponding to degenerate modules for each pair of Cartan elements. The full parameter space of generic modules thus has a chamber-like structure with the modules degenerating at the walls. At the intersection of more walls, we expect further degeneration to appear. These intersections correspond to larger Levi subgroups. In the case that GW parameters are maximally specialized, we have a trivial interface (there are no singularities in the bulk) and we expect the corresponding modules to be maximally degenerate. The corresponding modules are labeled by finite representations of gauge groups (labeling line operators at the interfaces).
 
Finally, let us note that throughout the discussion above, one needs to mod out Weyl groups of $U(N_i)$ since modules related by the Weyl transformations are gauge equivalent.

\subsection{$Y_{0,0,1}=\widehat{\mathfrak{gl}}(1)$ example}

Let us illustrate how above gauge theory elements fit nicely with the simplest example $Y_{0,0,1}=\widehat{\mathfrak{gl}}(1)$. This example is extremely important since all the other algebras can be obtained from a fusion (coproduct) combined with the triality transformation of this simple algebra.

\begin{figure}[h]
\centering
\includegraphics[width=0.72\textwidth]{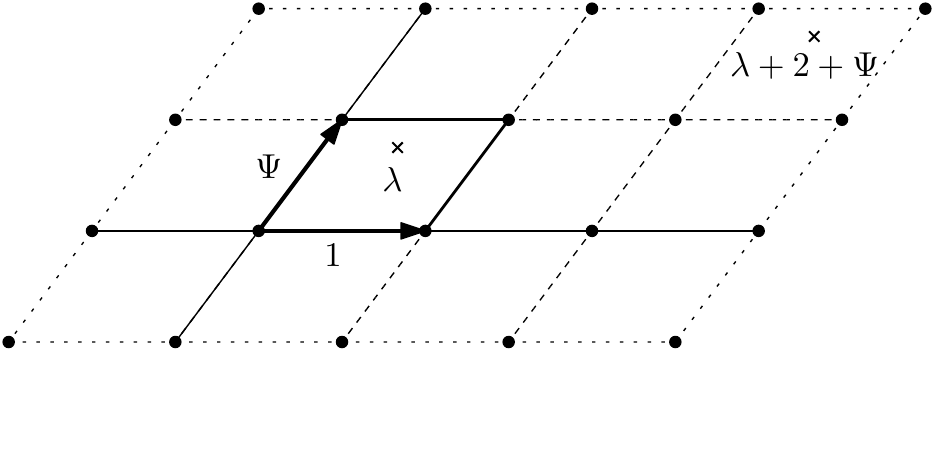}
\caption{The lattice structure of modules of the $\widehat{\mathfrak{gl}}(1)$ algebra. GW-defects are labeled by a point in the torus of modular parameter $\Psi$. Fusion with electric and magnetic modules of charges $n$ and $\Psi n$ lift the torus along the full complex plane parametrizing generic module of the algebra. Lattice points correspond to dyon modules of the algebra and the position in the fundamental domain corresponds to the GW-parameter. For example, the modules of charge $\tilde{\lambda}=\lambda+2+\Psi$ and $\lambda$ in the fundamental domain are related by fusion with the electric module of charge 2 and the magnetic module of charge 1.}
\label{u1modules}
\end{figure}

The insertion of the complexified gauge connection $\mathcal{A}$ at the corner can be identified with the $\widehat{\mathfrak{gl}}(1)$ current $J$ normalized as
\begin{eqnarray}
J(z)J(w)\sim \frac{\Psi}{(z-w)^2}.
\end{eqnarray}
 In \cite{Gaiotto:2017euk}, line operators supported at the NS5-boundary were identified with electric modules of charge $n\in \mathbb{Z}$ and conformal dimension $\frac{1}{2\Psi}n^2$. Line operators at the D5-boundary were identified with magnetic operators with monodromy $\Psi m\in \Psi \mathbb{Z}$ and conformal dimension $\frac{\Psi}{2}m^2$. On the other hand, GW defects are parametrized by a complex torus with the modular parameter $\Psi$ parametrizing the monodromy for the complexified gauge connection in the bulk.  If the GW defect ends at the NS5 boundary, one can fuse the end line of the defect with line operators supported at the boundary. Such a line operator shifts the charge by $1$ and lifts the torus of the Gukov-Witten defect in the real direction. Similarly, fusing with modules supported at the D5-boundary lifts it in the $\Psi$ direction tessellating $\mathbb{C}$ as shown in the figure \ref{u1modules}. The GW parameter $\lambda$ thus lifts to $\tilde{\lambda}\in \mathbb{C}$ that can be identified with the $J_0$ eigenvalue. The fusion with an electric module shifts it by one $\tilde{\lambda} \rightarrow \tilde{\lambda}+1$, whereas the fusion with a magnetic module shifts it by $\Psi$, i.e. $\tilde{\lambda} \rightarrow \tilde{\lambda}+\Psi$. The module coming from the GW defect has charge $\tilde{\lambda}$ and conformal dimesion $\frac{1}{2\Psi}\tilde{\lambda}^2$.

Note that the S-duality transformation exchanges NS5-brane and D5-brane and the orientation of the diagram gets reversed. The transformed level of the algebra is $1/\Psi$ and the transformed lifted GW parameter becomes $\tilde{\lambda}/\Psi$. This is consistent both with the transformation of degenerate modules and the unlifted GW parameter. Note that conformal dimension of the generic module is invariant under the S-duality transformation and so is the charge if we renormalize $\tilde{J}=\frac{1}{\sqrt{\Psi}}J$. The roles of $\alpha$ and $\eta$ interchange.

Let us show that transformations of parameters are also consistent with the triality relation
\begin{eqnarray}
Y_{0,0,1}\left [\Psi\right ]=Y_{0,1,0}\left [\tilde{\Psi}=1-\frac{1}{\Psi}\right ].
\end{eqnarray}
The insertion of $\mathcal{A}$ at the corner of $Y_{0,1,0} [\tilde{\Psi} ]$ leads to the $\widehat{\mathfrak{gl}}(1)$ Kac-Moody algebra normalized as
\begin{eqnarray}
J(z)J(w)\sim \frac{1-\tilde{\Psi}}{(z-w)^2}.
\end{eqnarray}
Consider a GW defect with the parameter $\tilde{\lambda}^{(2)}$. The charge of the corresponding module with respect to the normalized current $J/\sqrt{\tilde{\Psi}-1}$ equals
\begin{eqnarray}
\frac{\tilde{\lambda}^{(2)}}{\sqrt{1-\tilde{\Psi}}}= \frac{\tilde{\lambda}^{(2)}}{\sqrt{1-1+\frac{1}{\Psi}}}=\sqrt{\Psi}\tilde{\lambda}^{(2)}.
\end{eqnarray}
Comparing it with the charge with respect to the normalized current of $Y_{0,0,1}[\Psi]$ that equals $\tilde{\lambda}^{(3)}/\sqrt{\Psi}$, we see that the two GW parameters must be indeed related by $\tilde{\lambda}^{(3)}=\tilde{\lambda}^{(2)}/\Psi$ consistently with the above discussion.

\subsection{Reparametrization of GW defects}

The trivalent junction of interest is invariant under the $S_3$ subgroup of the $SL(2,\mathbb{Z})$ group of S-duality transformations. To get manifestly triality invariant parametrization of the algebra and its modules, let us introduce parameters $h_1, h_2, h_3$ by
\begin{eqnarray}
\Psi=-\frac{h_2}{h_1},\qquad h_1+h_2+h_3=0.
\end{eqnarray}
Note that the parameters $h_i$ are determined up to the overall rescaling. The VOA is independent on such a rescaling. Up to the rescaling, one can relate parameters $h_i$ and $\Psi$ for example as
\begin{eqnarray}
h_1=\frac{1}{\sqrt{\Psi}},\qquad h_2=-\sqrt{\Psi},\qquad h_3=\sqrt{\Psi}-\frac{1}{\sqrt{\Psi}}.
\end{eqnarray}
Instead of the lifted GW parameter $\tilde{\lambda}^{(3)}$ parametrizing defects in the third corner, one can consider the combination
\begin{eqnarray}
x^{(3)}=\frac{1}{\sqrt{\Psi}}\tilde{\lambda}^{(3)} =  h_1\alpha^{(3)} +h_2\eta^{(3)}
\end{eqnarray}
and similar combinations in the other two corners
\begin{eqnarray}\nonumber
x^{(2)}&=& h_3\alpha^{(2)} +h_1\eta^{(2)} \\
x^{(1)}&=& h_2\alpha^{(1)}+h_3\eta^{(1)}.
\end{eqnarray}

In the $Y_{0,0,1}$ example, we can identify the parameter $x^{(3)}$ with the coefficient in the exponent of the vertex operator\footnote{In the following we will drop the normal ordering symbols and we assume all the exponential vertex operators are normal ordered.}
\begin{eqnarray}
\exp \left [ x^{(3)} \phi(w)\right ]
\end{eqnarray}
in the free field realization of the module with the current $J^{(3)}=\partial \phi^{(3)}= J/\sqrt{\Psi}$ normalized as
\begin{eqnarray}
J^{(3)}(z)J^{(3)}(w)\sim -\frac{1}{h_1h_2}\frac{1}{(z-w)^2}.
\end{eqnarray}
In this parametrization, the electric module $M^2$ of unit charge corresponds to $\alpha^{(3)}=1$ whereas the magnetic module to $\eta^{(3)}=1$. 

In the other two frames $Y_{1,0,0}$ and $Y_{0,1,0}$ with currents $J^{(\kappa)}=\partial \phi^{(\kappa)}$ normalized as
\begin{eqnarray}
J^{(\kappa)}(z)J^{(\kappa)}(w)\sim -\frac{h_{\kappa}}{h_1h_2h_3}\frac{1}{(z-w)^2},
\end{eqnarray}
parameters $x^{(\kappa)}$ are again exponents of the corresponding vertex operator. We will later see that parameters $x^{(\kappa)}_i$ can be identified with shifts of exponents of $N_1+N_2+N_3$ vertex operators also for general $Y_{N_1,N_2,N_3}$.

In the parametrization using $h_i$, the triality tranformation simply permutes $h_\kappa$ together with parameters $\alpha^{(\kappa)},\eta^{(\kappa)}$. The invariance of the charge of the current normalized to identity is manifest.

\section{Y-algebras and free fields}

In this section, we review the definition of $Y_{N_1,N_2,N_3}$ in terms of truncations of the $\mathcal{W}_{1+\infty}$ algebra, the kernel of screening charges and the BRST reduction. Furthermore, we generalize the Miura transformation construction of the kernel of screening charges for $Y_{0,0,N}\equiv \mathcal{W}_N\times \widehat{\mathfrak{gl}}(1)$ to arbitrary $Y_{N_1,N_2,N_3}$.

\subsection{Three definitions of Y-algebras}

\paragraph{Truncations of $\mathcal{W}_{1+\infty}$}

The algebra $\mathcal{W}_{1+\infty}$ is an extension of the vertex operator algebra of the stress-energy tensor $T$ by primary fields $W_i$ of spin $i=1,3,4,\dots$. Jacobi identities fix all the structure constants \cite{Gaberdiel:2012aa,Prochazka:2014aa,Linshaw:2017tvv} of the algebra up to three parameters $(\lambda_1,\lambda_2,\lambda_3)$ subject to the constraint
\begin{eqnarray}
\frac{1}{\lambda_1}+\frac{1}{\lambda_2}+\frac{1}{\lambda_3}=0.
\end{eqnarray}

It was argued in \cite{Prochazka:2015aa,Prochazka:2017qum} that for each triple of non-negative integers $(N_1,N_2,N_3)$, the algebra $\mathcal{W}_{1+\infty}$ contains an ideal $\mathcal{I}_{N_1,N_2,N_3}$ generated by a singular vector at level $(N_1+1)(N_2+1)(N_3+1)$, whenever $\lambda_i$ satisfy
\begin{eqnarray}
\frac{N_1}{\lambda_1}+\frac{N_2}{\lambda_2}+\frac{N_3}{\lambda_3}=1.
\end{eqnarray}
For these values of $\lambda_i$, one can define the quotient\footnote{Some truncations of this sort have been recently discussed in \cite{Linshaw:2017tvv}.}
\begin{eqnarray}
Y_{N_1,N_2,N_3}[\Psi]=\mathcal{W}_{1+\infty}/\mathcal{I}_{N_1,N_2,N_3}\qquad \mbox{for}\qquad \Psi=-\frac{\lambda_1}{\lambda_2}.
\end{eqnarray} 

Apart from the primary basis mentioned above, there exists another useful basis (sometimes called the $U$-basis or the quadratic basis) related to the Miura transformation \cite{luk1988quantization, Prochazka:2014aa}. Generating fields of this basis are not quasi-primary and also explicitly depend on a choice of the triality frame (so there are in fact three different bases of this kind that are interchanged by the action of the triality). On the other hand, their main advantage is that operator product expansions in this basis have only quadratic non-linearity. This allowed to guess a closed form-formula for all OPEs \cite{Prochazka:2014aa}.

The transformation between the primary and the quadratic basis is not known in general but can be calculated spin by spin, i.e. we can construct primary fields in terms of the generators in the quadratic basis.\footnote{It is interesting that in the semiclassical limit, i.e. when the VOA simplifies to a Poisson vertex algebra, there is a closed-form determinantal formula for transformation between primary and quadratic basis \cite{DiFrancesco:1990qr} which has very similar form to the formula for Virasoro singular vectors.} Because of the presence of the composite primary fields, the primary basis is not uniquely determined by the primarity condition. Even if we decouple the spin one field from the rest of the algebra and work with the $\mathcal{W}_{\infty}$ subalgebra, one is still not able to uniquely fix the primary generators using only the condition of being primary. Starting at spin $6$, there appears the first primary composite $(W_3 W_3) + \ldots$ field. Generators $W_i$ can be the determined (with the normalization ambiguity still undetermined) by a further requirement of the orthogonality (vanishing two-point function) with the composite primaries. First few primary fields determined in this way are given in appendix \ref{primaryquadraticformulas}.

Identification between the triality-covariant parameters $\lambda_j$ and the parameters $N$ and $\alpha_0$ used in Miura transformation and the structure constants in the quadratic basis is \cite{Prochazka:2014aa}
\begin{equation}
\label{ubasisparam}
\lambda_3 = N, \quad\quad\quad \alpha_0^2 = -\frac{\lambda_1\lambda_2}{\lambda_3^2}.
\end{equation}
Note that there are indeed three possible identifications (and the corresponding $U$-bases) with $\lambda_i$ parameters permuted.

\paragraph{Affine Yangian}
The vertex operator algebra  $\mathcal{W}_{1+\infty}$ is isomorphic as an associative algebra (after a suitable completion) with the Yangian of affine $\mathfrak{gl}(1)$ as discussed in \cite{Maulik:2012rm,Schiffmann:2012gf,tsymbaliuk2017affine,Prochazka:2015aa,Gaberdiel:2017dbk}. The affine Yangian in the basis of \cite{tsymbaliuk2017affine} is generated by an infinite set of commuting generators $\psi_i$ together with an infinite set of raising $f_i$ and lowering $e_i$ ladder operators. As we will see below, the representation theory simplifies significantly using this Yangian point of view. The triality symmetry is manifest, but one loses the manifest locality and conformal field theory interpretation. The structure of Yangian depends on three complex parameters $h_1, h_2$ and $h_3$ constrained by
\begin{equation}
h_1 + h_2 + h_3 = 0.
\end{equation}
The map between these parameters and the $\lambda$-parameters of $\mathcal{W}_{1+\infty}$ is
\begin{equation}
\label{lambdatoh}
\lambda_j = -\frac{\psi_0 h_1 h_2 h_3}{h_j}
\end{equation}
where $\psi_0$ is the first (central) element of the commutative Cartan subalgebra of the Yangian. Specializations of the affine Yangian of $\mathfrak{gl}(1)$ are isomorphic to VOAs $Y_{N_1,N_2,N_3}$ as proved in \cite{Schiffmann:2012gf} for $\mathcal{W}_N$ truncations and \cite{Next3} in general, based on the previous work of \cite{Schiffmann:2012gf,Prochazka:2015aa,Fukuda:2015ura,tsymbaliuk2017affine,Prochazka:2017qum}.

\paragraph{Free field realization}

Another definition of $Y_{N_1, N_2, N_3}$ studied in \cite{bershtein,Litvinov:2016mgi} is in terms of subalgebras of free bosons\footnote{We are grateful to Mikhail Bershtein and Alexey Litvinov for pointing out this relation.}. Consider a set of $N_1+N_2+N_3$ free bosons $\phi_i^{(\kappa)}$, where $\kappa=1,2,3$ labels a type of the boson and $i=1,\dots, N_1+N_2+N_3$ with $N_\kappa$ bosons of type $\kappa$. Let us pick a fixed ordering of $\phi_i^{(\kappa)}$. To each neighbouring pair of free bosons, we can associate a screening charge according to \cite{bershtein,Litvinov:2016mgi} and reviewed later. The explicit form of the screening charge depends on the type of the neighbouring pair of free bosons. $Y_{N_1,N_2,N_3}$ can be then defined as a commutant of all such $N_1+N_2+N_3-1$ screening charges. The resulting algebra is independent of the choice of ordering but the way it is embedded in the corresponding Fock space depends on the ordering.

Below, we give an alternative way to construct the free field realization of $Y_{N_1,N_2,N_3}$. The construction is based on a generalization of the standard quantum Miura transformation for $Y_{0,0,N}=\mathcal{W}_N\times \widehat{\mathfrak{gl}}(1)$ \cite{Fateev:1987zh, Bouwknegt:1992wg}. For $Y_{0,0,N}$, one factorizes an $N$-th order differential operator as a product of first order operators. Replacing these elementary first order operator by certain pseudo-differential operators which we describe below, we obtain the desired free field realization. One can check that the two constructions give the same free field realization by comparing the results for $Y_{0,0,2}$ and $Y_{1,1,0}$ and realizing that both constructions are essentially local, i.e. involve only operations on the pairs of neighbouring bosons.

\paragraph{BRST construction}

Y-algebras were originally introduced in \cite{Gaiotto:2017euk} in terms of a BRST reduction translating the boundary conditions in $\mathcal{N}=4$ SYM \cite{Gaiotto:2008aa,Gaiotto:2008ab,Gaiotto:2008ac}. They were defined as a combination of the Drinfeld-Sokolov reduction and the BRST coset reduction of a pair of Kac-Moody super-algebras.  We refer reader to the original work \cite{Gaiotto:2017euk} and \cite{Prochazka:2017qum} for a summary.

\subsection{Miura transformation for $Y_{N_1,N_2,N_3}$}

Let us now give a generalization of the well-known Miura transformation for $Y_{0,0,N}$ of \cite{Fateev:1987zh, Bouwknegt:1992wg} to general $Y_{N_1,N_2,N_3}$ and relate it to the free field realization of \cite{bershtein,Litvinov:2016mgi}.

\subsubsection{Review of $Y_{0,0,N}$}

Firstly, we review the standard Miura transformation for $Y_{0,0,N}$. Consider a set of $N$ $\mathfrak{gl}(1)$ currents $J_j(z)$ with OPEs
\begin{equation}
J_j(z) J_k(w) \sim \frac{\delta_{jk}}{(z-w)^2} 
\end{equation}
and define operators $U_k(z)$ via
\begin{equation}
(\alpha_0 \partial + J_1(z)) \cdots (\alpha_0 \partial + J_N(z)) \equiv \prod_{j=1}^N R_j^{(3)}(z) = \sum_{k=0}^N U_k(z) (\alpha_0 \partial)^{N-k},
\end{equation}
where the parameter $\alpha_0$ is related to the parameters of $\mathcal{W}_{1+\infty}$ by (\ref{ubasisparam}). Operators $U_k$ and their normal ordered products and derivatives form a closed algebra under operator product expansion \cite{luk1988quantization}.

\subsubsection{The general case}

One can extend the Miura transformation to the case where there are nodes of different types. For that it is important to remember that we have three types of nodes corresponding to three different free field representations of $\mathcal{W}_{1+\infty}$ corresponding to $\lambda_1 = 1$, $\lambda_2 = 1$ or $\lambda_3 = 1$ (as well as their conjugates). The usual Miura transformation in our conventions has all nodes of type $3$ with $\lambda_3 = 1$. We will see that the usual procedure works even in the case of $\lambda_1 = 1$ or $\lambda_2 = 1$ but we have to replace the elementary factor
\begin{equation}
R^{(3)}(z) \equiv \alpha_0 \partial + J^{(3)}(z)
\end{equation}
by a pseudo-differential operator with an infinite number of coefficients which are local fields. This generalization is common in the context of integrable hierarchies of differential equations (e.g. KdV or KP hierarchies), \cite{miwa2000solitons,Khesin:1993ru}.

\begin{figure}[H]
\centering
\includegraphics[width=0.61\textwidth]{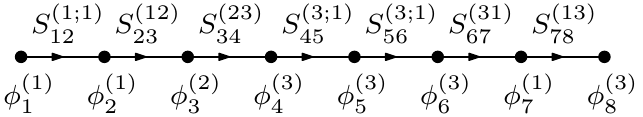}
\caption{An example of the ordering of free bosons for $Y_{3,1,4}$. The algebra can be found by multiplying the Miura pseudo-differential operators in the order $R_1^{(1)}(z)R_2^{(1)}(z)\dots R_7^{(1)}(z)R_8^{(3)}(z)$ as shown in the figure. Alternatively, one can construct the free field realization as an intersection of kernels of screening charges $S^{(1;1)}_{12},S^{(12)}_{23},\dots S_{78}^{13}$ associated to the lines of the chain of free bosons.}
\label{figscreening}
\end{figure}

Let us first consider what happens in the case that $\lambda_1= 1$. In this situation, there exists a free field representation of $\mathcal{W}_{1+\infty}$ associated to a single free boson $\phi^{(1)}$, but in the quadratic $U$-basis (which is itself associated to the third direction), there is an infinite number of non-trivial $U_j$ generators, all expressed in terms of $\phi^{(1)}$. Choosing for convenience the parametrization as in \cite{Prochazka:2015aa}
\begin{eqnarray}
\nonumber
h_1 & = & h \\
\label{hjtohparam}
h_2 & = & -\frac{1}{h} \\
\nonumber
h_3 & = & \frac{1}{h} - h = \alpha_0 \\
\nonumber
\psi_0 & = & \lambda_3 = N
\end{eqnarray}
we need to require
\begin{equation}
1 = \lambda_1^{(1)},
\end{equation}
i.e.
\begin{equation}
N^{(1)} = \lambda_3^{(1)} = -\frac{h^2}{h^2-1} = -\frac{h_1}{h_1 h_2 h_3}.
\end{equation}
From the Miura transformation point of view, this $N^{(1)}$ is the order of the pseudo-differential operator corresponding to the $\phi^{(1)}$ representation. In the following, it will be useful to choose the normalization coefficient of the two-point function of the current $J^{(1)} \equiv \partial\phi^{(1)}$ to be $N^{(1)}$,
\begin{equation}
J^{(1)}(z) J^{(1)}(w) \sim \frac{N^{(1)}}{(z-w)^2}.
\end{equation}
Having fixed all the parameters of algebra, we can now find the expressions for $U^{(1)}_j(z)$ fields in terms of $J^{(1)}$, requiring just the commutation relations spelled out in \cite{Prochazka:2014aa}. They are uniquely determined up to the conjugation $J^{(1)} \leftrightarrow -J^{(1)}$ symmetry. Fixing a positive sign, the expressions for the first few fields are
\begin{eqnarray}\nonumber
U^{(1)}_1 & = & J^{(1)} \\ \nonumber
U^{(1)}_2 & = & \left(2-\frac{1}{h^2}\right) \left( \frac{(J^{(1)} J^{(1)})}{2} + \frac{h\partial J^{(1)}}{2} \right) \\ \nonumber
U^{(1)}_3 & = & \left(2-\frac{1}{h^2}\right)\left(3-\frac{2}{h^2}\right) \Bigg( \frac{(J^{(1)}(J^{(1)} J^{(1)}))}{6} + \frac{h(\partial J^{(1)} J^{(1)})}{2} + \frac{h^2\partial^2 J^{(1)}}{6} \Bigg) \\ \nonumber
U^{(1)}_4 & = & \left(2-\frac{1}{h^2}\right)\left(3-\frac{2}{h^2}\right)\left(4-\frac{3}{h^2}\right) \Bigg( \frac{(J^{(1)}(J^{(1)}(J^{(1)}J^{(1)})))}{24} + \\
& & + \frac{h (\partial J^{(1)}(J^{(1)}J^{(1)}))}{4} + \frac{h^2 (\partial J^{(1)} \partial J^{(1)})}{8} + \frac{h^2(\partial^2 J^{(1)} J^{(1)})}{6} + \frac{h^3 \partial^3 J^{(1)}}{24} \Bigg)
\end{eqnarray}
The expressions for higher $U^{(1)}_j$ fields are uniquely determined from the OPE of $U^{(1)}_3 U^{(1)}_{j-1}$. But even the general pattern is not very difficult to understand: first of all, each $U^{(1)}_j$ has an overall multiplicative factor
\begin{equation}
\prod_{k=1}^{j-1} \left[1+k\left(1-\frac{1}{h^2}\right)\right] = \prod_{k=1}^{j-1} \left(1-\frac{k}{N^{(1)}}\right).
\end{equation}
Next, there is a sum of all dimension $j$ operators that we can construct out of a free boson. The power of $h$ in each term counts the number of derivatives appearing in the operator and the combinatorial factors can be most easily seen using the operator-state correspondence:
\begin{eqnarray}\nonumber
U_1^{(1)} & \to & a_{-1} \\ \nonumber
U_2^{(1)} & \to & \frac{a_{-1}^2}{2} + \frac{h a_{-2}}{2} \\ \nonumber
U_3^{(1)} & \to & \frac{a_{-1}^3}{6} + \frac{h a_{-1} a_{-2}}{2} + \frac{h^2 a_{-3}}{3} \\
U_4^{(1)} & \to & \frac{a_{-1}^4}{24} + \frac{h a_{-1}^2 a_{-2}}{4} + \frac{h^2 a_{-2}^2}{8} + \frac{h^2 a_{-1} a_{-3}}{3} + \frac{h^3 a_{-4}}{4}
\end{eqnarray}
These are exactly the coefficients appearing in Newton's identities if we think of $U_j$ to be the homogeneous symmetric polynomials and $a_j$ to be the power sum symmetric polynomials. One can thus also write a closed-form formula
\begin{equation}
U_j^{(1)} = \prod_{k=1}^{j-1} \left(1-\frac{k}{N^{(1)}}\right) \sum_{m_1 + 2m_2 + \ldots + jm_j = j} \prod_{k=1}^j \frac{1}{m_k! k^{m_k}} \left(\frac{h^{k-1}}{(k-1)!} \partial^{k-1} J^{(1)} \right)^{m_k}
\end{equation}
where everything is normal ordered. The total Miura operator representing the $\phi^{(1)}$ node of the diagram (see figure \ref{figscreening}) is thus given by the pseudo-differential operator
\begin{equation}
R^{(1)}(z) \equiv (\alpha_0 \partial)^{\frac{h_1}{h_3}} + \sum_{j=1}^{\infty} U_j^{(1)}(z) (\alpha_0 \partial)^{\frac{h_1}{h_3}-j}.
\end{equation}
In the case of representation of type $\phi^{(2)}$ the calculation is entirely analogous and in fact we can just make a replacement $h \leftrightarrow -\frac{1}{h}$. We require $\lambda_2^{(2)} = 1$ and so in this case
\begin{equation}
N^{(2)} = \lambda_3^{(2)} = \frac{1}{h^2-1} = -\frac{h_2}{h_1 h_2 h_3}.
\end{equation}
The current is again normalized such that the quadratic pole of the $J^{(2)} J^{(2)}$ OPE is equal to this value of $N^{(2)}$. Choosing the sign of $U^{(2)}_1$, all other $U^{(2)}_j$ operators are uniquely determined and we find
\begin{eqnarray}\nonumber
U_1^{(2)} & = & J^{(2)} \\ \nonumber
U_2^{(2)} & = & (2-h^2) \left( \frac{(J^{(2)} J^{(2)})}{2} - \frac{\partial J^{(2)}}{2h} \right) \\ \nonumber
U_3^{(2)} & = & (2-h^2)(3-2h^2) \Bigg( \frac{(J^{(2)}(J^{(2)} J^{(2)}))}{6} - \frac{(\partial J^{(2)} J^{(2)})}{2h} + \frac{\partial^2 J^{(2)}}{6h^2} \Bigg) \\ \nonumber
U_4^{(2)} & = & (2-h^2)(3-2h^2)(4-3h^2) \Bigg( \frac{(J^{(2)}(J^{(2)}(J^{(2)}J^{(2)})))}{24} - \\
& & - \frac{(\partial J^{(2)}(J^{(2)}J^{(2)}))}{4h} + \frac{(\partial J^{(2)} \partial J^{(2)})}{8h^2} + \frac{(\partial^2 J^{(2)} J^{(2)})}{6h^2} - \frac{\partial^3 J^{(2)}}{24h^3} \Bigg)
\end{eqnarray}
The formula for $U_j^{(2)}$ is now
\begin{equation}
U_j^{(2)} = \prod_{k=1}^{j-1} \left(1-\frac{k}{N^{(2)}}\right) \sum_{m_1 + 2m_2 + \ldots + jm_j = j} \prod_{k=1}^j \frac{1}{m_k! k^{m_k}} \left(\frac{(-1)^{k-1}}{(k-1)! h^{k-1}} \partial^{k-1} J^{(2)} \right)^{m_k}
\end{equation}
and the Miura pseudo-differential operator representing a node of type $\phi^{(2)}$ is
\begin{equation}
R^{(2)}(z) \equiv (\alpha_0 \partial)^{\frac{h_2}{h_3}} + \sum_{j=1}^{\infty} U_j^{(2)}(z) (\alpha_0 \partial)^{\frac{h_2}{h_3}-j}.
\end{equation}

We can use these newly constructed building blocks to find a free field representation of any $Y_{N_1, N_2, N_3}$ algebra: pick an arbitrary ordering of $N_j$ bosons of type $\phi^{(j)}$ as shown in the figure \ref{figscreening} for a particular ordering of the $Y_{3,1,4}$ algebra and multiply the corresponding Miura operators $R_j^{(\kappa_j)}$. Commuting all the derivatives to the right (recall that even for non-integer powers of derivative the generalization of Leibniz rule still applies), we find in the end a pseudo-differential operator of the form
\begin{equation}
R(z) = (\alpha_0 \partial)^{\frac{N_1 h_1 + N_2 h_2 + N_3 h_3}{h_3}} + \sum_{j=1}^{\infty} U_j(z) (\alpha_0 \partial)^{\frac{N_1 h_1 + N_2 h_2 + N_3 h_3}{h_3}-j}
\end{equation}
where $U_j$ are certain normal ordered differential polynomials in the free boson fields. The statement is that the fields $U_j(z)$, their normal ordered products and derivatives form a closed subalgebra of the algebra of $N_1+N_2+N_3$ free bosons which represents $Y_{N_1, N_2, N_3}$ in terms of free bosons. Furthermore, OPEs of these $U_j$ fields are still those of the quadratic $U$-basis with structure constants given in \cite{Prochazka:2014aa}. Examples will be discussed in later sections.

\subsubsection{Miura versus screening}

To each ordering of $N_\kappa$ free bosons $\phi^{(\kappa)}_i$ of type $\kappa$ with the corresponding currents $J_i^{(\kappa)}=\partial \phi^{(\kappa)}_i$ normalized as
\begin{eqnarray}
J_{i_1}^{(\kappa_1)}(z)J_{i_2}^{(\kappa_2)}(w)\sim -\frac{h_\kappa}{h_1 h_2 h_3}\frac{\delta^{\kappa_1,\kappa_2}}{(z-w)^2},
\label{normalization}
\end{eqnarray}
we have the associated free field realization of the algebra $Y_{N_1,N_2,N_3}$. On the other hand, the authors of \cite{bershtein,Litvinov:2016mgi} construct a free field realization of the same algebra as a kernel of $N_1+N_2+N_3-1$ screening charges acting on the tensor product of the current algebras above. Let us define screening charges for each such ordering and check that they are of the form of \cite{Litvinov:2016mgi}.

Consider a fixed ordering of free bosons such as the one in the figure \ref{figscreening}. One associates a screening charge to each neighboring free bosons (lines connecting two nodes of the chain). If the two free bosons are of the same type, say $\kappa_i=\kappa_{i+1}=3$, the corresponding screening current can be chosen to be either
\begin{eqnarray}
S_{i,i+1}^{(3;1)}=\oint dz\exp \left [-h_{1}\phi^{(3)}_i+h_{1}\phi^{(3)}_{i+1}\right ]
\label{screen1}
\end{eqnarray}
or
\begin{eqnarray}
S_{i,i+1}^{(3;2)}=\oint dz\exp \left [ -h_{2}\phi^{(3)}_i+h_{2}\phi^{(3)}_{i+1}\right ].
\label{screen2}
\end{eqnarray}
These two can be determined from the requirement that the zero mode of the exponential vertex operator commutes with the free field realization of the spin one and the spin two fields in the Virasoro algebra $Y_{0,0,2}$. One gets similar expressions for the other three types with the $h_i$ parameters permuted. To a pair of free bosons of different type (say ordering $\phi_i^{(3)}\times \phi_{i+1}^{(2)} $), one associates instead the screening charge\footnote{The commutation with the spin one and the spin two field gives two possible solutions as in the case of the Virasoro algebra but only one is preserved by the requirement of commutativity with the spin three generator.}
\begin{eqnarray}
S_{i,i+1}^{(32)}=\oint dz\exp \left [-h_{2}\phi^{(3)}_i+h_{3}\phi^{(2)}_{i+1}\right ]
\label{screen3}
\end{eqnarray}
and similarly for the other five combinations. 

The screening charge $S_{i,i+1}$ maps the vacuum representation of the product of the current algebras generated by $J^{(\kappa)}_i=\partial \phi^{(\kappa)}_i$ to a module with the highest weight vector $j_{i,i+1}(0) |0\rangle$, where $j_{i,i+1}$ is the screening current associated to the screening charge $S_{i,i+1}$. The algebra $Y_{N_1, N_2, N_3}$ can be defined as an intersection of kernels of screening charges
\begin{eqnarray}
\label{screeningcommutant}
Y_{N_1, N_2, N_3}=\bigcap_{i=1}^{N_1+N_2+N_3-1}\ker S_{i,i+1}.
\end{eqnarray}

Consider now a triple of free bosons neighbouring in the chain and let us compute the matrix of inner producs of the corresponding two exponents of the screening currents with respect to the metric given by the normalization of two-point function
\begin{eqnarray}
g_{jk}=-\frac{h_{\kappa_j}}{h_1 h_2 h_3}\delta_{jk}
\end{eqnarray}
We will see that the different choices of ordering and different choices of the screening currents (\ref{screen1}) and (\ref{screen2}) lead to different matrices from \cite{Litvinov:2016mgi}.

If all the three free bosons are of the same type $\phi_1^{(3)}\times\phi_2^{(3)}\times\phi_3^{(3)}$, one can pick either both screening charges to be of the same type (\ref{screen1}) or (\ref{screen2}) or one of the first type and the second one of the second type. In these four cases, one gets respectively the following two matrices
\begin{eqnarray}
-\frac{h_1}{h_2} \begin{pmatrix}
2 & -1 \\
-1 & 2
\end{pmatrix},
\qquad
-\frac{h_1}{h_2} \begin{pmatrix}
2 & -\frac{h_2}{h_1} \\
-\frac{h_2}{h_1} & 2
\end{pmatrix},
\end{eqnarray}
together with matrices with the parameters $h_1\leftrightarrow h_2$ interchanged.
These two matrices are of the form 1 and 2 from (2.24) of \cite{Litvinov:2016mgi}.

If one of the three free bosons is of a different type than the other two, say 332, one has two possible orderings. In the first case, $\phi_1^{(3)}\times\phi_2^{(3)}\times\phi_3^{(2)}$, one has again a choice between the screening currents (\ref{screen1}) and (\ref{screen2}) leading to the following two overlap matrices
\begin{eqnarray}
\begin{pmatrix}
-2\frac{h_1}{h_2} &1\\
1&1
\end{pmatrix},
\qquad
\begin{pmatrix}
-2\frac{h_2}{h_1} &\frac{h_2}{h_1}\\
\frac{h_2}{h_1}&1
\end{pmatrix}
\end{eqnarray}
that are of the form 4 and 3 of \cite{Litvinov:2016mgi}. The last, symmetric ordering $\phi_1^{(3)}\times\phi_2^{(2)}\times\phi_3^{(3)}$ gives an overlap matrix of the form 
\begin{eqnarray}
\begin{pmatrix}
1 &\frac{h_3}{h_1}\\
\frac{h_3}{h_1}&1
\end{pmatrix}
\end{eqnarray}
which is of the form 5. Finally, if all the bosons are of a different type, one gets the matrix of overlaps
\begin{eqnarray}
\begin{pmatrix}
1 &1\\
1&1
\end{pmatrix}.
\end{eqnarray}
Comparing the free field realizations of $Y_{0,0,2}$ and $Y_{0,1,1}$ from the Miura transformation and from the kernel of screening charges together with the triality symmetry permuting the Y-algebra labels, one can see that the two free field realizations are the same.

\section{Generic modules}

Let us turn to the discussion of generic modules of $Y_{N_1,N_2,N_3}$ associated to Gukov-Witten defects. We start with a review of the algebra of zero modes and how to parametrize modules of a VOA induced from modules of the zero mode algebra. In the section \ref{generic2}, we review a compact way to parametrize highest weights in terms of Yangian generating functions $\psi(u)$. The section \ref{generic3} describes a general structure of the variety of highest weights parametrizing generic representations of $Y_{N_1,N_2,N_3}$ in the primary basis. The next two sections state the generating function for such representations and related its parameters with Gukov-Witten parameters and parameters of Fock modules in the corresponding free field realization of modules. Finally, we give two examples of variety of highest weights in the section \ref{examples}.

\subsection{Zero modes and generic modules}
\label{generic1}
 
A rich class of $Y_{N_1,N_2,N_3}$ representations can be induced from representations of the subalgebra of zero modes
\begin{eqnarray}
X_0=\frac{1}{2\pi i}\oint dz z^{h(X)-1}X(z)
\end{eqnarray}
for $X$ a field of spin $h(X)$. Starting with a highest-weight vector anihilated by all positive modes, one can show that the algebra of zero modes of truncations of $\mathcal{W}_{1+\infty}$ acting on the highest weight vector is commutative \cite{Linshaw:2017tvv}. We can thus define a one-dimensional module for the zero-mode algebra by prescribing how zero modes of the strong generators $W_j$ act. If there are relations in the space of fields (which show as singular vectors of the vacuum Verma module), the zero mode of the corresponding null fields must vanish when acting on the highest weight state. The existence of null fields thus constrains possible highest weights leading to a variety of highest weights.

Let us add few comments:
\begin{enumerate}
\item In the math literature, the algebra of zero modes acting on the highest weight state appears under the name of the Zhu algebra\footnote{The Zhu commutative product is defined as a modified normal ordered product $[X]\star[Y]=(X,Y)\ +$ corrections. The corrections are the commutators $[Y_{1},Y_{-1}]+[Y_{2},Y_{-2}]+\dots$ from the mode expansion of the normal ordered product acting on the highest weight state. For a more precise comparison see \cite{Brungs:1998ij}.} \cite{Zhu1995ModularIO}. If the Zhu algebra is commutative (as in the $Y_{N_1,N_2,N_3}$ case \cite{Linshaw:2017tvv}) the variety of highest weights is the spectrum of the Zhu algebra.
\item Not all the modules produced by gluing are induced from the algebra of zero modes with trivial action of the positive modes on the highest weight vectors. Gluing of highest weight modules of $Y_{N_1,N_2,N_3}$ leads in general to irregular modules of the glued algebra. We will later illustrate this phenomenon on the simplest example of the $\widehat{\mathfrak{gl}} (2)$ Kac-Moody algebra. 
\item Even in the case when the module of the glued algebra has a trivial action of positive modes on the space of highest weights, the space of highest weights itself generically forms an infinite-dimensional representation of the zero mode algebra.
\end{enumerate}

\paragraph{Example - Ising model}
As an illustration of possible restrictions that arise in the presence of null states that are quotiened out, let us consider the $c=1/2$ representation of the Virasoro algebra
\begin{equation}
T(z) T(w) \sim \frac{1/4}{(z-w)^4} + \frac{2T(w)}{(z-w)^2} + \frac{\partial T(w)}{z-w}.
\end{equation}
It is well-known \cite{francesco2012conformal} that the vacuum representation contains a singular vector at level $6$ with the corresponding primary field
\begin{equation}
\phi_6 = 128(T(TT)) + 186 (\partial T \partial T) - 264 (\partial^2 TT) - 9\partial^4 T.
\end{equation}
The requirement of vanishing of the null field in any correlator constrains possible modules for the VOA. In our case, let us consider a generic primary field $\chi(z)$ of dimension $h$,
\begin{equation}
T(z) \chi(w) \sim \frac{h \chi(w)}{(z-w)^2} + \frac{\partial \chi(w)}{z-w},
\end{equation}
and require the operator product expansion of $\phi_6$ and $\chi$ to vanish. The most singular (sixth order) term is precisely the zero mode of $\phi_6$ acting on the highest weight discussed above
\begin{eqnarray}
\frac{1}{2\pi i}\oint dz z^{5}\phi_6(z) |h\rangle =4h(2h-1)(16h-1)|h\rangle=0
\end{eqnarray}
and the variety of highest weights consists of three points $h = 1/2$, $h = 1/16$, and $h=0$. These are the allowed primary fields of the Ising model.

It is interesting to look also at the conditions following from the vanishing of the lower order poles in the OPE
\begin{eqnarray}
\phi_6(z) \chi(w) & \sim & \frac{4h(2h-1)(16h-1) \chi(w)}{(z-w)^6} + \frac{12(2h-1)(16h-1) \partial \chi(w)}{(z-w)^5} \\
\nonumber
& & + \frac{48h(8h-17)(T\chi)(w)}{(z-w)^4} + \frac{6(64h+7)\partial^2\chi(w)}{(z-w)^4} + \mathcal{O}((z-w)^{-3}).
\end{eqnarray}
The quintic pole vanishes for $h = 1/2$ and $h = 1/16$, while for $h=0$ it requires $\partial \chi_0 = 0$ which is the usual singular vector of the vacuum representation at level $1$ (translation invariance of the vacuum).

Let us look at the  quartic pole more closely. For $h=0$ it does not give us anything new while for $h=1/2$ it requires
\begin{equation}
4(T\chi_{1/2})(z) - 3\partial^2 \chi_{1/2}(z)
\end{equation}
to be zero and for $h=1/16$
\begin{equation}
3(T\chi_{1/16})(z) - 4\partial^2 \chi_{1/16}(z)
\end{equation}
to be zero. These are just the singular vectors of $h_{2,1}$ and $h_{1,2}$ Virasoro primaries. We could proceed further and find other relations coming from the lower order poles.
 
From this simple example we see that the singular vectors of the vacuum representation carry interesting information that constrains the spectrum of primaries of the theory. If we impose that $\phi_6$ vanishes in all the correlation functions (which we would need to do for example in a unitary theory), we find that there are only three possible primary fields and we also find their singular descendants.

\subsection{Generating function of highest weights for $Y_{N_1,N_2,N_3}$}
\label{generic2}

Generic highest weight modules of a VOA with a commutative algebra of zero modes are parametrized by the action of such zero modes on the highest weight state. For example, modules of the $\widehat{\mathfrak{gl}}(1)\times\mathcal{W}_N \equiv Y_{0,0,N}$ algebra are labeled by $N$ highest weights, i.e. eigenvalues of $W_i$ zero modes for $1,2, 3, \ldots N$. Analogously, a generic representation of $\mathcal{W}_{1+\infty}$ is specified by an infinite set of higher spin charges of the highest weight state, one for each independent generator of spin $1,2, 3, \ldots$. To label a generic highest weight representation of $\mathcal{W}_{1+\infty}$ and its truncations, it is convenient to introduce a generating function of the highest weight charges.

\paragraph{Generating function of highest weights}

We will not be able to write down explicitly the generating function of highest weights in the primary basis of the algebras. Instead, we will see that the modules can be easily parametrized using the Yangian description in temrs of generators $\psi_i,f_i,e_i$ from \cite{tsymbaliuk2017affine,Prochazka:2015aa}. We will specify the module by the eigenvalues of the commuting $\psi_i$ generators on the highest weight state encoded in the generating function
\begin{equation}
\psi(u) = 1 +  h_1 h_2 h_3 \sum_{j=0}^{\infty} \frac{\psi_j}{u^{j+1}}.
\end{equation}

Another possibility to encode the highest weight charges is in terms of the generating function of $U$-charges of the quadratic basis\footnote{OPEs of the $\mathcal{W}_{1+\infty}$ algebra in the U-basis contain only quadratic non-linearities with all the structure constants fixed in \cite{Prochazka:2014aa}.} of $\mathcal{W}_{1+\infty}$.
$U$-basis is particulary usefull for description of $Y_{0,0,N}$ with the generating function given by
\begin{equation}
\mathcal{U}(u) = \sum_{k=0}^N \frac{u_k}{(-u)(-u+\alpha_0)\cdots(-u+(k-1)\alpha_0)}
\end{equation}
where $u_j$ are the eigenvalues of zero modes of the $U_j$-generators of $Y_{0,0,N}$ and $u_0 \equiv 1$. The generating function is a ratio of two $N$-th order polynomials in $u$-plane, so we may factorize it and write
\begin{equation}
\mathcal{U}(u) = \prod_{j=1}^N \frac{u-\Lambda_j-(j-1)\alpha_0}{u-(j-1)\alpha_0}.
\label{Ugen}
\end{equation}
As shown in \cite{Prochazka:2015aa}, the transformation between generating function $\mathcal{U}(u)$ and $\psi(u)$ is given by
\begin{equation}
\psi(u) = \frac{u-N\alpha_0}{u} \frac{\mathcal{U}(u-\alpha_0)}{\mathcal{U}(u)}
\end{equation}
if we identify the parameters as
\begin{equation}
h_1 h_2 = -1, \quad\quad h_3 = \alpha_0, \quad\quad \psi_0=N.
\end{equation}
These relations allow us to translate between $\psi_j$ charges of the highest weight state and the corresponding $u_j$ charges.

Plugging in the product formula for $\mathcal{U}$, we find
\begin{equation}
\psi(u) = \frac{(u-\Lambda_1-\alpha_0)(u-\Lambda_2-2\alpha_0)\cdots(u-\Lambda_N-N\alpha_0)}{(u-\Lambda_1)(u-\Lambda_2-\alpha_0)\cdots(u-\Lambda_N-(N-1)\alpha_0)}.
\end{equation}
Defining
\begin{equation}
x_j = \Lambda_j + (j-1) h_3
\end{equation}
we can rewrite this as
\begin{equation}
\psi(u) = \prod_{j=1}^N \frac{u-x_j-h_3}{u-x_j},
\label{psiw}
\end{equation}
i.e. the parameters $x_j$ specify the positions of poles of $\psi(u)$ in the spectral parameter plane while the zeros are at positions $x_j+h_3$. Using the variables $x_j$, we have a manifest permutation symmetry of the generating function, while the shifted variables $\Lambda_j$ are chosen such that the vacuum representation has $\Lambda_j = 0$.

\subsection{Zero mode algebra of $Y_{N_1,N_2,N_3}$}
\label{generic3}

The algebras $Y_{N_1,N_2,N_3}$ are finitely (generically non-freely) generated vertex operator algebras by fields $W_1,W_2,\dots,W_n$, where
\begin{eqnarray}
\label{n1n2n3level}
n=(N_1+1)(N_2+1)(N_3+1)-1.
\end{eqnarray}
The finite generation can be seen from the structure of null states of the algebra. The first state of $\mathcal{W}_{1+\infty}$ that needs to be removed in order to get the algebra $Y_{N_1,N_2,N_3}$ appears at level $n+1$. Assuming that the coefficient in front of $W_{n+1}$ does not vanish, one can use this null field to eliminate the $W_{n+1}$ field from the OPEs. At the next level, three more null fields appear. Two of them are the derivative of the null field at level $n+1$ and its normal ordered product with $W_1$ but one also gets one extra condition. This condition can be used to remove the field $W_{n+2}$. One can continue this procedure and (assuming that there are enough conditions at each level) one can remove all $W_i$ for $i>n$ from OPEs.

In this way, one can solve many null state conditions by restricting to a finite number of W-generators but generically (apart from the case of $Y_{N,0,0},Y_{0,N,0},Y_{0,0,N}$) some null states remain. These are going to be composite primary fields formed by the restricted set of $W$-generators and need to be removed as well. The first constraint appears generically already at level $n+2$. For large enough values of $N_i$, one can see from the box-counting that there are be 12 null states at this level but only $\partial^2 W_n, (W_n\partial J),(J,\partial W_n),(J,(J,W_n)), (T,W_n),\partial W_{n+1},(J,W_{n+1}),W_{n+2}$ are removed by the above argument. One has still 4 constrains that lead to a non-trivial conditions on the algebra of zero modes. Note that for small values of of $N_i$, there will be less states at this level as can be easily seen from the box-counting and as we will see in examples below. We will also see that some constraints will be trivially satisfied and only some of them are actually non-trivial.

One can see that for generic values of $N_1,N_2,N_3$ the problem outlined above becomes rather complex. The null states have been fully identified only in the case $Y_{0,1,1}$ and $Y_{0,1,2}$ in the literature \cite{Wang:1998bt,1207.3909} and lead to nontrivial constraints on the allowed highest weights\footnote{The cases $Y_{N,N,0}$, $Y_{N,N-1,0}$ and $Y_{N,N-2,0}$ have been considered rigorously in math literature \cite{linshaw2008invariant,Creutzig:2014lsa,Linshaw:2015bqv,Arakawa:2017rrn,Linshaw:2017tvv}}. From the discussion above, one can still draw the conclusion what will be the general structure of the variety of highest weights. As argued above, the possible highest weights are given by a subvariety inside the space of the highest weights of zero modes 
\begin{eqnarray}
(W_i)_0|w_i\rangle =w_i|w_i\rangle .
\end{eqnarray}
The highest weights are constrained by the existence of null states $X^i_{null}$ and we conjecture that the resulting variety of highest weights of the algebra of zero modes 
\begin{eqnarray}
(X^i_{null})_0|w_i\rangle=f^i (w_i)|w_i\rangle =0
\end{eqnarray}
is $N_1+N_2+N_3$ dimensional subvariety inside $\mathbb{C}^{n}$. Although we will not be able to explicitly construct the null states in general in terms of primary fields, we will give an explicit parametrization of the variety by generalizing the generating function of $\psi_i$ charges of the $\mathcal{W}_N$ algebra. The conjecture for the dimensionality comes from the existence of $N_1+N_2+N_3$ continuous parameters of surface defects available in the configuration. The number $N_1+N_2+N_3$ can be also guessed from the free field realization of the algebra $Y_{N_1,N_2,N_3}$ inside $Y_{1,0,0}^{\otimes N_1}\otimes Y_{0,1,0}^{ \otimes N_2}\otimes Y_{0,0,1}^{\otimes N_3}$, where modules of each of the factors are parametrized by $N_1, N_2$ and $N_3$ parameters respectively. The dimensionality indeed matches in examples of $Y_{0,1,1}$ and $Y_{0,1,2}$  from the literature.

Note that the above discussion also implies that the character of the module with generic highest weights counts $N_1+N_2+N_2$-tuples of partitions, i.e.
\begin{eqnarray}
\chi_{N_1+N_2+N_3} (q)=\prod_{n=1}^{\infty}\frac{1}{(1-q^n)^{N_1+N_2+N_3}}.
\end{eqnarray}
A general state of a generic module of the algebra can be constructed by an action of negative modes $W_i$ on the highest weight state subject to the null state conditions. As in the case of zero modes, where the null states were used to carve out an $N_1+N_2+N_3$ dimensional subvariety, one can use negative modes of the null conditions to remove appropriate states at higher levels. Only $N_1+N_2+N_2$ of the modes at each level are independent, giving rise to the above character. 

\subsection{Generating function for $Y_{N_1,N_2,N_3}$}

As we have just seen, truncations $Y_{N_1,N_2,N_3}$ are finitely generated by $W_1,\dots,W_n$ where $n$ is given by (\ref{n1n2n3level}). In particular, generic representations have a finite number of states at level one. Following the usual notion of quasi-finite representations of linear $\mathcal{W}_{1+\infty}$ \cite{Kac:1993zg,Kac:1995sk}, it was argued in \cite{Prochazka:2015aa} that a highest weight representation of $\mathcal{W}_{1+\infty}$ has a finite number of states at level $1$ if and only if generating function $\psi(u)$ equals a ratio of two Drinfeld polynomials of the same degree.  This is indeed true for $Y_{0,0,N}$. We will now generalize the formula (\ref{psiw}) to a generating function $\psi (u)$ that parametrize generic representations for all $Y_{N_1,N_2,N_3}$. In particular, we conjecture that the complicated variety parametrizing modules of the algebra $Y_{N_1,N_2,N_3}$ can be simply parametrized. 

Such a parametrization of the variety of highest weights is natural the from point of view of the coproduct structure of the affine Yangian, but also from free field realization viewpoint and the gauge theory perspective. After stating these motivations, we write down an explicit formula for the generating function of $\psi_i$ charges for arbitrary $Y_{N_1,N_2,N_3}$ in \ref{psihw}. A parametrization of the variety of highest weights can be recovered after changing the variables from the affine Yangian generators $\psi_i$ to the zero modes of $W_i$ generators according to the appendix \ref{primaryquadraticformulas}.

\paragraph{Free field realization}

Both the Miura transformation for $Y_{N_1,N_2,N_3}$ and the definition in terms of a kernel of screening charges give an embedding of the algebras of the form
\begin{equation}
Y_{N_1,N_2,N_2} \subset Y_{N_1,0,0} \times Y_{0,N_2,0} \times Y_{0,0,N_3} \subset Y^{\otimes N_1}_{1,0,0} \times Y^{\otimes N_2}_{0,1,0} \times Y^{\otimes N_3}_{0,0,1}.
\end{equation}
Each factor $Y_{0,0,1}$ in the free field realization above can be identified with one multiplicative factor in (\ref{psiw}). The full free field realization therefore suggests that the generating function of a generic module of $Y_{N_1,N_2,N_3}$ should be simply a product of three $W_N$ factors corresponding to $Y_{N_1,0,0}$, $Y_{0,N_2,0}$ and $Y_{0,0,N_3}$. Note that the parameter $\alpha_0$ remains the same in the fusion procedure. From the formula (\ref{ubasisparam}), we see that this requires $(\lambda_1^{(1)},\lambda_2^{(1)},\lambda_3^{(1)})$ and $(\lambda_1^{(2)},\lambda_2^{(2)},\lambda_3^{(2)})$ to be proportional and the fusion is simply additive in $\lambda$-parameters.

\paragraph{Yangian point of view}
Using the map between $\mathcal{W}_{1+\infty}$ modes and Yangian generators \cite{Prochazka:2015aa}, we can translate the fusion to Yangian variables. The coproduct of $\psi_j$ generators with $j \geq 3$ is no longer a finite linear combination of other generators and their products, but involves an infinite sum. This is related to the non-local terms that enter the map between VOA description and the Yangian description. Fortunately, when acting on a highest weight state (corresponding to a primary field via the operator-state correspondence) these additional terms drop out and we obtain a simple formula
\begin{equation}
\psi(u) = \psi^{(1)}(u) \psi^{(2)}(u)
\end{equation}
analogous to the usual ones in finite Yangians.\footnote{Since the Yangian has a non-trivial automorphisms, like the spectral shift automorphism translating the parameter $u$, we can precompose this with the coproduct if needed to obtain slightly more general coproducts. This is actually what is needed if we want the fusion of two vacuum representations to produce a vacuum representation.} This coproduct of the affine Yangian also suggests a simple form of the generating function in terms of a product of three $W_{N_i}$ factors associated to each corner. The compatibility of parameters in this case requires that $h_1, h_2$ and $h_3$ parametrizing the algebra are the same while the $\psi_0$ is additive under the fusion. In terms of $\lambda$-parameters this is the same condition as found above.

\paragraph{Gauge theory and brane picture}

The gauge theory setup suggests that the modules should be parametrized linearly. The GW parameters that label modules live in the $N_1+N_2+N_3$ dimensional tori (modulo Weyl group) that we expect to be lifted to $\mathbb{C}^{N_1+N_2+N_3}$ by boundary conditions imposed on the GW defect ending at the interfaces. Moreover, this picture suggests that generically the contribution from GW-parameters in each corner should be independent.

The coproduct from the point of view of the gauge theory corresponds to increasing the rank of gauge groups in the three corners of the diagram. One can look at it as an inverse process to Higgsing the theory that corresponds to separation of D3-branes and reduces the gauge group. This procedure can be performed in each corner suggesting that the coproduct of $\mathcal{W}_N$ should have a natural generalization for $Y_{N_1,N_2,N_3}[\Psi]$. The process is independent on the gauge coupling suggesting that $\Psi$ is constant in agreement with the other pictures discussed above.

\paragraph{Generating function}

The discussion above motivates us to write down an explicit formula for the generating function of $\psi_i$ charges for $Y_{N_1, N_2, N_3}$ acting on the highest weight state by simply multiplying contributions from $\mathcal{W}_N$ algebras from each corner
\begin{equation}
\label{psihw}
\psi(u) = \prod_{j=1}^{N_1} \frac{u-x^{(1)}_{j}-h_1}{u-x^{(1)}_j} \prod_{j_2=1}^{N_2} \frac{u-x^{(2)}_{j}-h_2}{u-x^{(2)}_j} \prod_{j_3=1}^{N_3} \frac{u-x^{(3)}_{j}-h_3}{u-x^{(3)}_j}.
\end{equation}
Note that the expression is manifestly triality invariant, depends on the correct number of parameters and the truncation curves are reproduced correctly. In particular, extracting $\psi_0$ from the expression above, one gets
\begin{eqnarray}
h_1h_2h_3\psi_0=-N_1h_1-N_2h_2-N_3h_3.
\end{eqnarray}
Identifying the scaling-independent combinations\footnote{The algebra is invariant under the simultaneous rescaling of $\psi_0$ and $h_i$, see \cite{tsymbaliuk2017affine,Prochazka:2015aa}.}
\begin{eqnarray}
\lambda_1=-\psi_0 h_2 h_3,\qquad \lambda_2=-\psi_0 h_1 h_3,\qquad \lambda_3=-\psi_0 h_1 h_2,
\end{eqnarray}
one gets the correct expression
\begin{eqnarray}
\frac{N_1}{\lambda_1}+\frac{N_2}{\lambda_2}+\frac{N_3}{\lambda_3}=1
\end{eqnarray}
satisfied by parameters of $Y_{N_1,N_2,N_3}$.

Parameters $x^{(\kappa)}_i$ can be identified with the lifted Gukov-Witten parameters in the third corner. This can be seen from the comparison of the $U(1)$ charge for $Y_{0,0,1}$ and the fact that each multiplicative factor corresponds to one such factor. The unlifted Gukov-Witten parameters themselves can be identified  by modding out by the lattice $h_1 n + h_2 m$ for $n,m\in \mathbb{Z}$. We will later see that that $x^{(3)}_i=h_1 n + h_2 m$ corresponding to the trivial GW defect (and a possibly non-trivial line operator) corresponds to a degenerate module. We will also see that the fusion of a degenerate module with a generic module labeled by a parameter $x^{(3)}$ amounts to a shift of $x^{(3)}$ by a lattice vector.

Note also that the generating function is manifestly invariant under the Weyl group associated to the three gauge groups $U(N_i)$.

\paragraph{Applications}
To illustrate the power of the simple generalization of  (\ref{psihw}) let us discuss one simple application and find primaries of the Ising model once again. The Ising model, being a $c=1/2$ minimal model of Virasoro algebra, lies on two intersection curves. Its $\lambda$-parameters are $(2/3, -1/2, 2)$ and can be thought of simultaneously as a truncation of $Y_{002}$ algebra as well as $Y_{210}$ algebra. Choosing the $h_j$ parameters to be integers,
\begin{equation}
h_1 = 3, \quad\quad\quad h_2 = -4, \quad\quad\quad h_3 = 1
\end{equation}
and $\psi_0 = 1/6$ so that (\ref{lambdatoh}) holds, the formula (\ref{psihw}) implies that it should be possible to write $\psi(u)$ as product of two zero-pole pairs separated by distance $h_3=1$ ($Y_{002}$ point of view) or alternatively as two zero-pole pairs separated by distance $h_1=3$ and one zero-pole pair separated by distance $h_2=-4$ ($Y_{210}$ point of view). Up to an overall translation in the $u$-space (spectral shift) there are only three possible solutions:
\begin{equation}
\psi_0(u) = \frac{u-2}{u}, \quad\quad \psi_{1/16}(u) = \frac{(u-1/2)(u-5/2)}{(u-3/2)(u+1/2)}, \quad\quad \psi_{1/2}(u) = \frac{(u-4)(u+1)}{(u-3)(u+2)}.
\end{equation}
Extracting the conformal dimensions \cite{tsymbaliuk2017affine,Prochazka:2015aa}, we find them to be $h=0$, $h=1/16$ and $h=1/2$ which are exactly the conformal dimensions of the Ising model.

\subsection{Relation to the free boson modules}

The parameters $x^{(\kappa)}_i$ from the generating function of $\psi_i$ charges that have been already related to the Gukov-Witten parameters can be also related to exponents in the expression for the vertex operators in free field realization. A highest weight vector in free field representation with generic charges can be obtained by acting on the vacuum state with the vertex operator
\begin{equation}
\ket{q^1,\ldots,q^N} = \exp \left(\sum_{j=1}^N q^j \phi_j \right) \ket{0}.
\end{equation}
Acting on this state with the zero mode of current $J_j = \partial \phi_j$, we find
\begin{equation}
J_{j,0} \ket{q^1,\ldots,q^N} = g_{jk} q^k \ket{q^1,\ldots,q^N} \equiv q_j \ket{q^1,\ldots,q^N}
\end{equation}
where $g_{jk}$ is the metric extracted from the two-point functions of the currents,
\begin{equation}
J_j(z) J_k(w) \sim \frac{g_{jk}}{(z-w)^2} \sim -\frac{h_{\kappa(j)}}{h_1 h_2 h_3} \frac{\delta_{jk}}{(z-w)^2}.
\end{equation}
Our conventions for charges are such that $q^j$ are the charges that appear in the exponents of vertex operators (and in positions of zeros and poles of $\psi(u)$) while $q_j$ are the coefficients of the first order poles of OPE with currents $J_j$. We reintroduce the $-h_1 h_2$ factors in order to make the expressions manifestly triality invariant and also of definite scaling dimension under the scaling symmetry of the algebra \cite{Prochazka:2015aa}.

The $U(1)$ current of $\mathcal{W}_{1+\infty}$ whose zero mode is $\psi_1$ is given by
\begin{equation}
U_1(z) = \sum_{j=1}^N J_j(z)
\end{equation}
so $\psi_1$ acts on the highest weight state by
\begin{equation}
\psi_1 \ket{q^1,\ldots,q^N} = \left(\sum_{j=1}^N q_j\right) \ket{q^1,\ldots,q^N}.
\end{equation}
To find the total stress-energy tensor of $\mathcal{W}_{1+\infty}$, we first use the Miura transform to find the free field representation of $U_2(z)$:
\begin{equation}
U_2(z) = \frac{1}{2} \sum_{j \geq 1} \left(1 - \frac{h_3}{h_{\kappa_j}}\right) \left( (J_j J_j)(z) + h_{\kappa_j} \partial J_j(z) \right) + \sum_{j<k} (J_j J_k)(z) + \sum_{j<k} h_{\kappa_j} \partial J_k
\end{equation}
from which we can find the total $\mathcal{W}_{1+\infty}$ stress-energy tensor
\begin{eqnarray}
T_{1+\infty}(z) & = & -\frac{1}{2} \sum_{j} \frac{h_1 h_2 h_3}{h_{\kappa_j}} (J_j J_j)(z) + \frac{1}{2} \sum_{j<k} h_{\kappa_k} \partial J_j - \frac{1}{2} \sum_{j>k} h_{\kappa_k} \partial J_j
\end{eqnarray}

Let us now Consider one free boson $\phi^{(\kappa)}(z)$ in $\kappa$-th direction associated to elementary Miura factor $R^{(\kappa)}$. It is easy to verify that the state created by the vertex operator
\begin{equation}
:\exp \left( q \phi^{(\kappa)} \right):
\end{equation}
from the vacuum is a highest weight state with the generating function of highest weight charges $\psi(u)$ equal to
\begin{equation}
\psi^{(\kappa)}(u) = \frac{u-q-h_\kappa}{u-q}.
\end{equation}
For a longer chain with more free bosons, we have an analogous product of the corresponding simple factors, but the spectral parameter is shifted between the nodes: $\psi(u)$ corresponding to $Y_{0,0,2}$ with ordering of fields $R(z)=R_1^{(3)}(z) R_2^{(3)}(z)$
\begin{equation}
\psi(u) = \frac{u-q^1-h_3}{u-q^1} \frac{u-q^2-2h_3}{u-q^2-h_3}.
\end{equation}
Analogously, $\psi(u)$ corresponding to $Y_{1,1,0}$ with ordering of fields $R(z)=R_1^{(1)}(z) R_2^{(2)}(z)$ has
\begin{equation}
\psi(u) = \frac{\left(u - q^1 - h_1\right)}{\left(u - q^1\right)} \frac{\left(u - q^2 - h_1 - h_2\right)}{\left(u - q^2 - h_1 \right)}.
\end{equation}
In other words, the Miura factor on the left affects the factors that come on the right of it by shifting the $u$-parameter. The general formula for an arbitrary ordering
\begin{equation}
R(z) = R_1^{(\kappa_1)}(z) \cdots R_{N_1+N_2+N_3}^{(\kappa_{N_1+N_2+N_3})}(z)
\end{equation}
has the generating function of charges equal to
\begin{equation}
\psi(u) = \prod_{j=1}^{N_1+N_2+N_3} \frac{u-q^j - \sum_{k \leq j} h_{\kappa_k}}{u-q^j - \sum_{k < j} h_{\kappa_k}}.
\end{equation}
We see that up to constant shifts and rescalings (depending on ordering of free fields) the zeros and poles of the generating function $\psi(u)$ of highest weight state correspond to zero modes $q^j$ of the free bosons, in particular
\begin{eqnarray}
x_j^{(\kappa(j))}=q^j + \sum_{k < j} h_{\kappa_k}.
\end{eqnarray}

\subsection{Two examples of varieties of highest weights}
\label{examples}

Finally, we are ready to show how the generating function (\ref{psihw}) nicely parametrizes the variety of highest weights in the examples $Y_{1,1,0}$ and $Y_{2,1,0}$ studied in \cite{Wang:1998bt,1207.3909}. Using the free field realization, we can construct all modules (there are no further restrictions on the variety of highest weights). The knowledge of the generating function allows to determine the variety for all the other $Y_{N_1,N_2,N_3}$ examples without the necessity of going through the tedious calculation of the null constraints in the primary basis and translating them to the constraints on the zero modes of the null fields.

\subsubsection{$Y_{1,1,0}$ - singlet algebra of symplectic fermion}

The algebra $Y_{1,1,0}$ is the simplest truncation of $\mathcal{W}_{1+\infty}$ which is not a $\mathcal{W}_N$ algebra, although as we will see, it can be understood as (a simple quotient of) $\mathcal{W}_{3}$ algebra at a special value of the central charge. First of all, the $Y_{1,1,0}$ truncation requires
\begin{equation}
\frac{1}{\lambda_1} + \frac{1}{\lambda_2} = 1
\end{equation}
as well as the usual constraint
\begin{equation}
\frac{1}{\lambda_1} + \frac{1}{\lambda_2} + \frac{1}{\lambda_3} = 0.
\end{equation}
From these constraints, we learn that $\lambda_3 = -1$. Plugging this into the central charge formula, we find
\begin{equation}
c_{\infty} = -2
\end{equation}
independently of the value of $\lambda_1$.

Considering $Y_{1,1,0}$ algebra as truncation of $\mathcal{W}_{1+\infty}$, the first singular vector in the vacuum representation appear at level $4 = 2 \cdot 2 \cdot 1$. Generically, starting from spin $4$ we can use these singular vectors to eliminate the higher spin generators of spin $4, 5, \ldots$, obtaining an algebra that is generated by fields of spins $1$, $2$ and $3$. Therefore we identify $Y_{1,1,0}$ with a quotient of the $\mathcal{W}_3$ algebra at $c=-2$ times a free boson as further discussed at the level of generating functions $\psi(u)$ in appendix \ref{w3quotient}. The OPEs of $\mathcal{W}_3$ are given by the Virasoro algebra coupled to a spin $3$ current which has OPE
\begin{eqnarray}
\nonumber
W_3(z) W_3(w) & \sim & C_{33}^0 \Bigg( \frac{\mathbbm{1}}{(z-w)^6} - \frac{3T(w)}{(z-w)^4} - \frac{3\partial T(w)}{2(z-w)^3} - \frac{4(TT)(w)}{(z-w)^2} \\
& & + \frac{3\partial^2 T(w)}{4(z-w)^2} - \frac{4(\partial T T)(w)}{z-w} + \frac{\partial^3 T(w)}{6(z-w)} \Bigg).
\end{eqnarray}
We kept the normalization of $W_3$ generator free for later convenience. We could absorb the structure constant $C_{33}^0$ by rescaling the $W_3$ generator. 

We are now interested in constraints on generic representations of $Y_{1,1,0}$. From the physical reasoning as well as from the free field representations, we would expect the generic representation of $Y_{1,1,0}$ to be parametrized by two continuous parameters, while the $U(1) \times \mathcal{W}_3$ algebra have in general three highest weights. We thus need to find a singular vector in $\mathcal{W}_3$ that would reduce the number of parameters by one. From the general reasoning, we expect the first relation to appear at level $6$. In fact, there are two singular primaries at level $6$. We can see this by looking at characters: the character of the vacuum representation of $U(1) \times \mathcal{W}_3$ is
\begin{equation}
\prod_{s=1}^3 \prod_{j=0}^{\infty} \frac{1}{1-q^{s+j}} \simeq 1 + q + 3q^2 + 6q^3 + 12q^4 + 21q^5 + 40q^6 + 67q^7 + 117q^8 + \ldots
\end{equation}
while the vacuum representation of $Y_{1,1,0}$ has
\begin{eqnarray}
\label{y110char}
\chi_{vac}(q) & = & \sum_{j=0}^{\infty} \frac{q^j}{\prod_{k=1}^j (1-q^k)^2} = \frac{\sum_{j=0}^{\infty} (-1)^j q^{j(j+1)/2}}{\prod_{k=1}^{\infty} (1-q^k)^2} \\
\nonumber
& \simeq & 1 + q + 3q^2 + 6q^3 + 12q^4 + 21q^5 + 38q^6 + 63q^7 + 106q^8 + 170q^9 + \ldots
\end{eqnarray}
We see that at level $6$ there are two null states in $Y_{1,1,0}$ compared to the situation in $U(1) \times \mathcal{W}_3$ at the generic value of the central charge. The first null state is the even quadratic primary composite field
\begin{equation}
\label{y110null6e}
N_{6e} = (W_3 W_3) + C_{33}^0 \left( \frac{8}{9} (T(TT)) + \frac{19}{36} (\partial T \partial T) + \frac{7}{9} (\partial^2 T T) - \frac{2}{27} \partial^4 T\right)
\end{equation}
and the second one is the odd field
\begin{equation}
\label{y110null6o}
N_{6o} = 8(T \partial W_3) - 12(\partial T W_3) - \partial^3 W_3.
\end{equation}
Requiring that the action of the zero mode of $N_{6o}$ on the generic highest weight state vanishes gives us identical zero while the similar requirement for $N_{6e}$ gives us a non-trivial constraint
\begin{equation}
\label{y110constraint}
0 = w_3^2 + \frac{C_{33}^0}{9} h^2(8h+1).
\end{equation}
This is the constraint we were looking for. It reduces the dimension of the space of generic primaries from three to two which is in accordance with what we expect. In principle, we could proceed further by studying the singular vectors at higher levels and possibly discover new (independent) constraints. In order to show that (\ref{y110constraint}) is necessary and sufficient, we will construct a free field realization of $Y_{1,1,0}$ and check that the generic modules can indeed by realized.

\paragraph{Free field realization}

From the general fusion ideology we expect that $Y_{1,1,0} \subset Y_{1,0,0} \times Y_{0,1,0}$, i.e. that there exists a representation of $Y_{1,1,0}$ in terms of two $U(1)$ currents $J_1$ and $J_2$ with OPE
\begin{equation}
J_j(z) J_k(w) \sim \frac{\delta_{jk}}{(z-w)^2}.
\end{equation}
With this normalization, we are still free to make $O(2)$ rotations in the space of free bosons so we may with no loss of generality align the $U(1)$ current of $Y_{110}$ to be in $J_1+J_2$ direction, \footnote{In this section we are temporarily using a different normalization of $U(1)$ currents than in the rest of the paper.}
\begin{equation}
J = J_1 + J_2.
\end{equation}
Denoting the normalized orthogonal combination
\begin{equation}
J_- \equiv \frac{1}{\sqrt{2}} (J_1-J_2),
\end{equation}
the unique stress-energy tensor $T_{\infty}(z)$ commuting with $J(z)$ and with central charge $c_{\infty} = -2$ is
\begin{equation}
T_{\infty}(z) = \frac{1}{2} (J_-,J_-)(z) + \frac{1}{2} \partial J_-(z).
\end{equation}
We can also find one spin $3$ primary field commuting with $J$,
\begin{equation}
W_3(z) = (J_-(J_-J_-))(z) + \frac{3}{2} (J_- \partial J_-)(z) + \frac{1}{4} \partial^2 J_-(z).
\end{equation}
The normalization coefficient is now $C_{33}^0 = -9$. We can verify that there are no fields other than descendants of the identity in the OPE of $W_3$ current with itself and also that the dimension $6$ singular primaries vanish identically. Note that we did not need to require $c_\infty=-2$ and the requirement of existence of spin $3$ primary constructible from $J_-$ would force us to choose $c_\infty=-2$ anyway.

Consider now the highest weight representation of the $U(1) \times U(1)$ algebra such that the $J_-$ charge is $q_-$. We find that the conformal dimension with respect to $T_{\infty}$ and the spin $3$ charge of $W_3$ are
\begin{equation}
h_\infty = \frac{q_-(q_--1)}{2}, \quad\quad\quad w_3 = \frac{q_-(q_--1)(2q_--1)}{2}
\end{equation}
and the relation (\ref{y110constraint}) is satisfied if and only if $C_{33}^0 = -9$ which is indeed the case. This means that all the generic representations of $Y_{1,1,0}$ with (\ref{y110constraint}) are realizable in terms of two free bosonic currents.

\paragraph{Free field realization from Miura}

Let us see what free field representation we find by applying the Miura transformation explained above. The total Miura operator is a product of two basic Miura factors associated to first and second asymptotic direction
\begin{eqnarray}\nonumber
R(z) & = & \partial^{-1} + U_1(z) \partial^{-2} + U_2(z) \partial^{-3} + \ldots \\ \nonumber
& = & \left[ \mathbbm{1} + U_1^{(1)} (\alpha_0 \partial)^{-1} + U_2^{(1)} (\alpha_0 \partial)^{-2} + \ldots \right] (\alpha_0 \partial)^{h_1/h_3} \times \\
& & \times \left[ \mathbbm{1} + U_1^{(2)} (\alpha_0 \partial)^{-1} + U_2^{(2)} (\alpha_0 \partial)^{-2} + \ldots \right] (\alpha_0 \partial)^{h_2/h_3}
\end{eqnarray}
By commuting the derivatives to the right we find
\begin{eqnarray}
\nonumber
U_1 & = & U_1^{(1)} + U_1^{(2)} \\
U_2 & = & U_2^{(1)} + U_2^{(2)} + U_1^{(1)} U_1^{(2)} + h_1 \partial U_1^{(2)} \\
\nonumber
U_3 & = & U_3^{(1)} + U_3^{(2)} + U_1^{(1)} U_2^{(2)} + U_2^{(1)} U_1^{(2)} + (h_1-h_3) U_1^{(1)} \partial U_1^{(2)} \\
\nonumber
& & + h_1 \partial U_2^{(2)} + \frac{h_1(h_1-h_3)}{2} \partial^2 U_1^{(2)}
\end{eqnarray}
Plugging in expressions for $U_j$ in terms of free bosons, we find
\begin{eqnarray}
\nonumber
U_1 & = & J_1 + J_2 \\
\nonumber
U_2 & = & \frac{2h^2-1}{2h^2} (J_1 J_1) + (J_1 J_2) + \frac{2-h^2}{2} (J_2 J_2) + \frac{2h^2-1}{2h} \partial J_1 + \frac{3h^2-2}{2h} \partial J_2 \\
\nonumber
U_3 & = & \frac{(2h^2-1)(3h^2-2)}{6h^4} (J_1(J_1 J_1)) + \frac{2h^2-1}{2h^2} (J_1(J_1 J_2)) - \frac{h^2-2}{2} (J_1(J_2 J_2)) \\
& & + \frac{(h^2-2)(2h^2-3)}{6} (J_2(J_2 J_2)) + \frac{(2h^2-1)(3h^2-2)}{2h^3} (\partial J_1 J_1) \\
\nonumber
& & + \frac{2h^2-1}{2h} (\partial J_1 J_2) + \frac{5h^2-4}{2h} (J_1 \partial J_2) - \frac{(h^2-2)(4h^2-3)}{2h} (\partial J_2 J_2) \\
\nonumber
& & + \frac{(2h^2-1)(3h^2-2)}{6h^2} \partial^2 J_1 + \frac{11h^4-16h^2+6}{6h^2} \partial^2 J_2
\end{eqnarray}
in the normalization
\begin{equation}
J_1(z) J_1(w) \sim -\frac{h_1/h_1 h_2 h_3}{(z-w)^2}, \quad\quad\quad J_2(z) J_2(w) \sim -\frac{h_2/h_1 h_2 h_3}{(z-w)^2}
\end{equation}
and with conventions in (\ref{hjtohparam}). There is an infinite number of non-zero $U_j$ operators with $j \geq 4$ but they can all be read off from OPE of $U_j$ fields with $j=1,2,3$. Finally using the transformations of appendix \ref{primaryquadraticformulas} we find in the primary basis
\begin{eqnarray}
\nonumber
W_1 & = & -J_1 - J_2 \\
\nonumber
W_2 & = & \frac{1}{2h^2} (J_1 J_1) + (J_1 J_2) + \frac{h^2}{2} (J_2 J_2) - \frac{1}{2h} \partial J_1 - \frac{h}{2} \partial J_2 \\
W_3 & = & -\frac{h^2+1}{3h^4} (J_1(J_1 J_1)) - \frac{h^2+1}{h^2}(J_1(J_1 J_2)) - (h^2+1)(J_1(J_2 J_2)) \\
\nonumber
& & - \frac{h^2(h^2+1)}{3} (J_2(J_2 J_2)) + \frac{h^2+1}{2h^3} (\partial J_1 J_1) + \frac{h^2+1}{2h} (\partial J_1 J_2) \\
\nonumber
& & + \frac{h^2+1}{2h} (J_1 \partial J_2) + \frac{h(h^2+1)}{2} (\partial J_2 J_2) - \frac{h^2+1}{12h^2} \partial^2 J_1 - \frac{h^2+1}{12} \partial^2 J_2
\end{eqnarray}
with all other $W_j$ currents, $j \geq 4$ vanishing (as they should). To compare to the previous discussion, where the current $J$ was chosen to be $J_1+J_2$ with unit normalization and $T$ and $W_3$ were expressed in terms of the orthogonal combination, if we choose the orthogonal combination to be the current
\begin{equation}
J_- = -h^{-1} J_1 - h J_2
\end{equation}
we exactly reproduce the formulas of the previous section up to an overall normalization.

\subsubsection{$Y_{0,1,2}$ - Parafermions}
Another interesting truncation of $\mathcal{W}_{1+\infty}$ is the chiral algebra of parafermions $Y_{0,1,2}$. Bootstrap construction of $\mathcal{W}$-algebras generated by primaries of spin $3,4,5$ was carried out in \cite{Hornfeck:1992tm}, where two solutions were found. The first one is the standard $\mathcal{W}_5$ algebra. There exists one more solution corresponding to $Y_{0,1,2}$ with a singular vectors starting at level $8$ that need to be factorized in order the bootstrap equations to be satisfied\footnote{Coset representation of $Y_{0,1,2}$ as a quotient $SU(2)/U(1)$ was discussed in \cite{Blumenhagen:1994wg}. The algebra was further studied in \cite{dong2009w}.}.

Let us consider $\mathcal{W}$-algebra generated by fields of spin $3$, $4$ and $5$. We can also assume the $\mathbb{Z}_2$ symmetry under which the spin $3$ and $5$ fields are odd. The ansatz for operator product expansions of primary families is
\begin{eqnarray}
\nonumber
W_3 W_3 & \to & C_{33}^0 \mathbbm{1} + C_{33}^4 W_4 \\
\nonumber
W_3 W_4 & \to & C_{34}^3 W_3 + C_{34}^5 W_5 \\
\nonumber
W_4 W_4 & \to & C_{44}^0 \mathbbm{1} + C_{44}^4 W_4 + C_{44}^{(33)} [W_3 W_3] \\
W_3 W_5 & \to & C_{35}^4 W_4 + C_{35}^{(33)} [W_3 W_3] \\
\nonumber
W_4 W_5 & \to & C_{45}^3 W_3 + C_{45}^5 W_5 + C_{45}^{(34)} [W_3 W_4] + C_{45}^{(34)^\prime} [W_3 W_4]^\prime \\
\nonumber
W_5 W_5 & \to & C_{55}^0 \mathbbm{1} + C_{55}^4 W_4 + C_{55}^{(33)} [W_3 W_3] + C_{55}^{(33)^{\prime\prime}} [W_3 W_3]^{\prime\prime} \\
\nonumber
& & + C_{55}^{(44)} [W_4 W_4] + C_{55}^{(35)} [W_3 W_5] + C_{55}^{(35)^\prime} [W_3 W_5]^\prime
\end{eqnarray}
We only list the primary fields appearing in the OPE, the coefficients of descendants are always fixed by the Virasoro algebra. We also denote the primary composites by brackets, e.g. $[W_3 W_3]$ denotes the primary operator which is the leading regular term of $W_3 W_3$ OPE after subtracting descendants of primaries appearing in the singular part of the OPE. Analogously for the composites involving derivatives which we extract from subleading regular terms, i.e.
\begin{equation}
[W_3 W_4]^\prime(z) = -\frac{3}{7} (W_3 \partial W_4)(z) + \frac{4}{7} (\partial W_3 W_4)(z) + \ldots.
\end{equation}
We can conveniently extract these primaries using the function \textsf{OPEPPole} of $\textsf{OPEdefs}$ \cite{Thielemans:1991uw} which automatically performs the primary projection. Our next goal would be to fix the $20$ free coefficients appearing in the ansatz for OPE using the Jacobi identities. Not all of these coefficients can actually be fixed, because we still have the freedom to change the normalization of the generators, i.e. we have $3$-parametric gauge freedom. Since the first primary composite appears at spin $6$ while our algebra is generated by primaries of spins $3, 4$ and $5$, there are no additional redefinitions possible. We fix our conventions such that the coefficients $C_{33}^0$, $C_{33}^4$ and $C_{34}^5$ remain undetermined and we express all the remaining structure constants fixed by Jacobi identities in terms of these three constants and the central charge $c$. With this spin content, there are two algebras that solve the Jacobi identities. One of them is the algebra $\mathcal{W}_5$ which is freely generated by primaries of spin $3, 4$ and $5$. The other one is $Y_{012}$ which has for generic $c$ two singular vectors at level $8$ so the Jacobi identities are satisfied only up to these singular vectors. As a consequence of this, the constants $C_{45}^{(34)^\prime}, C_{55}^{(33)^{\prime\prime}}$ and $C_{55}^{(35)^\prime}$ are indeterminate. The expressions for other structure constants are given in appendix \ref{appendixy012}.

We are interested in the generic highest weight modules of $Y_{012}$. From the general discussion of the fusion procedure, we expect these to be parametrized by $3$ complex numbers. Including the $U(1)$ degree of freedom, we have $5$ generators of $Y_{012}$ with spin $1$ to $5$. We thus need to find two constraints that reduce the space of highest weight representations to expected $3$-dimensional subvariety.

The first constraint appears at level $8$. There are two singular vectors at this level, one even and one odd under $\mathbbm{Z}_2$ symmetry. The even spin $8$ field is the linear combination
\begin{eqnarray}
\nonumber
\mathcal{N}_{8e} & = & [W_3 W_3]^{\prime\prime} + \frac{3c(2c-1)(7c+114)C_{33}^4 C_{34}^5}{8(c+10)(29c^2+533c-870) C_{33}^0} [W_3 W_5] \\
& & - \frac{c(c+7)(2c-1)(7c+114) (C_{33}^4)^2}{2(c+10)^2)(29c^2+533c-870)C_{33}^0} [W_4 W_4].
\end{eqnarray}
Its zero mode when acting on the highest weight state gives us a constraint
\begin{eqnarray}
\label{y012rel1}
\nonumber
0 & = & 72h^2(c+10)^2(7c+114)(56h-3c-2)(17ch+2h-c^2-2c-4) C_{33}^0 \\
\nonumber
& & - 36c(c+10)^2(2c-1)(5c+22)^2(-31ch+218h+c^2-2c+12) w_3^2 \\
\nonumber
& & - 12ch(c+10)(2c-1)(5c+22)(7c+114) (-22ch-388h+c^2+19c+6) C_{33}^4 w_4 \\
& & -4c^2(c+7)(2c-1)^2(5c+22)^2(7c+114) \frac{(C_{33}^4)^2}{C_{33}^0} w_4^2 \\
\nonumber
& & + 3c^2(c+10)(2c-1)^2(5c+22)^2(7c+114) \frac{C_{33}^4 C_{34}^5}{C_{33}^0} w_3 w_5
\end{eqnarray}
on higher spin charges of the highest weight state. The odd spin $8$ singular field is $[W_3 W_4]^\prime$ and its action on the highest weight state is identically zero. This means that to find the second relation of the variety of highest weights we need to look at level $9$. Here we have again one odd and one even field. This time the even field $[W_3 W_5]^\prime$ gives identical zero if we act with its zero mode on the highest weight state. On the other hand, the odd field
\begin{eqnarray}
\nonumber
\mathcal{N}_{9o} & = & [W_3 W_3 W_3] + \frac{(c+13)(45c^2+1214c-2832) C_{33}^4}{18(c+10)(c+24)(7c+114)} [W_3 W_4]^{\prime\prime} \\
& & - \frac{c(2c-1) (C_{33}^4)^2 C_{34}^5}{54(c+10)^2 C_{33}^0} [W_4 W_5]
\end{eqnarray}
gives us the second algebraic relation
\begin{eqnarray}
\label{y012rel2}
\nonumber
0 & = & -36h(10+c)^2(1288c^2h^2+44264ch^2+3376h^2-117c^3h-4063c^2h-6392ch \\
\nonumber
& & -7468h+4c^4+118c^3+380c^2+464c+288) C_{33}^0 w_3 \\
& & +54c(c+7)(c+10)^2(2c-1)(5c+22)(7c+114) w_3^3 \\
\nonumber
& & -c^2(c+7)(2c-1)^2(5c+22)(7c+114) \frac{(C_{33}^4)^2 C_{34}^5}{C_{33}^0} w_4 w_5 \\
\nonumber
& & +6c(c+7)(c+10)(2c-1)(5c+22)(-314ch-3668h+9c^2+117c+138) C_{33}^4 w_3 w_4 \\
\nonumber
& & -3ch(c+7)(c+10)(2c-1)(7c+114)(-56h+2c+2) C_{33}^4 C_{34}^5 w_5.
\end{eqnarray}
Although the result is complicated, we achieved what we wanted: we reduced the dimension of the variety of possible higher spin charges from $(2+1)(1+1)-1 = 5$ to $2+1 = 3$ using two constraints coming from the singular vectors. In the next section, using an explicit free field representation of $Y_{0,1,2}$ we will actually check that there exists a three-parametric family of primaries whose higher spin charges satisfy our constraints, so there cannot be any other constraints on higher spin charges that could reduce the dimension further.

\paragraph{Free field representation}

We can construct a free field representation of $Y_{0,1,2}$ in terms of three bosons using the Miura operators. We will choose two Miura operators in 3rd direction and one associated to the second direction with the ordering
\begin{eqnarray}
R(z) & = & R_1^{(3)}(z) R_{2}^{(3)}(z) R_3^{(2)}(z) \\
\nonumber
& = & (\alpha_0 \partial + J_1(z)) (\alpha_0 \partial + J_2(z)) (\mathbbm{1} + U_1^{(2)} \alpha_0 \partial + \ldots) (\alpha_0 \partial)^{\frac{h_2}{h_3}}
\end{eqnarray}
The advantage of this ordering is that the differential operators on the left which we need to pass to the right are just first order, so it is simpler than other orderings. We use the same normalization as in the general discussion of Miura transformation. Commuting the derivatives to the right, we find for the first three $U$-currents
\begin{eqnarray}
\nonumber
U_1 & = & J_1 + J_2 + J_3 \\
\nonumber
U_2 & = & (J_1 J_2) + (J_1 J_3) + (J_2 J_3) + \frac{2-h^2}{2} (J_3 J_3) - \frac{h^2-1}{h} \partial J_2 - \frac{3h^2-2}{2h} \partial J_3 \\
\nonumber
U_3 & = & (J_1 (J_2 J_3)) + \frac{2-h^2}{2} (J_1 (J_3 J_3)) + \frac{2-h^2}{2} (J_2 (J_3 J_3)) \\
& & + \frac{(h^2-2)(2h^2-3)}{6} (J_3 (J_3 J_3)) - \frac{h}{2} (J_1 \partial J_3) - \frac{h}{2} (J_2 \partial J_3) \\
\nonumber
& & - \frac{h^2-1}{h} (\partial J_2 J_3) + \frac{(h^2-2)(2h^2-1)}{2h} (\partial J_3 J_3) + \frac{2h^2-1}{6} \partial^3 J_3
\end{eqnarray}
All the higher $U_j$ currents are non-zero but they can be calculated in a straightforward way by calculating the OPEs of $U_j$ with $j \leq 3$ (the OPEs of $U_j$ currents are those of $\mathcal{W}_{1+\infty}$ as discussed in \cite{Prochazka:2014aa}). If we are interested in primary fields, we can use the formulas given in appendix \ref{primaryquadraticformulas} or apply directly the orthogonalization procedure and the result for $W_j$ generators is
\begin{eqnarray}
\nonumber
W_1 & = & -J_1 - J_2 - J_3 \\
\nonumber
W_2 & = & \frac{h^2}{2(2h^2-1)} (J_1 J_1) + \frac{h^2}{2(2h^2-1)} (J_2 J_2) + \frac{(h^2-1)^2}{2h^2-1)} (J_3 J_3) \\
\nonumber
& & - \frac{h^2-1}{2h^2-1} (J_1 J_2) - \frac{h^2-1}{2h^2-1} (J_1 J_3) - \frac{h^2-1}{2h^2-1} (J_2 J_3) \\
& & - \frac{h}{2} \partial J_1 + \frac{h^2-2}{2h} \partial J_2 + \frac{h^2-1}{h} \partial J_3
\end{eqnarray}
The current $W_3$ has already quite a long expression which we don't need to write explicitly. It can be checked that $W_j$ generators of spin $6$ and higher are identically zero when expressed in terms of this free field representation. This is something that was expected to hold more generally from the discussion of $Y_{N_1, N_2, N_3}$ algebras and their singular vectors.

Acting on the highest weight state, the eigenvalues of $W_j$ zero modes are
\begin{eqnarray}
\nonumber
w_1 & = & -q_1-q_2-q_3 \\
\nonumber
w_2 & = & \frac{h^2}{2(2h^2-1)} q_1^2 + \frac{h^2}{2(2h^2-1)} q_2^2 + \frac{(h^2-1)^2}{2h^2-1} q_3^2 - \frac{h^2-1}{2h^2-1} q_1 q_2 \\
& & - \frac{h^2-1}{2h^2-1} q_1 q_3 - \frac{h^2-1}{2h^2-1} q_2 q_3 + \frac{h}{2} q_1 - \frac{h^2-2}{2h} q_2 - \frac{h^2-1}{h} q_3
\end{eqnarray}
The spin $3$ charges is given in the appendix \ref{appendixy012}. To verify the identities one needs also $w_4$ and $w_5$ charges, but their expressions are too long. What is important is that plugging these explicit formulas in equations (\ref{y012rel1}) and (\ref{y012rel2}), we find that they are identically satisfied (for any values of $q_j$). This means that we have an explicit parametrization of the variety of highest weights of allowed primary charges in terms of $3$ free boson charges. This parametrization linearizes the variety of highest weights, but is not one-to-one. For example, as we will see later, for the vacuum representation there are $(2+1)! = 6$ choices for $q_j$ charges which give vanishing $w_j$ charges. The situation is analogous to parametrization of the characteristic polynomial of a matrix in terms of its eigenvalues or parametrization of the Casimir elements of $\mathfrak{gl}(N)$ in terms of eigenvalues of the Cartan generators.

\section{Degenerate modules}

\subsection{Surface defects preserving Levi subgroups}

A generic Gukov-Witten defect breaks the gauge group at the defect to the maximal torus $U(1)^{N}$, but a larger symmetry group can be preserved if the GW-parameters are specialized. In particular, if the parameters $x^{(\kappa)}_i$ and $x^{(\kappa)}_j$ specifying the singularity of the $i$th and $j$th factors are equal $x^{(\kappa)}_i=x^{(\kappa)}_j$ (modulo the lattice), the next-to minimal Levi subgroup $U(2)\times U(1)^{N-2}$ is preserved by the configuration. On the VOA side, these specializations are going to correspond to degenerate modules. For a fixed value of the specialized GW parameters, one can still turn on a Wilson and 't Hooft oporator in some representation of the preserved $U(2)$ at each boundary. Different choice of the line operators will label different degenerate modules. Similarly, if parameters in different corners are specialized, $U(1|1)$ supergroup is preserved at the boundary Chern-Simons theory by the defect and one gets different classes of degenerate modules as we will see below.

We can see that the parameter space parametrizing generic modules is divided into domains with a degeneration appearing at the boundaries of the domains. At intersections of such domain walls (where more parameters are specialized), we expect further degeneration of the module. These more complicated representations correspond to larger Levi subgroups decorated by line operators in a representation of the preserved Levi subgroup on the gauge theory side. 

A maximal degeneration appears when $N-1$ parameters are specialized and the full gauge group $U(N)$ is preserved at the defect. Note that the value of the overall $U(1)$ charge does not affect the structure of modules and breaking of the gauge symmetry. On the other hand, maximally degenerate modules with generic values of the $U(1)$ charge still correspond to a nontrivial GW defect with a prescribed singularity for the $U(1)$ factor. Modules associated to line operators with a trivial GW defect correspond to maximal specializations of all the $N$ parameters with quantized values of the $U(1)$ charge.

\subsection{Minimal degenerations}

Let us start with the analysis of domain walls of minimal degenerations associated to the next-to-minimal Levi subgroup.

As we discussed in connection with (\ref{psihw}), the $N_1+N_2+N_3$ lifted GW parameters $x^{(\kappa)}_i$ correspond to positions of poles of the generating function $\psi(u)$ in the $u$-plane. The poles are determined up to a permutation of order of poles in each group. A natural question to ask is for which values of parameters $x_j^{(\kappa_j)}$ do we obtain a degenerate module.

\subsubsection{Singular vectors at level 1}
The discussion is easy at the first level. A generic module has $N_1+N_2+N_3$ states at this level. We can detect the appearance of a singular vector by studying the rank of the Shapovalov form
\begin{equation}
\bra{hw} f_k e_j \ket{hw} = - \bra{hw} \psi_{j+k} \ket{hw}
\end{equation}
(where we used the basic commutation relation between $e_j$ and $f_j$ generators of $Y$). The matrix on the right is a Hankel matrix and we can use a variant of the basic theorem by Kronecker which tells us that (in general infinite dimensional) Hankel matrix has a finite rank if and only if the associated generating function
\begin{equation}
\sum_j \psi_j z^j
\end{equation}
is a Taylor expansion of a rational function. Furthermore, the rank of the Hankel matrix is equal to one plus the degree of this rational function. In our case we have a slightly different version of this theorem because the coefficients $\psi_j$ are Taylor coefficients of
\begin{equation}
\frac{\psi(u)-1}{\sigma_3}
\end{equation}
but the result is the same: the number of vectors at level $1$ in the irreducible module with highest weight charges $\psi(u)$ is equal to the degree (i.e. number of zeros counted with multiplicities) of $\psi(u)$. This is automatically consistent with the form of the generating function (\ref{psihw}) which has generically $N_1+N_2+N_3$ zeros and poles. In this way we also rederive the result of \cite{Prochazka:2015aa} that the vacuum representation has exactly one zero and one pole. The distance between them is fixed by the parameters of the algebra. The absolute position of the zero in $u$-plane is determined by $U(1)$ charge of the highest weight vector and is translated under the spectral shift transformation.

This also refines the statement it \cite{Prochazka:2015aa} that the representation is quasi-finite (i.e. has only a finite number of states at each level) if and only if $\psi(u)$ is a rational function. In the case of $Y_{N_1,N_2,N_3}$ the quasi-finiteness is automatically satisfied.

Applying the results of the previous discussion to the highest weight vector of the generic $Y_{N_1,N_2,N_3}$ module with weights parametrized by (\ref{psihw}), we conclude that we have a singular vector at level $1$ if one of the following conditions is satisfied
\begin{equation}
x^{(\tau)}_j - x^{(\sigma)}_k = -h_{\tau},
\end{equation}
i.e. a zero of type $j$ collides with a pole of type $k$.

\subsubsection{Higher levels}
At higher levels the discussion is not so simple because the commutation relations used to evaluate the ranks of Shapovalov matrices become more involved. But from the structure of the Shapovalov matrices, we expect the highest singular vectors to appear only if the distance between a zero and a pole of (\ref{psihw}) is an integer linear combination of $h_j$ parameters. If this assumption of locality (i.e. pairwise interaction between zeros and poles) is satisfied, we can learn more about the relation between the level where such a singular vector appears and the corresponding distance between the zero-pole pair. It is then enough to look at the case of the zero-pole pair of the same type in the algebra $Y_{0,0,2}$ and of different type in the case of $Y_{1,1,0}$.

\subsubsection{Virasoro algebra}
The first case is simple - we are interested in singular vectors of the Virasoro algebra for which we have a known classification: for generic values of the central charge the Verma module has a singular vector at level $rs$ if and only if the highest weight equals $\Delta_{r,s}$ \cite{francesco2012conformal}. The generating function of charges $\psi(u)$ is
\begin{equation}
\psi(u) = \frac{(u-x_1^{(3)}-h_3)(u-x_2^{(3)}-h_3)}{(u-x_1^{(3)})(u-x_2^{(3)})}
\end{equation}
We can extract the conformal dimension $\Delta$ with respect to the $T_{\infty}$ Virasoro subalgebra (decoupled from the $U(1)$ field)
\begin{equation}
\Delta = \frac{h_3^2 - \left(x^{(3)}_1-x^{(3)}_2\right)^2}{4h_1 h_2}.
\end{equation}
This is equal to $\Delta_{r,s}$ if and only if
\begin{equation}
x^{(3)}_1 - x^{(3)}_2 = s h_1 + r h_2, \quad\quad \mathrm{or} \quad\quad x^{(3)}_2 - x^{(3)}_1 = s h_1 + r h_2.
\end{equation}
Therefore, a singular vector of the algebra $Y_{0,0,2}$ appears at level $rs$ if and only if the distance between two poles of the 3rd type is a positive or negative integer linear combination of $h_1$ and $h_2$. Similarly for the other two types of poles.

\subsubsection{$\mathcal{W}_N$ algebras}
The Kac determinant and singular vectors of $\mathcal{W}_N$ are known as well \cite{Mizoguchi:1988vk,Bouwknegt:1992wg}. The singular vectors (zeros of the Kac determinant) at level $rs$ (where $r,s \geq 1$ are integers) are labeled by roots of $SU(N)$. Choosing the standard ordering ($J_1$ the leftmost field in the Miura transformation), the equations for vanishing hyperplanes are
\begin{equation}
q^j - q^k + (j-k) h_3 = s h_1 + r h_2
\label{Wwalls}
\end{equation}
where $1 \leq j \neq k \geq N$ label the (positive and negative) roots of $SU(N)$. The poles of $\psi(u)$ are related to $U(1)$ charges $q^j$ (still assuming the standard ordering and using the conventions of (\ref{hjtohparam})) by
\begin{equation}
x^{(3)}_j = q^j + (j-1) h_3
\end{equation}
so we can rewrite the equations for vanishing hyperplanes as
\begin{equation}
\label{wndeg}
x^{(3)}_j - x^{(3)}_k = s h_1 + r h_2.
\end{equation}
This is exactly of the same form as the condition that we found in the case of the Virsoro algebra. We see is that the positive or negative roots in the $\mathcal{W}_N$ language determine which poles of $\psi(u)$ approach each other and the integers $s$ and $r$ determine the distance between these poles, quantized in the units of $h_1$ and $h_2$. Therefore, in the case of $\mathcal{W}_N$, we have an independent confirmation of the fact that the leading singular vectors in degenerate modules correspond to pairwise interactions between poles of $\psi(u)$.

In the gauge theory language, we see that (at least in the case of $\mathcal{W}_N$-algebras) degenerations appear when the GW parameters are specialized in such a way that a next-to-minimal Levi subgroup is preserved. The parameters $r,s$ then label representations of the preserved $SU(2)$ subalgebra associated to the corresponding line operators supported at the two interfaces.

\subsubsection{Algebra $Y_{1,1,0}$}
The remaining elementary case that we need to analyze is $Y_{1,1.0}$. In this case, the parameter space of generic modules is two-dimensional, so after decoupling the overall $U(1)$, we are left with a one-dimensional parameter space. Analogously to the case of the Virasoro algebra, there is no difference between minimally and maximally degenerate modules. We can look for degenerate modules in at least three possible ways: directly studying the Shapovalov form (Kac determinant), using box counting \cite{Prochazka:2015aa,Prochazka:2017qum} or using the BRST construction of the algebra \cite{Gaiotto:2017euk}.

A direct calculation (which we explicitly checked up to level 4) leads to the following condition: given $n \geq 1$, we have a leading singular vector at level $n$ if
\begin{equation}
\label{y110deg}
x^{(1)}_1 - x^{(2)}_2 = -h_1 - n h_3, \quad\quad \mathrm{or} \quad\quad x^{(1)}_1 - x^{(2)}_2 = h_2 + n h_3.
\end{equation}
Note that these two conditions are exchanged if we formally replace $n \leftrightarrow 1-n$. We can thus use only one of the conditions with $n$ running over all integers, but for non-positive values of $n$ the level at which corresponding singular vector appears is $1-n$.

In $Y_{110}$ there is no difference between the maximally degenerate and minimally degenerate modules. For the maximally degenerate modules we can use the box counting (plane partition) interpretation of modules.\footnote{In general the box counting works only for so called covariant modules which have asymptotics made of boxes (tensor products of the fundamental representation). In general it is important to consider a more general class of representations where there are both asymptotic boxes and anti-boxes. Fortunately in the case of $Y_{1,1,0}$ the anti-box in first direction is equivalent to a box in the second direction and vice versa, so the simple box counting picture is applicable.} The maximally degenerate modules of $Y_{110}$ in this picture correspond to plane partitions (with possible asymptotics) which have no box at position $(2,2,1)$. In other words, they can be thought of as pairs of partitions glued together by the first column (assuming for the moment that there is no asymptotics in 3rd direction). The degenerate modules are labeled by two integers, the heights of asymptotic Young diagrams in 1st and 2nd directions. But only the difference of these two integers matters, the modules with the same difference of heights differ only by the overall $U(1)$ charge. Finally, the parameter $n$ appearing in (\ref{y110deg}) can be identified with one plus the difference of the heights of the asymptotic Young diagrams. It is easy to check that this interpretation predicts the correct level of the null vector, the correct irreducible character and the conformal dimension.

Turning on a non-trivial asymptotics in 3rd direction decouples the pair of Young diagrams so the box counting predicts a generic module (i.e. character equal to the square of the free boson character). The conformal dimensions of these modules also don't produce any additional zero of the Shapovalov form, confirming the whole box-counting picture.

The same structure of maximally degenerate modules can also be seen from the BRST analysis of \cite{Gaiotto:2017euk}. In particular, the BRST analysis of the algebra have not found any other degenerate modules and the degenerate ones appear exactly for the above values of generic parameters. From the gauge theory point of view, the value $n$ can be identified with the difference of charges of the $U(1)$ line operators supported at the boundary $1$ and $2$. Turning on the Wilson line operator at the boundary $3$ lifts the degeneration.

\subsubsection{Summary}
The hyperplanes (\ref{wndeg}) and (\ref{y110deg}) that we found (together with their images under the triality) therefore divide the full space of $x_j$ parameters parameterizing the generic modules of $Y_{N_1,N_2,N_3}$ into domains. The points lying on the union of these hyperplanes correspond to degenerate modules while the remaining points label the generic modules. The degree of degeneration of a given module depends on the number of hyperplanes on which the corresponding $x_j$ lie. Translating all $x_j$ by a same constant corresponds to the spectral shift transformation that only changes the $U(1)$ charge of the whole module and in particular does not change the structure of the singular vectors.

\subsubsection{Free field representation of degenerate primaries}

Let us briefly comment on the realization of the degenerate modules of $Y_{1,1,0}$ in a given free field realization. The highest weight primaries of all the representations (including the generic ones) can be realized as simple exponential vertex operators with exponents given by the parameters $q^j$ (related to $x^{(\kappa_j)}_j$ by constant shifts). It turns out that a half of the degenerate modules associated to the degenerations (\ref{y110deg}) can be also realized in terms of a free boson descendant of an exponential vertex operator. For example, in the $ \phi^{(1)}_1 \times \phi^{(2)}_2$ ordering, the modules in the 2nd direction specialized to $n=1$ and $n=2$ can be realized as
\begin{eqnarray}\nonumber
\left (h_2 J^{(1)}_1-h_1 J^{(2)}_2\right )\exp \left [q \phi^{(1)}_1+(q+h_3)\phi^{(2)}_2 \right ],\\
-\frac{1}{2}\left  (\left  (h_2  J^{(1)}_1-h_1 J^{(2)}_2\right )^2-\partial \left (h_2  J^{(1)}_1-h_1 J^{(2)}_2\right )\right )\exp \left [q \phi^{(1)}_1+(q+2h_3)\phi^{(2)}_2\right ].
\end{eqnarray}
Similarly, for any $n>0$, one can realize the corresponding degenerate modules in terms of a level $n$ descendant. The descendants are generally given in terms of Bell polynomials
\begin{eqnarray}
\label{bellpoly}
\sum_{m_1+2m_2+\dots+nm_n=n}\prod_{k=1}^n \frac{1}{m_k!k^{m_k}}\left (\frac{-1}{(k-1)!}\partial^{k-1}J \right )^{m_k}\exp \left [q \phi^{(1)}_1+(q+nh_3)\phi^{(2)}_2 \right ]
\end{eqnarray}
for $J=h_2 J^{(1)}_1-h_1 J^{(2)}_2$. This is analogous to expressions for singular vectors in free field representations of Virasoro algebra which are given in terms of Jack polynomials \cite{mimachi1995singular,Awata:1995np}. In the case of $Y_{110}$ these reduce to Schur polynomials whose special case are the Bell polynomials (\ref{bellpoly}). Higher level specializations will be further discussed in the next section in the context of maximally degenerate representations but note that the issue is present already for the partially degenerate modules associated to specializations of GW parameters at different corners.

\subsection{Maximally degenerate modules}
\label{maxdeg}

In the previous section, we have discussed the general structure of degenerations of Y-algebra modules and concentrated mostly on the minimally degenerate ones. On the other hand, we will now discuss briefly free field realization of the maximally degenerate modules associated to line operators supported at the interfaces, i.e. trivial GW defects. These modules play an important role in the gluing construction that allows to engineer more complicated VOAs by extensions of tensor products of $Y_{N_1,N_2,N_3}$ algebras \cite{Gaiotto:2017euk,Prochazka:2017qum,Feigin:2018bkf}.

\subsubsection{Identity, box and anti-box}

In this section, we mostly concentrate on the free field realization of the identity operator together with the modules associated to the line operators in the fundamental and the anti-fundamental representation. All the other maximally degenerate representations can be obtained from the fusion of these two (and a shift of $U(1)$ charge). We will further restrict to the case when $N_3=0$. The general case is a bit more complicated because of the appearance of continuous families of free field realizations. We will briefly comment on this issue later. Let us start with writing down the generating function $\psi(u)$ for such representations.

The generating function for the vacuum representation has a single factor
\begin{eqnarray}
\psi_{\bullet}(u)=\frac{u+h_1 h_2 h_3 \psi_0}{u} = \frac{u - N_1 h_1 - N_2 h_2 - N_3 h_3}{u}
\label{psivac}
\end{eqnarray}
where we used the identity
\begin{eqnarray}
h_1 h_2 h_3 \psi_0 =-N_1h_1-N_2h_2-N_3h_3.
\end{eqnarray}
On the other hand the generating function for the fundamental representation in the first direction can be written as
\begin{eqnarray}
\psi_{\square_1}(u)=\frac{(u+h_1 h_2 h_3 \psi_0)(u+h_1)}{(u-h_2)(u-h_3)}
\label{psibox}
\end{eqnarray}
and similarly for the fundamental representation in the other two directions \cite{Prochazka:2015aa,Prochazka:2017qum}.

The generating function of the anti-fundamental representation can be obtained from the formula for the generating function $\psi(u)$ of a conjugate representation \cite{Gaberdiel:2017hcn,Prochazka:2017qum}
\begin{equation}
\bar{\psi}(u)=\psi^{-1}\left(-u-h_1 h_2 h_3 \psi_0\right).
\end{equation}
This is a composition of the inverse anti-automorphism and the reflection in spectral parameter and produces an automorphism just as in the case of finite Yangians. The additional spectral shift is necessary in order to have self-conjugate vacuum representation. It is easy to verify that the effect of conjugation is to flip the sign of all odd primary highest weight charges. Note that there exists a conjugation automorphism of the whole affine Yangian (not just acting on the highest weight state), but the $\psi_j$ generators transform in a more complicated way, mixing with $e_j$ and $f_j$ generators.

Applying the conjugation to the generating function (\ref{psibox}), we get the generating function for the anti-fundamental representation
\begin{eqnarray}
\psi _{\bar{\square}_1}=\frac{(u+h_2+h_1 h_2 h_3 \psi_0)(u+h_3+h_1 h_2 h_3 \psi_0)}{u(u-h_1+h_1 h_2 h_3 \psi_0)}
\end{eqnarray}
and similarly for the other two directions.

\subsubsection{Analysis at the level of generating functions}

We will now identify the triple (identity, fundamental representation, anti-fundamental representation) for $Y_{N_1,N_2,0}$ in terms of a specialization of the parameters of the generic generating function
\begin{eqnarray}
\psi(u)&=&\prod_{i=1}^{N_1}\frac{u-x_{i}^{(1)}-h_1}{u-x_{i}^{(1)}} \prod_{i=N_1+1}^{N_1+N_2}\frac{u-x_{i}^{(2)}-h_2}{u-x_{i}^{(2)}}.
\label{generic}
\end{eqnarray}

For such an analysis, it is useful to introduce a diagrammatic picture for zeros and poles in the spectral parameter plane. Each factor in the generating function $\psi(u)$ contributes by a pole at position $x^{(\kappa_i)}_i$ and a zero at  $x^{(\kappa_i)}_i+h_{\kappa_i}$. One can draw such a combination as a diatomic polar molecule (with a circle corresponding to the pole and a full dot corresponding to zero) separated by the distance $h_{\kappa_i}$. In the generating function $\psi(u)$, we have $N_1$ molecules of length $h_1$ and $N_2$ molecules of length $h_2$. To realize the generating function of the vacuum, the fundamental or the anti-fundamental representation, most of the factors of the generic generating function must cancel. At the level of the interaction of molecules, such a cancellation appears when a circular node coincides with a filled node of a different molecule. In our diagrams, we denote such a zero-pole pair by a cross. The study of realizations of various degenerate representations thus translates into the analysis of paths between zeros and poles in the $u$-plane.

\paragraph{Identity representation}

\begin{figure}
  \centering
      \includegraphics[width=0.4\textwidth]{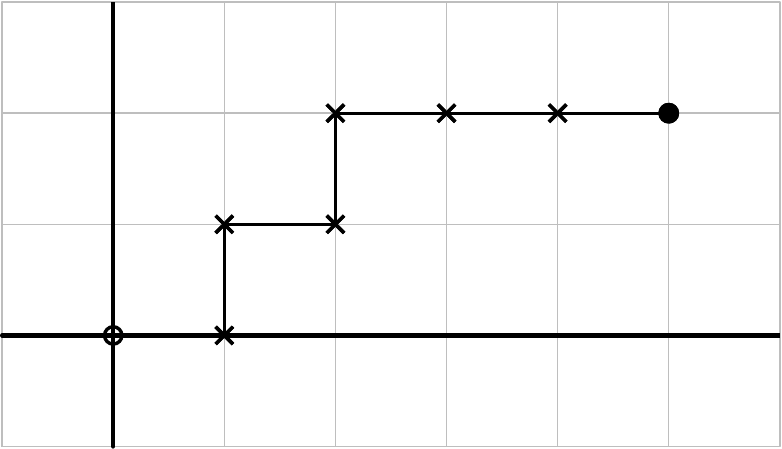}
  \caption{The structure of zeros and poles of the generating function $\psi_0 (u)$ for the vacuum representation of $Y_{5,2,0}[\Psi]$. We draw one example of the snake configuration connecting the zero and the pole.}
\label{vacuum_snake}
\end{figure}

Let us start with a discussion of possible realizations of the identity representation (\ref{psivac}). In the molecular picture, we want to connect the pole at zero with the zero at $N_1 h_1+N_2 h_2$ by $N_1$ steps in the $h_1$ direction and  $N_2$ steps in the $h_2$ direction. There are clearly 
\begin{eqnarray}
\begin{pmatrix}
N_1+N_2\\
N_1
\end{pmatrix}=\frac{(N_1+N_2)!}{N_1!N_2!}
\end{eqnarray}
such paths (note that the molecules of one type are indistinguishable if we mod out by the Weyl group in each corner, i.e. if we identify the permutations of $x_j^{(\kappa_j)}$ with the same value of $\kappa_j$). One such path for the algebra $Y_{5,2,0}$ is drawn in the figure \ref{vacuum_snake}. If we treat the molecules to be distinguishable, we get $(N_1+N_2)!$ solutions which nicely corresponds to $(N_1+N_2)!$ possible orderings in the Miura transformation.

Turning on all three parameters $Y_{N_1,N_2,N_3}$ leads to a more complicated story. In such cases, we expect to obtain continuous families of realizations corresponding to the continuous centre of mass of the triple $(x^{(1)}_i=\alpha,x^{(2)}_j=\alpha+h_1,x^{(3)}_k=\alpha+h_1+h_2)$ whose contribution cancels in any generating function $\psi (u)$ for any choice of $\alpha$. In the picture of molecules, such a factor corresponds to a triangular loop that can be freely moved in the $u$-plane. In general, we get as many of these continuous moduli as is the minimum of $N_j$.

\paragraph{Fundamental representation}

\begin{figure}
  \centering
      \includegraphics[width=0.4\textwidth]{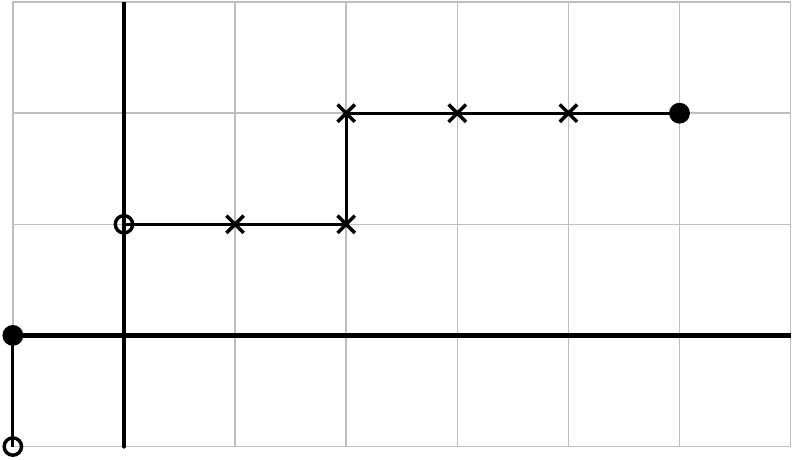}
  \caption{The structure of zeros and poles of the generating function $\psi (u)$ for the fundamental representation in direction $h_3$ of $Y_{5,2,0}[\Psi]$. We draw one example of the snake configuration.}
\label{fundamental_snake}
\end{figure}

Let us now move to the more complicated discussion of the fundamental representation in the direction $h_1$ with the generating function (\ref{psibox}). The discussion of the anti-fundamental representation completely mirrors the fundamental one.  To find the realization in terms of the generating function (\ref{generic}), we need to connect the two poles and two zeros with two snakes composed of $N_1$ molecules of length $h_1$ and $N_2$ molecules of length $h_2$. The only possibility is to connect the pole at $u=h_3$ with the zero at $u=-h_1$ and then draw a snake connecting the other zero and the pole using $N_1$ segments of length $h_1$ but only $N_2-1$ segments of length $h_2$. One has
\begin{eqnarray}
\begin{pmatrix}
N_1+N_2-1\\
N_1
\end{pmatrix}
\end{eqnarray}
possible configurations. An example of such a realization for the algebra $Y_{5,2,0}$ is shown in the figure \ref{fundamental_snake}.

\begin{figure}
  \centering
      \includegraphics[width=0.39\textwidth]{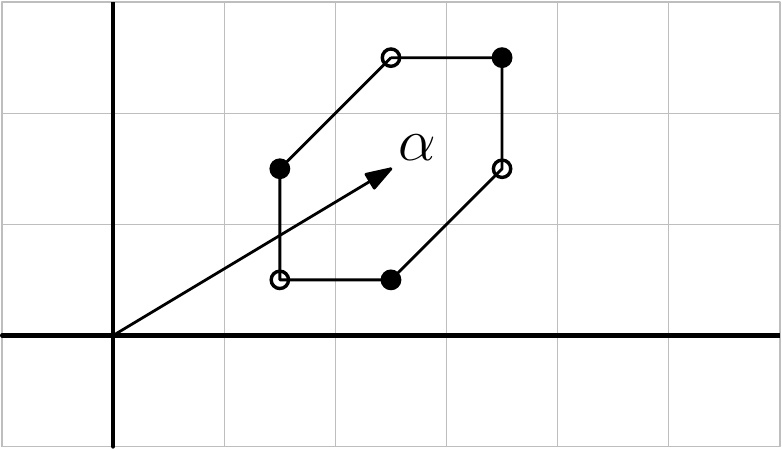}
  \caption{hexagon factor from (\ref{hexagon}).}
\label{hexagonpic}
\end{figure}
It turns out that compared to the vacuum representation, the fundamental representation can be generically realized at higher levels in the bosonic Fock space. Recall that when acting on the states in $\mathcal{W}_{1+\infty}$ with ladder operators $e_j$ and $f_j$, the eigenvalue of the generating function $\psi(u)$ changes by products of elementary factors
\begin{equation}
\label{hexagon}
\varphi (u-\alpha) = \frac{(u-\alpha+h_1)(u-\alpha+h_2)(u-\alpha+h_3)}{(u-\alpha-h_1)(u-\alpha-h_2)(u-\alpha-h_3)}.
\end{equation}
These factors play a role of the structure constants of the affine Yangian \cite{tsymbaliuk2017affine,Prochazka:2015aa}. The possible values of parameter $\alpha$ depend on the state on which we act. The number of such factors determines the level of the descendant. Here we will call this factor a \emph{hexagon factor} is because it forms a hexagon in the $u$-plane as illustrated in the figure \ref{hexagonpic}.

In the generic modules the states that we get by acting with ladder operators $e_j$ are never highest weight states. But in the case of degenerate modules, the action of ladder operators $e_j$ can produce singular vectors which are annihilated by all raising operators $f_j$, i.e. they are primary. In the irreducible modules we identify these states with zero. In the free field representation, these singular vectors can be mapped either to an identical zero, or to a non-trivial primary state which is a free boson descendant of the exponential vertex operator. We can therefore find free field representatives of primary states which are not pure exponential vertex operators but are dressed by the action of bosonic ladder operators.

\begin{figure}
  \centering
      \includegraphics[width=0.4\textwidth]{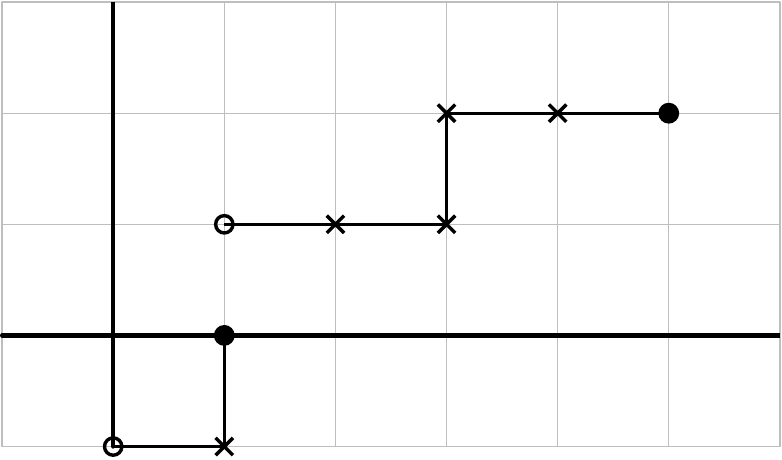}
  \caption{Generating function of the exponential factor in the realization of the fundamental representation in the first direction at level one.}
\label{fundamental_snake_lev1}
\end{figure}
\begin{figure}
  \centering
      \includegraphics[width=0.4\textwidth]{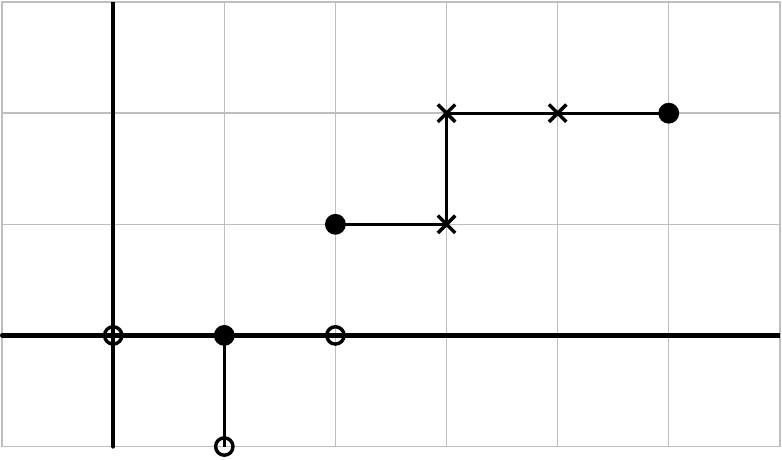}
   \caption{Generating function of the exponential factor in the realization of the fundamental representation in the first direction at level two.}
\label{fundamental_snake_lev2}
\end{figure}
Even though the attaching hexagon factors by itself does not ensure that the descendant field is singular, we will now sketch few configurations for $Y_{5,2,0}$ that we expect to be realized for some ordering of free bosons. If we place the (the inverse) hexagon factor at the origin, three nodes cancel out and three new ones are created. The new configuration is shown in the figure \ref{fundamental_snake_lev1}.  One can now realize this new configuration in various ways in terms of the generic generating function of the exponential factor (\ref{generic}). Going to one level higher, one gets after the division by the two hexagon factors a combination from the figure and similarly at higher levels. We will see that all the solutions at level zero are realized for each ordering of the free bosons but only some of the realizations at higher levels are present for a given ordering of free fields.

Finally, let us discuss the situation of the fundamental representation in the third direction. It turns out that there are $(N_1+N_2)!$ realizations and one can find them already at level zero. There are two possibilities. The first option is to connect the pole at $h_1$ with the zero at $-h_3$ by a molecule of type $h_2$ and create a snake between the pole at $h_2$ and the zero at $N_1h_1+N_2h_2$. The other possibility is to connect the pole at $h_2$ with the zero at $-h_3$ by a molecule of type $h_1$ and create a snake connecting the pole at $h_1$ with the zero at $N_1h_1+N_2h_2$. The two possible snake configurations give the correct number of realizations
\begin{eqnarray}
\begin{pmatrix}
N_1+N_2-1\\
N_1
\end{pmatrix}+
\begin{pmatrix}
N_1+N_2-1\\
N_1-1
\end{pmatrix}=
\begin{pmatrix}
N_1+N_2\\
N_1
\end{pmatrix}.
\end{eqnarray}

\subsubsection{Free field realization and OPE of modules}

After the identification of possible values of parameters $x^{(\kappa)}_j$ for the identity and the fundamental and the anti-fundamental representation, let us discuss how are these different possibilities realized in the context of the free field representation.

There are $(N_1+N_2)!$ free field realizations of any $Y_{N_1,N_2,0}$ algebra associated to different orderings of the free bosons. It turns out that not all the possibilities discussed above at the level of generating functions are realized for any given choice of ordering. 

Moreover, it might be puzzling that we find more than one free field realization of the same $Y_{N_1,N_2,0}$ module since it is not clear that all of these have the correct fusion and braiding properties and lead eaquivalent OPEs of degenerate modules. Following \cite{Dotsenko:1984nm,Felder:1988zp}, it turns out that all the realizations seem to be equivalent if we work modulo insertions of screening charges (contour integrals of screening currents) in all the examples bellow.

To determine the structure constants (and braiding and fusion in particular), one needs to determine three-point functions of all the degenerate modules. Choosing a particular free field realization of degenerate modules within a given three-point function leads to a zero value if we do not insert a correct number of the screening charges. After such an insertion, one can show that (in all the examples that we considered) all the free field realizations lead to the same OPEs (if the correlator was non-vanishing). Note also that the free field realization gives an explicit construction of all the conformal blocks in terms contour integrals of meromorphic functions with possible branch-cuts.

\paragraph{A simple realization}

Before discussing the fusion and braiding and checking the independence on the choice of the free field representative, let us mention one simple realization of the identity-box-anti-box triple that exists for every free field realization.

As discussed above, the relation between parameters $x^{(\kappa_j)}_j$ and exponents $q^j$ of the pure exponential realization of a module is given for a fixed ordering of free bosons by shifts
\begin{eqnarray}
\label{shift}
x^{(\kappa_j)}_j=q^j+\sum_{k<j}h_{\kappa_k}
\end{eqnarray}
We are summing over all free fields that appear to the left of $\phi_j$ in the Miura transformation. All the $(N_1+N_2)!$ solutions for $x_j$ for the vacuum representation can be identified with $(N_1+N_2)!$ representations of the free boson vacuum $q^j=0$ corresponding to $(N_1+N_2)!$ orderings. The free boson vacuum $q^j=0$ is the simplest realization of the vacuum of the corresponding $Y_{N_1,N_2,0}$ algebra.

There's a similar story also for the fundamental and the anti-fundamental representation with some extra complications since some of the realizations are not in terms of pure exponential vertex operators but in terms of their free field descendants. Based on examples, we conjecture that one can realize the fundamental representation in the first direction as a descendant of the exponential $\exp [h_3\phi_i^{(2)}]$, where $\phi_i^{(2)}$ is the left-most free boson of the second type in a given ordering. The level of the descendant equals the number of free bosons of the first type on the left of such $\phi_i^{(2)}$. The anti-fundamental field is given by a descendant of $\exp [-h_3\phi_j^{(2)}]$, where $\phi_j^{(2)}$ is the right-most free boson of the second type and the level is given by the number of free bosons of the first type on the right of $\phi_j^{(2)}$. Similar simple realizations can be found also for representations in the second and third direction: a simple box in the second direction is associated to the left-most free boson of the first or third type and the level is given by the number of bosons of the second type on the left of it. For $N_3=0$ the box and antibox in the third direction correspond to the first and last boson and are always on level $0$ (there are no obstructions since we have no bosons of the third type). The charge $q$ appearing in the exponential is given by $h_\sigma$ for box and $-h_\sigma$ for the anti-box and $\sigma$ is such that the triple $(\sigma,\tau,\pi)$ in $h_\sigma, \phi^{(\tau)}$ and the direction $\pi$ is a permutation of $(123)$.


\subsubsection{$Y_{0,0,2}$ example}

Let us start by an illustration how things work in the case of the Virasoro algebra in ordering $R_1^{(3)} R_2^{(3)}$. The two available screening currents are
\begin{eqnarray}\nonumber
j_1&=&\exp \left [-h_1 \left(\phi^{(3)}_1-\phi^{(3)}_2\right)  \right ]\\
j_2&=&\exp \left [-h_2 \left(\phi^{(3)}_1-\phi^{(3)}_2\right)  \right ]
\end{eqnarray}
The two realizations of the identity, the fundamental representation and the anti-fundamental representation in the first and the second direction are
\begin{eqnarray}
\nonumber
&M^1_{\mathds{1}}=\mathds{1},\qquad &M^2_{\mathds{1}}=\exp \left [h_3 \left(\phi^{(3)}_1-\phi^{(3)}_2\right) \right ],\\
\nonumber
&M^1_{\square_1}=\exp \left [ h_2\phi^{(3)}_1 \right ],\qquad &M^2_{\square_1}=\exp \left [h_2\phi^{(3)}_2 + h_3 \left(\phi^{(3)}_1-\phi^{(3)}_2\right) \right ],\\
&M^1_{\bar{\square}_1}=\exp \left [-h_2\phi^{(3)}_2 \right ],\qquad &M^2_{\bar{\square}_1}=\exp \left [-h_2\phi^{(3)}_1 + h_3 \left(\phi^{(3)}_1-\phi^{(3)}_2\right) \right ], \\
\nonumber
&M^1_{\square_2}=\exp \left [ h_1\phi^{(3)}_1 \right ],\qquad &M^2_{\square_2}=\exp \left [h_1\phi^{(3)}_2 + h_3 \left(\phi^{(3)}_1-\phi^{(3)}_2\right) \right ],\\
\nonumber
&M^1_{\bar{\square}_2}=\exp \left [-h_1\phi^{(3)}_2 \right ],\qquad &M^2_{\bar{\square}_2}=\exp \left [-h_1\phi^{(3)}_1 + h_3 \left(\phi^{(3)}_1-\phi^{(3)}_2\right) \right ].
\end{eqnarray}
We see that there indeed exists the simple free field realization of the identity, the fundamental and the anti-fundamental representation.

Let us now check that two-point functions of different realizations of the identity and the two-point function of the fundamental with the anti-fundamental field are independent of the choice of the free field realization. To check all the three-point functions, one would have to relate normalizations of different realizations of all the degenerate modules and then compare all the three point funcions. Because we do not aim to do the comparison here, we disregard such normalizations and only check the braiding properties. 

The charge of the identity realized by $M^2_{\mathds{1}}$ cannot be subtracted by insertions of the screening charges and thus vanishes. The true identity $\mathds{1}$ is the only realization of the vacuum module giving a non-zero one-point function.

The only combination that gives a non-vanishing two-point function of the fundamental and the anti-fundamental representation comes from the first realizations and give
\begin{eqnarray}
\nonumber
\langle M^1_{\square_2}(z)M^1_{\bar{\square}_2}(w) \rangle_{Y_{0,0,2}} &\propto & \oint_z d\tilde{z}\langle j_1(\tilde{z})  M^1_{\square_2}(z)M^1_{\bar{\square}_2}(w)\rangle \\
\nonumber
& \propto & \oint_z d\tilde{z}\ (\tilde{z}-z)^{\frac{h_1}{h_2}}(\tilde{z}-w)^{\frac{h_1}{h_2}}\propto\oint_0 d\tilde{z}\ \tilde{z}^{\frac{h_1}{h_2}}(\tilde{z}+z-w)^{\frac{h_1}{h_2}} \\
& \propto & (z-w)^{2\frac{h_1}{h_2}}\oint_0 d\tilde{z}\ \left ( \frac{\tilde{z}}{w-z}\right )^{\frac{h_1}{h_2}}\left (1-\frac{\tilde{z}}{w-z}\right )^{\frac{h_1}{h_2}} \\
\nonumber
& \propto & (z-w)^{2\frac{h_1}{h_2}+1},
\end{eqnarray}
where $\langle\dots \rangle_{Y_{N_1,N_2,N_3}}$ denotes the correlation function with possible insertions of the screening charges of $Y_{N_1,N_2,N_3}$ that cancel the charge of the exponential factors. The exponent is exactly (up to the minus sign) the sum of conformal dimensions of the fundamental and the anti-fundamental representation which is the expected $z$-dependence of the two-point function.

\subsubsection{$Y_{1,1,0}$ example}

The second example is the first non-trivial case that contains free field realizations of degenerate modules at higher levels and at the same time there is a mismatch between the number of free field realizations of the fundamental and the anti-fundamental representation. One gets the following realizations of the identity, the fundamental and the anti-fundamental field in the first and second direction for the ordering $\phi^{(1)}_1\times \phi^{(2)}_2$ of the free bosons
\begin{eqnarray}\nonumber
&M^1_{\mathds{1}}=\mathds{1},\qquad &M^2_{\mathds{1}}=\exp \left [h_2 \phi^{(1)}_1-h_1 \phi^{(2)}_2 \right ]\\  \nonumber
&M^1_{\square_1}=\exp \left [ h_2\phi^{(1)}_1+(h_3-h_1)\phi^{(2)}_2 \right ],\qquad &M^2_{\square_1}=(h_2J^{(1)}_1-h_1J^{(2)}_2)\exp \left [ h_3\phi^{(2)}_2 \right ],\\
&M_{\bar{\square}_1}=\exp \left [ -h_3\phi^{(2)}_2 \right ], \\
\nonumber
&M_{\square_2}=\exp \left [ h_3\phi^{(1)}_1 \right ], \\
\nonumber
&M^1_{\bar{\square}_2}=\exp \left [ (h_2-h_3) \phi^{(1)}_1 -h_1\phi^{(2)}_2 \right ],\qquad &M^2_{\bar{\square}_2}=(h_2J^{(1)}_1-h_1J^{(2)}_2)\exp \left [ -h_3\phi^{(1)}_1 \right ].
\end{eqnarray}
and the following screening current
\begin{eqnarray}
j=\exp \left [-h_2\phi^{(1)}_1+h_1\phi^{(2)}_2 \right ].
\end{eqnarray}
Note that there is only a single realization of the fundamental field and one of the realizations (the simple one) of the anti-box is at level one.

Let us first check that one-point function of the identity realized as $M_{\mathds{1}}^2$ equals the vacuum amplitude
\begin{eqnarray}
\left \langle M_{\mathds{1}}^2(z)\right \rangle_{Y_{0,1,1}} \propto \oint_z d\tilde{z}\ \langle j(\tilde{z})M_{\mathds{1}}^2(z) \rangle \propto \oint_z d\tilde{z}\ (\tilde{z}-z)^{\frac{h_2}{h_3}+\frac{h_1}{h_3}}\propto 1.
\end{eqnarray}
Similarly for the two-point function with two contour integrations, one gets
\begin{eqnarray}\nonumber
\left \langle M_{\mathds{1}}^2(z)M_{\mathds{1}}^2(w)\ \right \rangle_{Y_{0,1,1}} &\propto&\oint_z d\tilde{z}_2\oint_w d\tilde{z}_1\langle j(\tilde{z}_1)j(\tilde{z}_2)M_{\mathds{1}}^2(z)M_{\mathds{1}}^2(w) \rangle \\ \nonumber
&\propto&\oint_z d\tilde{z}_2\oint_w d\tilde{z}_1 \frac{(\tilde{z}_1-\tilde{z}_2)(z-w)}{(\tilde{z}_1-z)(\tilde{z}_1-w)(\tilde{z}_2-z)(\tilde{z}_2-w)}\\
&\propto&\oint_z d\tilde{z}_2 \frac{(w-\tilde{z}_2)}{(\tilde{z}_2-z)(\tilde{z}_2-w)}\propto 1.
\end{eqnarray}

Let us now show that the two-point function of both realizations of the anti-fundamental representation with the fundamental representations are also equal
\begin{eqnarray}
\langle M_{\square_1}^1(z)M^1_{\bar{\square}_1}(w)\rangle_{Y_{0,1,1}}&\propto&\oint_z d\tilde{z}\ \langle j(\tilde{z}) M_{\square_1}^1(z)M_{\bar{\square}_1}(w)\rangle \\
\nonumber
& \propto &\oint_z d\tilde{z}\ (\tilde{z}-z)^{-2}(\tilde{z}-w)(z-w)^{\frac{h_1}{h_3}-1}=(z-w)^{\frac{h_3}{h_1}-1}.
\end{eqnarray}
One gets the same expression from the other realization
\begin{eqnarray}
\nonumber
\left \langle M_{\square_1}^2(z)M_{\bar{\square}_1}(w)\right \rangle_{Y_{0,1,1}} & \propto & \left \langle J_2^{(2)}\exp \left [h_3\phi^{(2)}_2 \right ](z)\exp \left [-h_3\phi^{(2)}_1 \right ](w)\right  \rangle \\
& \propto & (z-w)^{\frac{h_1}{h_3}-1},
\end{eqnarray}
where the $-1$ factor comes from the contraction with $J_2^{(2)}$.

\subsubsection{$Y_{2,1,0}$ example}

\begin{figure}
    \centering
    \begin{subfigure}[b]{0.27\textwidth}
        \includegraphics[width=\textwidth]{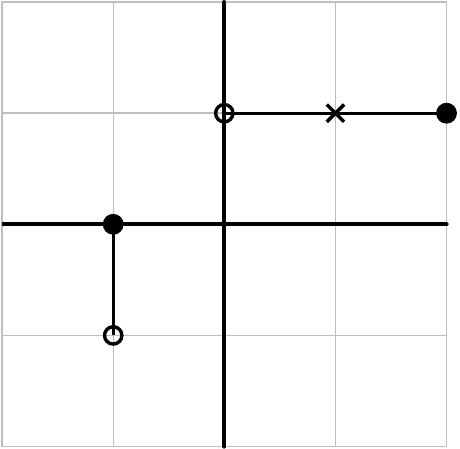}
        \caption{Level zero}
\label{box_zero}
    \end{subfigure}
 \quad
    \begin{subfigure}[b]{0.27\textwidth}
        \includegraphics[width=\textwidth]{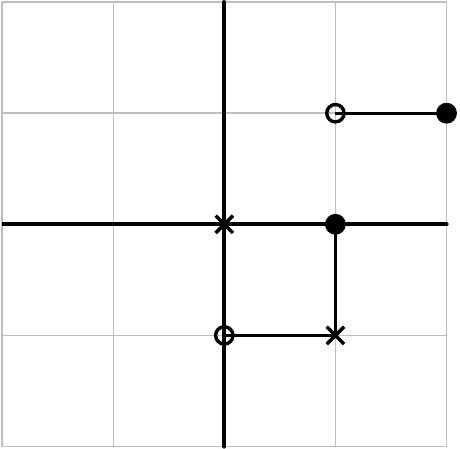}
        \caption{Level one}
\label{box_one}
    \end{subfigure}
 \quad
    \begin{subfigure}[b]{0.27\textwidth}
        \includegraphics[width=\textwidth]{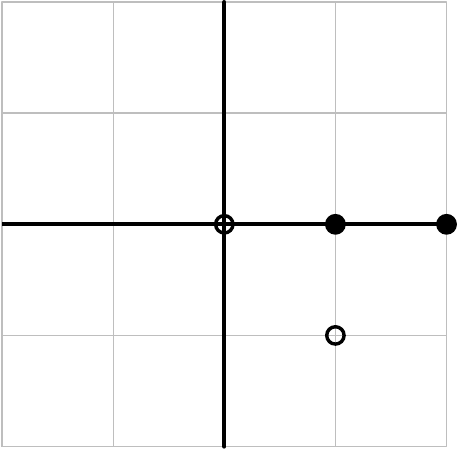}
        \caption{Level two}
\label{box_two}
    \end{subfigure}
    \caption{Exponential factors of the fundamental representation in direction 1. The two crosses in the middle figure correspond to two possible paths.}
\end{figure}

To illustrate the existence of the simple realization of the fundamental and the anti-fundamental representation, let us now discuss the algebra $Y_{2,1,0}$. Firstly, the generating function of $\psi_i$ charges for the vacuum, the fundamental and the anti-fundamental representation in the first asymptotic direction is
\begin{eqnarray}\nonumber
\psi_{\bullet}(u)&=&\frac{u-2h_1-h_2}{u}\\ \nonumber
\psi_{\square_1}(u)&=&\frac{u-2h_1-h_2}{u-h_2}\frac{u+h_1}{u-h_3},\\
\psi_{\bar{\square}_1}(u)&=&\frac{u-2h_1}{u}\frac{u-3h_1-2h_2}{u-3h_1-h_2}.
\end{eqnarray}
The structure of the generating function in terms of zeros (dots) and poles (circles) is captured in the figure  \ref{box_zero}. From this figure, one can read of possible values of parameters $x_i$ parametrizing the modules
\begin{eqnarray}
x_1^{(1)}=h_1+h_2,\quad x_2^{(1)}=h_2,\quad x_3^{(2)}=-h_1-h_2
\end{eqnarray}
as well as the solution with $x_1^{(1)}$ and $x_1^{(2)}$ permuted. 

One can also realize the same highest weight state as a free field descendant of a different exponential primary. Charges of such a module can be deduced from the composition with hexagon factors $\varphi (u-\alpha)$ for some shift $\alpha$. If we multiply the configuration \ref{box_zero} by a hexagon with the center at zero, we get a structure of zeros and poles from the figure \ref{box_one}. One can thus realize the same module as a level one descendant of the module with charges\footnote{Note that there exists also a configuration with $x^{(1)}_1=-h_2$ and $x^{(2)}_3=h_1-h_2$ that does not seem to correspond to any simple realization. Similarly, we get one extra solution also at level two.}
\begin{eqnarray}
x_1^{(1)}=0, \quad x_2^{(1)}=h_1+h_2, \quad x_3^{(2)}=-h_2
\end{eqnarray}
or with $x_1^{(1)}$ and $x_2^{(1)}$ possibly permuted. Finally, composing \ref{box_one} with yet another hexagon factor with the center at $h_1$, one gets the generating function from the figure \ref{box_two} with parameters
\begin{eqnarray}
x_1^{(1)}=0, \quad x_2^{(1)}=h_1, \quad x_3^{(2)}=h_1-h_2.
\end{eqnarray}

\begin{figure}
    \centering
    \begin{subfigure}[b]{0.27\textwidth}
        \includegraphics[width=\textwidth]{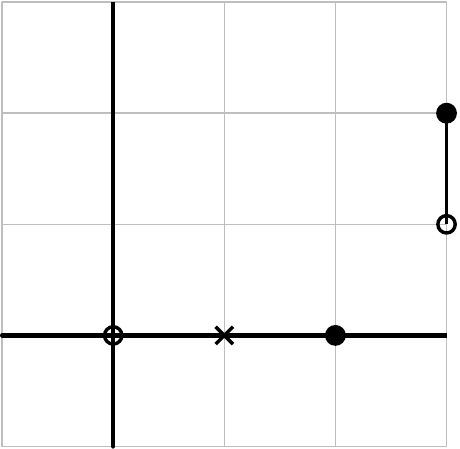}
        \caption{Level zero}
    \end{subfigure}
 \quad
    \begin{subfigure}[b]{0.27\textwidth}
        \includegraphics[width=\textwidth]{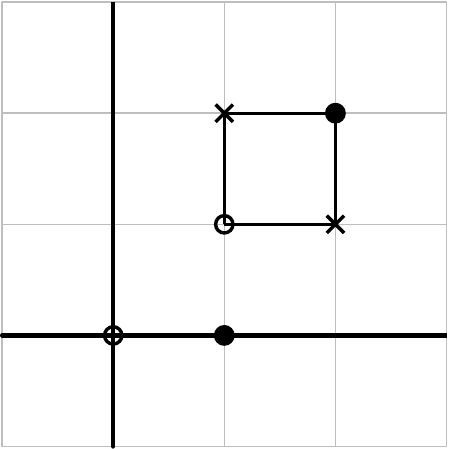}
        \caption{Level one}
    \end{subfigure}
 \quad
    \begin{subfigure}[b]{0.27\textwidth}
        \includegraphics[width=\textwidth]{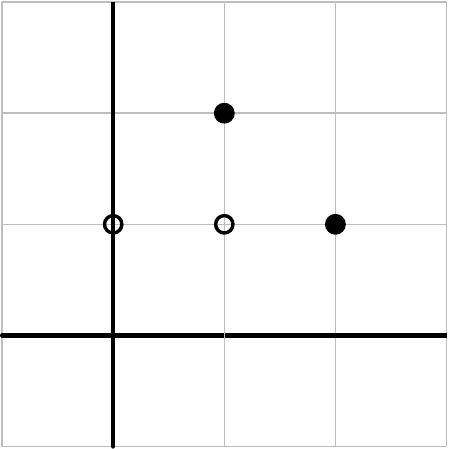}
        \caption{Level two}
    \end{subfigure}
    \caption{Exponential factors of the anti-fundamental representation in direction 1. The two crosses in the middle figure correspond to two possible paths.}
\label{anti_box}
\end{figure}

One can do a similar investigation for the anti-fundamental representation and recover the structure from the figure \ref{anti_box}. In terms of the standard parameters $x^{(\kappa)}_i$, one gets
\begin{eqnarray}
\nonumber
x_1^{(1)}&=&0,\quad x_2^{(1)}=h_1,\quad x_3^{(2)}=3h_1+h_2 \\
x_1^{(1)}&=&0,\quad x_2^{(1)}=h_1+h_2,\quad x_3^{(2)}=2h_1+h_2 \\
\nonumber
x_1^{(1)}&=&h_1+h_2,\quad x_2^{(1)}=h_2,\quad x_3^{(2)}=h_1+h_2
\end{eqnarray}

In total, we found two realizations in terms of level zero, level one and level two descendants. We will now see that different solutions can be identified with simple realizations of the $3!$ free field realizations. Shifting the simple exponentials according to the corresponding ordering of free bosons, on recovers:
\begin{center}
\begin{tabular}{| c | c | c |}
\hline
Representation & Level & $(x_1^{(1)},x_2^{(1)},x_3^{(2)})$ charges\\
\hline
 \multicolumn{3}{|c|}{$R^{(1)}_1 R^{(1)}_2 R^{(2)}_3 $}\\
\hline
 $\square$ & 2& $(0,0,h_3)+(0,h_1,2h_1)=(0,h_1,h_1-h_2)$ \\
 $\bar{\square}$ & 0 & $(0,0,-h_3)+(0,h_1,2h_1)=(0,h_1,3h_1+h_2)$ \\
\hline
 \multicolumn{3}{|c|}{$R^{(1)}_1 R^{(2)}_2 R^{(1)}_3 $}\\
\hline
 $\square$ & 1 & $(0,0,h_3)+(0,h_1+h_2,h_1)=(0,h_1+h_2,-h_2)$ \\
 $\bar{\square}$ & 1 & $(0,0,-h_3)+(0,h_1+h_2,h_1)=(0,h_1+h_2,2h_1+h_2)$ \\
\hline
 \multicolumn{3}{|c|}{$R^{(2)}_1 R^{(1)}_2 R^{(1)}_3 $}\\
\hline
 $\square$ & 0 & $(0,0,h_3)+(h_1+h_2,h_2,0)=(h_1+h_2,h_2,-h_1-h_2)$ \\
 $\bar{\square}$ & 2 & $(0,0,-h_3)+(h_1+h_2,h_2,0)=(h_1+h_2,h_2,h_1+h_2)$ \\ \hline
\end{tabular}
\end{center}
Note that the level agrees with our proposal about the number of free bosons of the other type on the left and right of the left-most and the right-most free boson of the second type.

\section{Gluing and generic modules}

\subsection{Gluing using free fields}
Starting with $Y_{N_1,N_2,N_3}$ as a building block, one can construct more complicated VOAs associated to an arbitrary $(p,q)$ web of five-branes and D3-branes attached to them at various faces. The resulting vertex operator algebra is an extension of tensor product of Y-algebras associated to each vertex by bi-modules (and their fusion) associated to each internal line of the web diagram. Existence of such an extension was conjectured in \cite{Prochazka:2017qum} but no explicit construction of OPEs between gluing bi-modules was proposed.  The free field realization discussed above seems to provide us with a way to determine OPEs of such bi-modules. In the two explicit examples bellow, we will indeed see that this is indeed the case. Note also that such construction leads to an algorithmic way to determine a free field realization of the glued algebra. We expect some of the free field realizations to be related via bosonisation to well-known free-field realizations, such as the Wakimoto realization of Kac-Moody algebras \cite{Wakimoto:1986gf,Feigin:1990jc}. 

\begin{figure}
  \centering
      \includegraphics[width=0.65\textwidth]{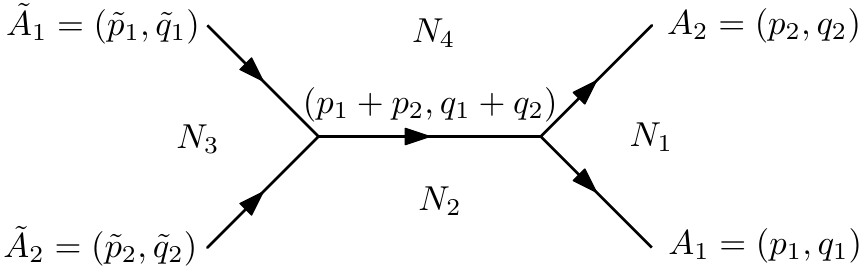}
  \caption{Gluing of two vertices.}
\label{gluing}
\end{figure}

Let us briefly review the gluing construction in the case of a single edge. The generalization to more complicated configurations is straightforward and will be briefly discussed later. Consider a $(p,q)$-brane configuration from the figure \ref{gluing}. The resulting VOA is an extension of the product
\begin{eqnarray}
Y_{N_2,N_4,N_3}^{-\tilde{A}_1,-\tilde{A}_2,\tilde{A}_1+\tilde{A}_2}[\Psi] \otimes Y_{N_4,N_2,N_1}^{A_1,A_2,-A_1-A_2}[\Psi]
\end{eqnarray}
where $Y_{N_1,N_2,N_3}^{A_1,A_2,A_3}[\Psi]$ is related to the standard algebra $Y_{N_1,N_2,N_3}[\Psi]$ by an $SL(2,\mathbb{Z})$ transformation of parameters
\begin{eqnarray}
Y_{N_1,N_2,N_3}^{A_1,A_2,A_3}[\Psi] = Y_{N_1,N_2,N_3}\left [ -\frac{q_2\Psi-p_2}{q_1\Psi-p_1}\right ].
\end{eqnarray}
The parameters $h_{i}$ of the algebra can be easily determined from
\begin{eqnarray}
h_i=A_i\cdot \epsilon,
\end{eqnarray}
where we have introduced the vector $\epsilon=(\epsilon_1,\epsilon_2)$ and $A_i$ are the $(p,q)$ charges of the $i$th interface with the arrow pointing out of the vertex. Note that $\epsilon_i$ are universal parameters and in the case of the standard trivalent junction of NS5, D5 and $(1,1)$ branes, one has the identification $h_i=\epsilon_i$ with $\epsilon_3=-\epsilon_1-\epsilon_2$\footnote{If we consider gluing of vertices, we need to distinguish $\epsilon$-parameters and $h$-parameters. The $\epsilon$-parameters are determined by $\Psi$ while the $h$-parameters are associated to each vertex and they are related to $\epsilon_j$ by $SL(2,\mathbbm{Z})$-transformation which brings the vertex to the standard one. \cite{Prochazka:2017qum}}. The extension is then generated by fusions of the tensor product of the fundamental representation associated to the first vertex and anti-fundamental representation associated to the second vertex and vice versa. 

In the free field realization, the fundamental and the anti-fundamental representation have a simple realization in terms of an exponential vertex operator and its descendant. For simplicity of the discussion, we will restrict to the case $N_4=0$ and identify only the simple realization of the fundamental and the anti-fundamental representation for the following ordering
\begin{eqnarray}
R_1^{(2)} \cdots R_{N_2}^{(2)} \, R_{N_2+1}^{(3)} \cdots R_{N_1+N_2}^{(3)}
\end{eqnarray}
of free bosons in the right vertex and 
\begin{eqnarray}
\tilde{R}_1^{(3)} \cdots \tilde{R}_{N_2}^{(3)} \, \tilde{R}_{N_2+1}^{(1)} \cdots \tilde{R}_{N_2+N_3}^{(1)}
\end{eqnarray}
in the left vertex. The generalization for $N_4\neq 0$, a general ordering and `non-simple' realizations is straightforward but the formulas become more involved.

The gluing fields are the generated from the fundamental and the anti-fundamental representation associated to lines supported at the internal interface generated by
\begin{eqnarray}
M_{\square}=M^{3}_{\square}\otimes \tilde{M}^{3}_{\bar{\square}},\qquad M_{\bar{\square}}=M^{3}_{\bar{\square}}\otimes \tilde{M}^{3}_{\square}
\end{eqnarray}
where $M^{3}_{\square}$ and $M^{3}_{\bar{\square}}$ are the primaries associated to the fundamental and the anti-fundamental module associated to the third direction of the right vertex and $\tilde{M}^{3}_{\square}$ and $\tilde{M}^{3}_{\bar{\square}}$ associated to the left vertex. The simple free field realizations in the given ordering are of the form
\begin{eqnarray}\nonumber
M^{3}_{\square}=\exp \left [ h_1\  \phi_{1}^{(2)} \right ]\qquad M^{3}_{\bar{\square}}=f(J)\exp \left [-h_1\ \phi_{N_2}^{(2)} \right ]\\
\tilde{M}^{3}_{\square}=\exp \left [ -\tilde{h}_2\  \tilde{\phi}_{1}^{(3)} \right ]\qquad \tilde{M}^{3}_{\bar{\square}}=f(\tilde{J})\exp \left [\tilde{h}_2\ \tilde{\phi}_{N_2}^{(3)} \right ]
\end{eqnarray}
where $f(J)$ is a level $N_1$ and $f(\tilde{J})$ is a level $N_3$ field of the free boson. Even though we lack a closed form expression for $f(J)$ and $f(\tilde{J})$, they can be easily determined from the requirement that $M^{3}_{\square}$ and $\tilde{M}^{3}_{\bar{\square}}$ are primary fields of correct $W$-charges. All the other bi-fundamental fields can be constructed from the fusion of $M_{\square}$ and $M_{\bar{\square}}$. 

In configurations with more internal finite interfaces, one can introduce corresponding fundamental and anti-fundamental representations associated to each finite segment and extend the tensor product of Y-algebras by fusion of all such generators. We will illustrate the gluing procedure on two examples bellow.

\subsection{Gluing and generic modules}

Let us discuss how to glue generic modules and its interpretation in terms of the physics of GW defects. The highest weight vector of a generic module of a Y-algebra can be realized as an exponential vertex operator $\exp \left [Q^\mu \Phi_\mu\right ]$, where we introduced a vector of free fields and a dual vector of charges
\begin{eqnarray}\nonumber
\Phi_\mu&=&\left (\phi_1^{(2)}, \dots, \phi_{N_2}^{(2)},  \phi_{N_2+1}^{(3)}, \dots, \phi_{N_2+N_1}^{(3)}\right )\\
Q^\mu&=&\left (q^1, \dots, q^{N_2},  q^{N_2+1}, \dots, q^{N_2+N_1}\right )
\end{eqnarray}
and similarly for the other vertex
\begin{eqnarray}\nonumber
\tilde{\Phi}_\mu&=&\left (\tilde{\phi}_1^{(1)}, \dots, \tilde{\phi}_{N_2}^{(1)},  \tilde{\phi}_{N_2+1}^{(3)}, \dots, \tilde{\phi}_{N_2+N_1}^{(3)}\right )\\
\tilde{Q}^\mu&=&\left (\tilde{q}^1, \dots, \tilde{q}^{N_2},  \tilde{q}^{N_2+1}, \dots, \tilde{q}^{N_2+N_1}\right )
\end{eqnarray}
A generic module of a glued algebra can be then realized as a tensor product of such exponentials associated to each vertex in the diagram. 

Note that the parameters $q^i$ and $\tilde{q}^i$ correspond to the same GW defect and the gauge theory setup suggests that they must be identified (up to shifts induced by line operators supported at the boundary $A_1$ and $\tilde{A}_2$), in particular
\begin{eqnarray}
q^{i}\pm\tilde{q}^{i}=n^i h_3
\label{identification}
\end{eqnarray}
for some integers $n^i$ and $h_3=A_3\cdot \epsilon=-\tilde{A}_3\cdot \epsilon=(-A_1-A_2)\cdot \epsilon$. The relative sign depends on the relative orientation of the two glued vertices. In \cite{Prochazka:2017qum}, we defined the orientation of a vertex $Y_{N_1,N_2,N_3}^{A_1,A_2,A_3}$ as a sign given by $(-1)^{p_1p_2+q_1q_2+p_2q_1}$. The relative orientation and the sign in the above equation\footnote{The sign would be opposite if we have glued the fundamental representation of the first vertex with the fundamental representation of the second vertex and similarly for the anti-fundamental representation.} is given by a product of such factors in the two vertices. In particular, one gets $-1$ for the resolved conifold diagram and $+1$ for the toric diagram of $\mathbb{C}/\mathbb{Z}_2 \times \mathbb{C}$. We will see later in examples that this condition is necessary for the gluing bi-modules to be local with the GW modules. 

Note that inclusion of bi-fundamental fields might change the algebra of zero modes that might become non-commutative. Moreover, we will see later that the modules are in general not even modules induced from the modules of the zero-mode algebra.  GW modules associated to the commutative zero-mode algebra of $Y_{N_1,N_2,N_3}$ are thus building blocks of modules for more complicated algebras with non-commutative algebra of zero modes.

\subsection{Gluing two $\widehat{\mathfrak{gl}}(1)$'s}

\begin{figure}
  \centering
      \includegraphics[width=0.18\textwidth]{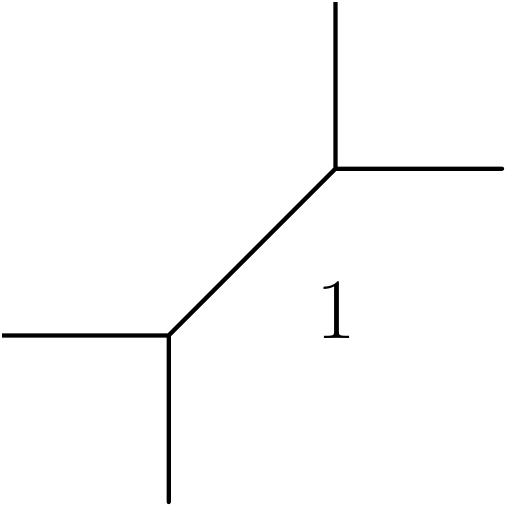}
  \caption{The simplest example of gluing of two $\widehat{\mathfrak{gl}}(1)$ Kac-Moody algebras.}
\label{u1glue}
\end{figure}
Let us consider the first example of gluing of two $\widehat{\mathfrak{gl}}(1)_\Psi$ Kac-Moody algebras as shown in the figure \ref{u1glue}. Let $\phi\equiv \phi^{(2)}_1$ be the free boson associated to the right vertex and $\tilde{\phi}\equiv \tilde{\phi}^{(1)}_1$ be the one associated to the second one. We normalize them such that $J=\partial \phi^{(2)}_1$ and  $\tilde{J}=\partial \tilde{\phi}^{(2)}_1$ have the following OPE
\begin{eqnarray}
J(z)J(w)\sim -\frac{1}{\epsilon_1 \epsilon_3}\frac{1}{(z-w)^2},\qquad \tilde{J}(z)\tilde{J}(w)\sim -\frac{1}{\epsilon_2 \epsilon_3}\frac{1}{(z-w)^2}.
\end{eqnarray}
Generators that need to be added to the algebra can be identified with the fusion of the following vertex operators realizing the fundamental and anti-fundamental representation
\begin{eqnarray}
M_{\square}=\exp \left [\epsilon_1 \phi -\epsilon_2 \tilde{\phi} \right],\qquad M_{\bar{\square}}=\exp \left [-\epsilon_1 \phi +\epsilon_2 \tilde{\phi} \right].
\end{eqnarray}
One can easily check that the two generators have correct charges with respect to the two $\mathfrak{gl}(1)$ subalgebras and that the conformal weight with respect to the sum of the two stress-energy tensors is $1/2$. Moreover, the free field realization gives also an explicit realization of the OPE between the added fields $M_{\square}$ and $ M_{\bar{\square}}$ that has the following simple form
\begin{eqnarray}
M_{\square}(z) M_{\bar{\square}}(w)\sim \frac{1}{z-w}
\end{eqnarray}
with all the other OPEs trivial. The exponent was determined from the product of the two exponents (with the metric determined by the normalization of the free bosons)
\begin{eqnarray}
-\frac{(-\epsilon_1)\epsilon_1}{\epsilon_1 \epsilon_3}-\frac{\epsilon_2(-\epsilon_2)}{\epsilon_2 \epsilon_3}=-1.
\end{eqnarray}

One can immediately see that the BRST definition of the algebra is reproduced. In particular, the added fields $M_{\square}$ and $M_{\bar{\square}}$ form the free fermion pair and the combination $J+\tilde{J}$ can be identified with the decoupled $\widehat{\mathfrak{gl}}(1)$ Kac-Moody algebra. The relation between free fermions and the vertex operators $M_{\square}$, $M_{\bar{\square}}$ is the well known bosonization.

Having an explicit description of the glued algebra in terms of free fields, we would like to discuss generic modules of the glued algebra. According to the discussion above, we expect the correct GW-defect module to be generated by descendants of
\begin{eqnarray}
M\left [q,\tilde{q}\right ]= \exp  \left [ q\phi +\tilde{q}\tilde{\phi} \right],
\end{eqnarray}
where the parameters $\beta,\tilde{\beta}$ are related by (\ref{identification}), i.e.
\begin{equation}
\label{const1}
q-\tilde{q}=n \epsilon_3
\end{equation}
for some integer $n$. Note that this is exactly the condition following form the locality of $M[q,\tilde{q}]$ with the gluing bi-modules $M_{\square}$ and $M_{\bar{\square}}$. In particular, requiring the OPE to be of the following form
\begin{equation}
M_{\square}(z)M[q,\tilde{q}](w)\sim \frac{\exp \left [q-\epsilon_1,\tilde{q}+\epsilon_2\right ](w)}{(z-w)^{n}}+\dots
\end{equation}
where $n$ is an integer, one gets a constraint
\begin{equation}
\frac{\epsilon_1q}{\epsilon_1 \epsilon_3}-\frac{\epsilon_2\tilde{q}}{\epsilon_2 \epsilon_3}=n
\label{constraint_glue1}
\end{equation}
which is the same constraint as (\ref{const1}).

Note that the fusion with gluing fields preserve the constraint (\ref{const1}) and only shifts the coefficient $n$. Fields $M[q-\epsilon_1,\tilde{q}+\epsilon_2]$ and $M[q,\tilde{q}]$  are actually vectors of a single module. The only parameter of the module is thus the $\mathfrak{gl}(1)$ charge of the decoupled current $J+\tilde{J}$.

\subsection{$\widehat{\mathfrak{gl}}(2)$ from gluing}

\begin{figure}
  \centering
      \includegraphics[width=0.16\textwidth]{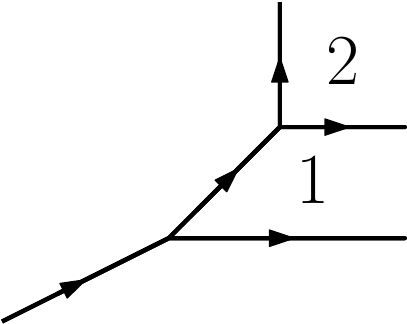}
  \caption{The web diagram associated to the $\widehat{\mathfrak{gl}}(2)$ Kac-Moody algebra.}
\label{ExF2}
\end{figure}

Let us now discuss the structure of glued generic modules for the algebra $\widehat{\mathfrak{gl}}(2)$ associated to the figure \ref{ExF2}. This example will serve as a prototype for a more general configuration whose GW-defects give rise to VOA modules induced from the Gelfand-Tsetlin modules of the zero-modes algebra.

\paragraph{$Y_{0,1,2}$ vertex}

First, let us construct the free field realization of the algebra $Y_{0,1,2}$. The algebra has a free field realization in terms of three free bosons normalized as
\begin{eqnarray}
\nonumber
J^{(2)}_1(z) J^{(2)}_1(w)&\sim& -\frac{1}{\epsilon_1 \epsilon_3}\frac{1}{(z-w)^2}, \\
J^{(3)}_2(z) J^{(3)}_2(w)&\sim& -\frac{1}{\epsilon_1 \epsilon_2}\frac{1}{(z-w)^2}, \\
\nonumber
J^{(3)}_3(z) J^{(3)}_3(w)&\sim& -\frac{1}{\epsilon_1 \epsilon_2}\frac{1}{(z-w)^2}.
\end{eqnarray}

The generators of the algebra $Y_{0,1,2}$ were already found previously. For the purpose of our discussion, let us recall the $\widehat{\mathfrak{gl}}(1)$ field
\begin{eqnarray}
J&=&J_1+J_2+J_3.
\label{first_vertex}
\end{eqnarray}

Let us now discuss the free field realization of the fundamental and the anti-fundamental module in the third direction that will play the role of $J^+$ and $J^-$ generators after tensoring with the corresponding modules of the other vertex. The fundamental field can be realized as
\begin{eqnarray}
M_{\square}=\exp \left [ \epsilon_1\phi^{(2)}_1\right ].
\end{eqnarray}
Note that $\epsilon_1$ is precisely the charge predicted by the generating function of the $\psi$-charges and all the $W$-charges of the representation match. The anti-fundamental field is more complicated since it appears at level two (there are two of free bosons of the third type to the left of $\phi^{(3)}_3$). One finds the following expression for the fundamental field
\begin{eqnarray}
M_{\bar{\square}}=\left ( -\frac{\epsilon_1\epsilon_2}{\epsilon_3}J_3J_2+\epsilon_1 J_1 (J_2+J_3)-\frac{\epsilon_1\epsilon_3}{\epsilon_2}J_1J_1-\partial J_2+\frac{\epsilon_3}{\epsilon_2}\partial J_1 \right )\exp \left [ -\epsilon_1\phi^{(2)}_1\right ].
\end{eqnarray}
Note that this asymmetric form of the fundamental and the anti-fundamental field is related to our asymmetric choice of the free-boson ordering. The symmetric choice would to lead to both $J^+$ and $J^-$ at level one. We expect the two choices to correspond to the symmetric and the asymmetric Wakimoto realizations. The symmetric Wakimoto realization is a free field realization of $\widehat{\mathfrak{gl}}(2)$ in terms of two free bosons and parafermionic fields. Parafermionic fields can be bosonized and we expect to find our symmetric free field realization. Similarly, one can bosonize the $\beta,\gamma$ system of the Wakimoto realization in terms of two free bosons and a $\beta,\gamma$ system and we expect to recover our non-symmetric free field realization. Detailed discussion of the relation with Wakimoto realization is beyond the scope of this paper.

\paragraph{$Y_{0,0,1}$ vertex}

Let us normalize the free boson $\tilde{J}=\partial \tilde{\phi}^{(2)}_1$ of the second vertex as
\begin{eqnarray}
\tilde{J} (z)\tilde{J} (w)\sim \frac{1}{\epsilon_1 \epsilon_3}\frac{1}{(z-w)^2}.
\end{eqnarray}

The fundamental and the anti-fundamental representations associated to the second direction are then
\begin{eqnarray}
\tilde{M}_{\square}=\exp \left [\epsilon_1 \phi (z)\right ],\qquad \tilde{M}_{\bar{\square}}=\exp \left [-\epsilon_1 \phi (z)\right ].
\end{eqnarray}

\paragraph{Glued algebra}

Having identified the fields and the relevant fundamental and the anti-fundamental representation of each vertex, one can now easily construct the glued VOA. The Cartan elements of the $\widehat{\mathfrak{gl}}(2)$ Kac-Moody algebra can be fixed by requiring the correct OPE between them and with the fields $J_{12}\propto M_{\bar{\square}}$ and $J_{21}\propto M_{\square}$. One finds
\begin{eqnarray}
J_{11}=\epsilon_3\tilde{J},\qquad J_{22}=-\frac{\epsilon_2}{\epsilon_3}J+\epsilon_1\tilde{J}.
\end{eqnarray}
The normalization of generators $J_{12}$ and $J_{21}$ can be found from their OPE. One finds
\begin{eqnarray}\nonumber
J_{12}=\frac{\epsilon_2\epsilon_3}{\epsilon_1}M_{\bar{\square}},\qquad J_{21}= M_{\square}.
\end{eqnarray}
Note that the OPE of the exponential factors is trivial and both the second order and the first order pole come from the OPE of the $J_i$ fields with the exponential factor of the anti-fundamental field. All the OPEs of $\widehat{\mathfrak{gl}}(2)$ Kac-Moody algebra are reproduced. 

\paragraph{Generic modules}

Generic modules can be now constructed from
\begin{eqnarray}
M[q^1,q^2,q^3,q^4]=\exp \left [ q^1\phi^{(2)}_1 +q^2\phi^{(3)}_2+q^3\phi^{(3)}_3+q^4\tilde{\phi}^{(2)}_1 \right ]
\end{eqnarray}
where $q^1$ and $q^4$ are constrained by the condition
\begin{eqnarray}
q^1+q^4= \epsilon_3n
\end{eqnarray}
for some integer $n$.

For each such module, it is simple to compute the action of the $\widehat{\mathfrak{gl}}(2)$ generators on each such vector. Depending on the number $n$ in the constraint above, one gets different structure of the modules. For example, for $n>1$, one gets
\begin{eqnarray}\nonumber
J_{21}(z)M[q^1,q^2,q^3,q^4](w)&\propto& \frac{M[q^1+\epsilon_1,q^2,q^3,q^4-\epsilon_1]}{(z-w)^{n}}+\dots\\
J_{12}(z)M[q^1,q^2,q^3,q^4](w)&\propto& \mathcal{O}((z-w)^{n-2}).
\end{eqnarray}
For $n<1$, the singularity is present in the OPE with $J_{12}$ instead. We expect corresponding modules to be a special type of the irregular modules discussed in \cite{Gaiotto:2013rk}.

The most interesting situation appears when $n=1$. In such a case both $M_{\square}$ and $M_{\bar{\square}}$ have a simple pole in the OPE with generic modules and one obtains
\begin{eqnarray}\nonumber
J_{11}(z)M[-q^4+\epsilon_3,q^2,q^3,q^4](w)&\sim& \frac{q^4}{\epsilon_1} \frac{M[-q^4+\epsilon_3,q^2,q^3,q^4]}{z-w} \\ \nonumber
J_{22}(z)M[-q^4+\epsilon_3,q^2,q^3,q^4](w)&\sim& -\frac{q^1+q^2+q^4+\epsilon_2}{\epsilon_1}\frac{M[-q^4+\epsilon_3,q^2,q^3,q^4]}{z-w} \\ \nonumber
J_{12}(z)M[-q^4+\epsilon_3,q^2,q^3,q^4](w)&\sim& -\frac{(q^1+q^4)(q^2+q^4-\epsilon_3)}{\epsilon_1^2}\frac{M[-q^4+\epsilon_3-\epsilon_1,q^2,q^3,q^4+\epsilon_1]}{z-w}\\
J_{21}(z)M[-q^4+\epsilon_3,q^2,q^3,q^4](w)&\sim& \frac{M[-q^4+\epsilon_3+\epsilon_1,q^2,q^3,q^4-\epsilon_1]}{z-w}.
\label{gtgl}
\end{eqnarray}
We can see that the zero modes of $J_{12}$ and $J_{21}$ shift the exponent of $M[-q^4+\epsilon_3,q^2,q^3,q^4]$. The representation of the zero-mode subalgebra is thus spanned by $M[-q^4+\epsilon_3+n\epsilon_1,q^2,q^3,q^4-n\epsilon_1]$ for $n \in \mathbb{Z}$.

\paragraph{Gelfand-Tsetlin modules}

In this section, we show that the above action of zero modes generate a generic Gelfand-Tsetlin module of  $\widehat{\mathfrak{gl}}(2)$.

Gelfand-Tsetlin modules for $\mathfrak{gl}(2)$ are parametrized by a triple of  complex parameters
\begin{eqnarray}
\begin{pmatrix}
\lambda_{21}& \lambda_{22}\\
\multicolumn{2}{c}{\lambda_{11}}
\end{pmatrix}
\end{eqnarray}
where $\lambda_{11}$ and $\lambda_{11}+n$ are vectors of the same module. For generic values of parameters, the Gelfand-Tsetlin module is spanned by vectors with Gelfand-Tsetlin table of the form
\begin{eqnarray}
\begin{pmatrix}
\lambda_{21}& \lambda_{22}\\
\multicolumn{2}{c}{\lambda_{11}+n}
\end{pmatrix}
\end{eqnarray}
for each $n\in \mathbb{Z}$. Generators $J_{11},J_{22},J_{12},J_{21}$ act on such vectors as
\begin{eqnarray}\nonumber
J_{11}
\begin{pmatrix}
\lambda_{21}& \lambda_{22}\\
\multicolumn{2}{c}{\lambda_{11}}
\end{pmatrix}
&=&\lambda_{11}
\begin{pmatrix}
\lambda_{21}& \lambda_{22}\\
\multicolumn{2}{c}{\lambda_{11}}
\end{pmatrix},
\\ \nonumber
J_{22}
\begin{pmatrix}
\lambda_{21}& \lambda_{22}\\
\multicolumn{2}{c}{\lambda_{11}}
\end{pmatrix}
&=&(1+\lambda_{22}+\lambda_{21}-\lambda_{11})
\begin{pmatrix}
\lambda_{21}& \lambda_{22}\\
\multicolumn{2}{c}{\lambda_{11}}
\end{pmatrix},
\\ \nonumber
J_{12}
\begin{pmatrix}
\lambda_{21}& \lambda_{22}\\
\multicolumn{2}{c}{\lambda_{11}}
\end{pmatrix}
&=&-(\lambda_{11}-\lambda_{21})(\lambda_{11}-\lambda_{22})
\begin{pmatrix}
\lambda_{21}& \lambda_{22}\\
\multicolumn{2}{c}{\lambda_{11}+1}
\end{pmatrix},
\\
J_{21}
\begin{pmatrix}
\lambda_{21}& \lambda_{22}\\
\multicolumn{2}{c}{\lambda_{11}}
\end{pmatrix}
&=&
\begin{pmatrix}
\lambda_{21}& \lambda_{22}\\
\multicolumn{2}{c}{\lambda_{11}-1}
\end{pmatrix},
\end{eqnarray}

Comparing parameters $\lambda_{21},\lambda_{22},\lambda_{11}$ with the lifted Gukov-Witten parameters $q_i$ from (\ref{gtgl}), one gets\footnote{There are actually two solutions related by an exchange of $\lambda_{21}\leftrightarrow \lambda_{22}$.}
\begin{eqnarray}\nonumber
\lambda_{11}&=&\frac{q^4}{\epsilon_1},\\ \nonumber
\lambda_{21}&=&-\frac{q^3}{\epsilon_1},\\
\lambda_{22}&=&-\frac{q^2-\epsilon_3}{\epsilon_1}.
\end{eqnarray}
Note that fusion of a vector of the generic module with $J_{12}$ and $J_{21}$ shifts $q^4$ by an integral multiple of $\epsilon_1$, this corresponds exactly to the shift of parameter $\lambda_{11}$ by an integer as expected. Note also that the parameters associated to a given face of the toric diagram correspond to Gelfand-Tsetlin parameters of a given row of the Gelfand-Tsetlin table.

\subsection{Gelfand-Tsetlin modules for $\widehat{\mathfrak{gl}}(N)$ and their $\mathcal{W}$-algebras}

\paragraph{$\widehat{\mathfrak{gl}}(N)$ Kac-Moody Algebras}

\begin{figure}
  \centering
      \includegraphics[width=0.33\textwidth]{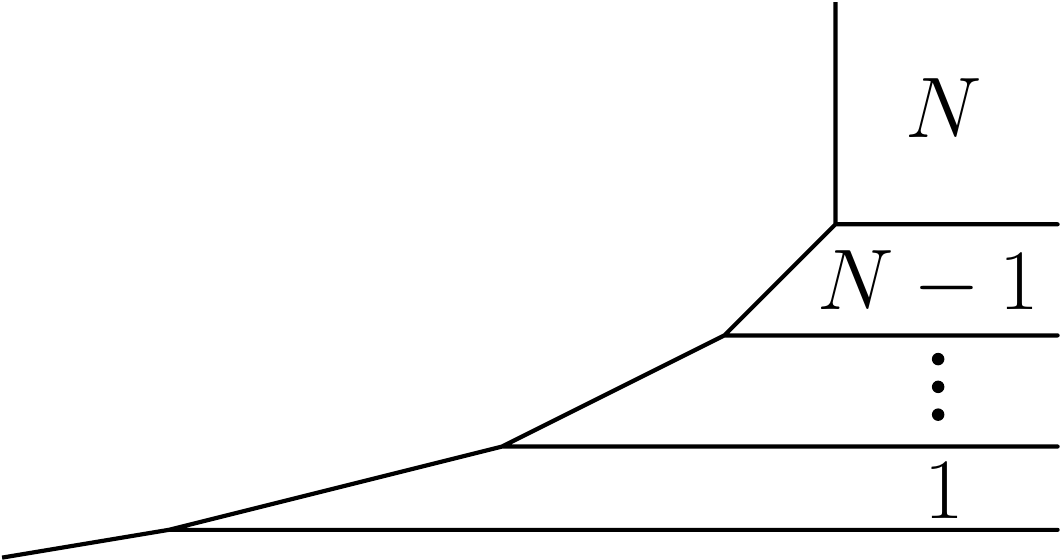}
  \caption{The web diagram associated to the $\widehat{\mathfrak{gl}}(N)$ Kac-Moody algebra.}
\label{un}
\end{figure}

In the previous section, we have described the structure of generic modules for the $\widehat{\mathfrak{gl}}(2)$ Kac-Moody algebra. Let us now comment on the structure of generic modules for any $\widehat{\mathfrak{gl}}(N)$ Kac-Moody algebra and $\mathcal{W}$-algebras associated to their Drinfeld-Sokolov reduction. 

The Kac-Moody algebra $\mathfrak{gl}(N)$ can be realized in terms of a web diagram in the figure \ref{un}. The lifted GW parameters associated to internal faces must be again equal up to  shifts induced by line operators supported at the $(1,0)$ interfaces, i.e. they differ by a multiple of $\epsilon_1+m \epsilon_2$, where $-(m,1)$ are charges of the finite interface of the given face. In the same way as in the case of the $\widehat{\mathfrak{gl}}(2)$ Kac-Moody algebra, one should be able to choose of the shifts of the lifted GW parameters such that the OPEs of $J_{ij}$ for $i>j$ with generic modules have OPE with a simple pole. Generic modules are then going to be parametrized by a Gelfand-Tsetlin table of $\frac{N(N+1)}{2}$ entries. For example, in the case of $\widehat{\mathfrak{gl}}(3)$, the Gelfand-Tsetlin table will be of the form
\begin{eqnarray}
\begin{pmatrix}
\lambda_{31}& \lambda_{32}& \lambda_{33}\\
\multicolumn{3}{c}{\lambda_{21}\ \  \lambda_{22}}\\
\multicolumn{3}{c}{\lambda_{11}}
\end{pmatrix}.
\end{eqnarray}
The parameters in each line will be shifted and renormalized GW parameters associated to a given face. The full modules is then spanned by the vectors 
\begin{eqnarray}
\begin{pmatrix}
\lambda_{31}&\ \ \  \ \lambda_{32}& \ \ \ \ \lambda_{33}\\
\multicolumn{3}{c}{\lambda_{21}+n_{1}\ \  \lambda_{22}+n_{2}}\\
\multicolumn{3}{c}{\lambda_{11}+n_{3}}
\end{pmatrix}
\end{eqnarray}
for any integers $n_1,n_2,n_3$. These shifts are generated by the fusion with bi-modules coming from line operators at each internal face. 

\paragraph{$\mathcal{W}$-algebras}

\begin{figure}
  \centering
      \includegraphics[width=0.18\textwidth]{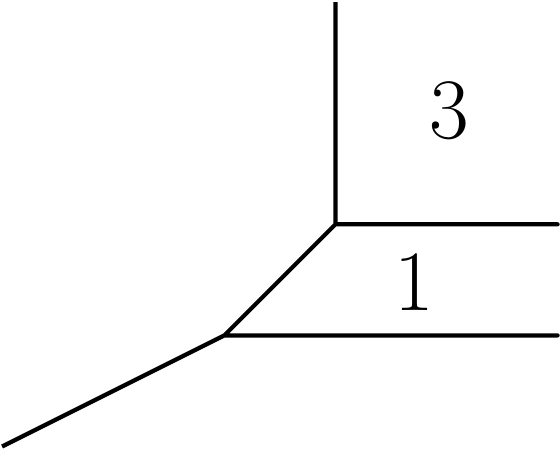}
  \caption{The web diagram associated to the $\mathcal{W}_3^{(2)}\times \widehat{\mathfrak{gl}}(1)$ algebra.}
\label{w32}
\end{figure}

The same structure of modules is expected also for similar configurations with different ranks of gauge groups. The corresponding algebra can be identified with a $\mathcal{W}$-algebra associated to the Drinfeld-Sokolov reduction of the $\widehat{\mathfrak{gl}}(N)$ Kac-Moody algebra possibly with extra symplectic bosons as discussed in \cite{Prochazka:2017qum}. The corresponding Gelfand-Tsetlin modules are parametrized by a generalized Gelfand-Tsetlin table with $N_i$ complex numbers associated to each face with gauge group $U(N_i)$. Except of the $N_1$ corner parameters in the upper-right face, all the other parameters can be shifted by fusion with bi-modules added to the algebra. 

For example the algebra $\mathcal{W}^{(2)}_3\times \widehat{\mathfrak{gl}}(1)$ associated to the diagram \ref{w32}. Have the following Gelfand-Tsetlin table parameterizing generic modules
\begin{eqnarray}
\begin{pmatrix}
\lambda_{31}&\lambda_{32}& \lambda_{33}\\
\multicolumn{3}{c}{\lambda_{11}}
\end{pmatrix}.
\end{eqnarray}
The full modules is spanned by such vectors with the parameter $\lambda_{11}$ shifted by any integer.

\section{Outlook}

Finally, let us mention few possible extensions of the discussion from this paper.

\paragraph{AGT for spiked instantons}

Recently, partition functions of theories coming from branes wrapping various four-cycles in Calabi-Yau four-folds have been considered in a series of papers \cite{Nekrasov:2015wsu,Nekrasov:2016qym,Nekrasov:2016gud,Nekrasov:2016ydq,Nekrasov:2017rqy,Nekrasov:2017gzb,Nekrasov:2017cih}. If one restricts to the toric three-fold case, one can relate the corresponding geometric setup to the one considered in this paper along the lines of \cite{Leung:1997tw,Prochazka:2017qum}. Generic modules discussed here can be then identified with the equivariant cohomology of the moduli space of spiked instantons with a geometric origin of the VOA action leading to AGT correspondence for spiked instantons. This issue, together with the relation with affine Yangians and the cohomological Hall algebra will be further discussed in \cite{Next3}.

\paragraph{Ramification of geometric Langlands}

Gukov-Witten defects were originaly introduced in the context of the ramification (inclusion of punctured Riemann surfaces) of the geometric Langlands program in \cite{Gukov:2006jk}. In the recenly introduced corner approach to the geometric Langlands program from \cite{Creutzig:2017uxh,Frenkel:2018dej}, Gukov-Witten defects play the role of generic modules for kernel VOAs. This work sets foundations for the inclusion of the ramification in this context.

\paragraph{More general VOA$[M_4]$}

The web algebras studied in this paper are associated to toric surfaces in toric Calabi-Yau threefolds. There are many hints that the story is much more general as discussed for example in \cite{Dedushenko:2017tdw,Feigin:2018bkf,Gadde:2013sca}. It would be interesting to extend the analysis also to non-toric surfaces.

\paragraph{Irregular modules}
Apart from modules induced from the algebra of zero-modes discussed here, one can consider more complicated irregular modules from \cite{Gaiotto:2012sf,Gaiotto:2013rk}. It would be interesting to extend the discussion to such modules. This might also play important role in the wild ramification of the Geometric Langlands program \cite{Witten:2007td}.

\paragraph{Rational levels}

As we have seen in the case of the Ising model, null-fields appear at rational values of parameters $h_i$. Considering quotients of the algebra by such null fields lead to new constraints on  modules. It would be be interesting to address this issue in our context. Many properties of the non-generic case might have a gauge theory explanation in terms of an existence of line operators in the bulk.

\paragraph{Ortho-symplectic algebras}

In this paper, we restrict our attention to algebras associated to $U(N)$ gauge theories. There should exist an analogous story associated to ortho-symplectic groups as briefly sketched in \cite{Gaiotto:2017euk}.

\paragraph{DIM algebra}

There exists a DIM algebra approach \cite{Ding:1996mq,Miki,Awata:2011dc,Awata:2011ce,Mironov:2016yue,Awata:2016riz,Awata:2016mxc,Awata:2016bdm,Bourgine:2016vsq,Bourgine:2017jsi} to the categorification of DT-invariants associated to toric Calabi-Yau three-folds. The DIM algebra itself is a q-deformation of the affine Yangian of $\mathfrak{gl}(1)$. The specializations of the affine Yangian are isomorphic to the $Y_{N_1,N_2,N_3}$ as proved in \cite{Next3} based on previous work of \cite{Prochazka:2015aa,Prochazka:2017qum}. It would be nice to find a precise relation between the construction of intertwining operators of the DIM algebra and the gluing of \cite{Prochazka:2017qum}.

\paragraph{Gelfand-Tsetlin modules}

We have sketched that Gelfand-Tsetlin modules of $\mathcal{W}$-algebras naturally appear in the context of gauge theories. It would be interesting to prove that the glued modules are indeed induced from Gelfand-Tsetlin modules of the zero-modes algebra and explore the relation further. 

\paragraph{Free field realization of degenerate modules}

Given a free field realization of the algebra $Y_{N_1,N_2,N_3}$, we have various realizations of maximally degenerate modules. When inserted in a correlator, they all give an equivalent result (if non-vanishing) in all the examples. It is desirable to explore this issue further as in \cite{Felder:1988zp}. Moreover, we do not give a closed-form expression for the descendant realization of maximally degenerate modules. It would be nice to find an explicit formula.

\paragraph{$R$-matrix}
The different free field representations corresponding to different ordering of free fields in the Miura transform are related by an $R$-matrix \cite{Maulik:2012rm, Smirnov:2013hh, Zhu:2015nha}. This $R$-matrix satisfies the Yang-Baxter equation which is the starting point of many developments in the quantum integrable models. More detailed exploration of this should strengthen the relation between the algebraic structures of two-dimensional quantum field theory on one hand and the theory of quantum integrable models and Yangians on the other hand.

\section*{Acknowledgements} 
We would like to thank Mikhail Bershtein, Thomas Creutzig, Davide Gaiotto, Sergei Gukov, Libor K\v{r}i\v{z}ka, Andrew Linshaw, Alexey Litvinov, Faroogh Moosavian, Ivo Sachs, Yan Soibelman, Yaping Yang, Ben Webster, Gufang Zhao for useful discussions. We thank Kris Thielemans for his Mathematica package OPEdefs. The research of TP was supported by the DFG Transregional Collaborative Research Centre TRR 33 and the DFG cluster of excellence Origin and Structure of the Universe. The research of MR was supported by the Perimeter Institute for Theoretical Physics. Research at Perimeter Institute is supported by the Government of Canada through the Department of Innovation, Science and Economic Development and by the Province of Ontario through the Ministry of Research, \& Innovation and Science.

\appendix

\section{Transformation between primary and quadratic bases}
\label{primaryquadraticformulas}
We list first few formulas relating the primary basis generators $W_j$ to the generators $U_j$ in quadratic basis.
\begin{eqnarray*}
W_1 & = & -U_1 \\
W_2 & = & -U_2 + \frac{N-1}{2N} (U_1 U_1) + \frac{(N-1)\alpha_0}{2} U_1^\prime \\
W_3 & = & -U_3 + \frac{N-2}{N} (U_1 U_2) - \frac{(N-1)(N-2)}{3N^2} (U_1 (U_1 U_1))- \frac{(N-1)(N-2) \alpha_0}{2N} (U_1^\prime U_1) \\
& & + \frac{(N-2)\alpha_0}{2} U_2^\prime - \frac{(N-1)(N-2)\alpha_0^2}{12} U_1^{\prime\prime} \\
W_4 & = & -U_4 + \frac{(N-3)(N-2)(N-1)(5N+6)\alpha_0(\alpha_0^2N^2-\alpha_0^2N-1)}{2N^2(5\alpha_0^2N^3-5\alpha_0^2N-5N-17)} (U_1^\prime (U_1 U_1)) \\
& & +\frac{(N-3)(N-2)(N-1)(\alpha_0^2N^2-\alpha_0^2N-1)(2\alpha_0^2N^2+3\alpha_0^2N-3)}{4N^2(5\alpha_0^2N^3-5\alpha_0^2N-5N-17)} (U_1^\prime U_1^\prime) \\
& & -\frac{(N-3)(N-2)(N-1)\alpha_0(5\alpha_0^2N^2+7\alpha_0^2N-5)}{2N(5\alpha_0^2N^3-5\alpha_0^2N-5N-17)} (U_1^\prime U_2) \\
& & +\frac{(N-3)(N-2)(N-1)(5N+6)(\alpha_0^2N^2-\alpha_0^2N-1)}{4N^3(5 \alpha_0^2N^3-5\alpha_0^2N-5N-17)} (U_1,(U_1,(U_1,U_1))) \\
& & -\frac{(N-3)(N-2)(5N+6)(\alpha_0^2N^2-\alpha_0^2N-1)}{N^2(5\alpha_0^2N^3-5\alpha_0^2N-5N-17)} (U_1 (U_1 U_2)) \\
& & +\frac{(N-3)(N-2)(5\alpha_0^2N^2+7\alpha_0^2N-5)}{2N(5\alpha_0^2N^3-5\alpha_0^2N-5N-17)} (U_2 U_2) \\
& & +\frac{(N-3)(N-2)(N-1)(2\alpha_0^4N^4-5\alpha_0^2N^3-2\alpha_0^4N^2-7\alpha_0^2N^2-4\alpha_0^2N+5N-2)}{4N^2(5\alpha_0^2N^3-5\alpha_0^2N-5N-17)} (U_1^{\prime\prime}U_1) \\
& & +\frac{(N-3)(N-2)(N-1)\alpha_0(\alpha_0^4N^4-10\alpha_0^2N^3-\alpha_0^4N^2-14\alpha_0^2N^2-2\alpha_0^2N+10N-1)}{24N(5\alpha_0^2N^3-5\alpha_0^2N-5N-17)} U_1^{(3)} \\
& & -\frac{(N-3)(N-2)(2\alpha_0^4N^4-2\alpha_0^4N^2-5\alpha_0^2N^2-11\alpha_0^2N+3)}{4N(5\alpha_0^2N^3-5\alpha_0^2N-5N-17)} U_2^{\prime\prime} \\
& & -\frac{(N-3)(N-2)\alpha_0}{2N} (U_1 U_2^\prime) +\frac{(N-3)}{N} (U_1,U_3) +\frac{(N-3)\alpha_0}{2} U_3^\prime
\end{eqnarray*}
We choose the normalization such that $W_j = -U_j + \ldots$. Since this choice of normalization is rather arbitrary, we should also specify the values of structure constants that fix the relative normalization of the charges:
\begin{eqnarray*}
C_{11}^0 & = & N \\
C_{22}^0 & = & \frac{1}{2}(N-1)(1-N(N+1)\alpha_0^2) \\
C_{33}^0 & = & \frac{(N-1)(N-2)(1-N(N+1)\alpha_0^2)(4-N(N+2)\alpha_0^2)}{6N} \\
C_{44}^0 & = & \frac{(N-1)(N-2)(N-3)(N+1)}{4N^2(5N^3\alpha_0^2-5N\alpha_0^2-5N-17)} \times (1-N(N+1)\alpha_0^2)(4-N(N+2)\alpha_0^2) \times \\
& & \times (9-N(N+3)\alpha_0^2)(1-N(N-1)\alpha_0^2) \\
C_{55}^0 & = & \frac{(N-1)(N-2)(N-3)(N-4)(N+1)}{10N^3(7N^3\alpha_0^2-7N\alpha_0^2-7N-107)} \times (1-N(N+1)\alpha_0^2)(4-N(N+2)\alpha_0^2) \times \\
& & \times (9-N(N+3)\alpha_0^2)(16-N(N+4)\alpha_0^2)(1-N(N-1)\alpha_0^2)
\end{eqnarray*}
Acting on the highest weight state, the relation between charges becomes somewhat simpler
\begin{eqnarray*}
w_1 & = & -u_1 \\
w_2 & = & -u_2 + \frac{N-1}{2N} u_1^2 - \frac{(N-1)\alpha_0}{2} u_1 \\
w_3 & = & -u_3 + \frac{N-2}{N} u_1 u_2 - \frac{(N-1)(N-2)}{3N^2} u_1^3 - (N-2)\alpha_0 u_2 \\
& & + \frac{(N-1)(N-2)\alpha_0}{2N} u_1^2 - \frac{(N-1)(N-2)\alpha_0^2}{6} u_1 \\
w_4 & = & -u_4 + \frac{N-3}{N} u_1 u_3 + \frac{(N-3)(N-2)(5\alpha_0^2N^2+7\alpha_0^2N-5)}{2N(5\alpha_0^2N^3-5\alpha_0^2N-5N-17)} u_2^2 \\
& & -\frac{(N-3)(N-2)(5N+6)(\alpha_0^2N^2-\alpha_0^2N-1)}{N^2(5\alpha_0^2N^3-5\alpha_0^2N-5N-17)} u_1^2 u_2 \\
& & + \frac{(N-3)(N-2)(N-1)(5N+6)(\alpha_0^2N^2-\alpha_0^2N-1)}{4N^3(5 \alpha_0^2N^3-5\alpha_0^2N-5N-17)} u_1^4 - \frac{3(N-3)\alpha_0}{2} u_3 \\
& & + \frac{(N-3)(N-2)\alpha_0(15\alpha_0^2N^3+2\alpha_0^2N^2-17\alpha_0^2N-15 N-29)}{2N(5\alpha_0^2N^3-5\alpha_0^2N-5N-17)} u_1 u_2 \\
& & - \frac{(N-3)(N-2)(N-1)(5N+6)\alpha_0(\alpha_0^2N^2-\alpha_0^2N-1)}{2N^2(5\alpha_0^2N^3-5\alpha_0^2N-5N-17)} u_1^3 \\
& & - \frac{(N-3)(N-2)(6\alpha_0^4N^4-6\alpha_0^4N^2-5\alpha_0^2N^2-19\alpha_0^2N-1)}{2N(5\alpha_0^2N^3-5\alpha_0^2N-5N-17)} u_2 \\
& & + \frac{(N-3)(N-2)(N-1)(\alpha_0^2N^2-\alpha_0^2N-1)(6\alpha_0^2N^2+7\alpha_0^2N+1)}{4N^2(5\alpha_0^2 N^3-5\alpha_0^2N-5N-17)} u_1^2 \\
& & - \frac{(N-3)(N-2)(N-1)\alpha_0(\alpha_0^2N^2-\alpha_0^2N-1)(\alpha_0^2N^2+\alpha_0^2N+1)}{4N(5\alpha_0^2N^3-5\alpha_0^2N-5N-17)} u_1
\end{eqnarray*}
As one can see from these expressions, they are becoming increasingly complicated and it is unfortunate that no closed-form expression for the primary charges is known.

\section{$Y_{1,1,0}$ from $\mathcal{W}_3$}
\label{w3quotient}
Since it turns out that $Y_{1,1,0}$ truncation of $\mathcal{W}_{1+\infty}$ is a special case of $\mathcal{W}_3$ algebra, it is instructive to have a look how this happens at the level of the generating function of higher spin charges. Let us start with $\mathcal{W}_3$ algebra with parameters
\begin{equation}
h_1 = \sqrt{\frac{2}{3}}, \quad\quad h_2 = -\sqrt{\frac{3}{2}}, \quad\quad h_3 = \frac{1}{\sqrt{6}}, \quad \psi_0 = 3
\end{equation}
and as usual
\begin{equation}
h_3 = \alpha_0, \quad\quad \psi_0 = N
\end{equation}
This choice in particular means that $\lambda_3 = 3$ and $c_{\infty} = -2$ which is the condition for $Y_{1,1,0}$. In $\mathcal{W}_3$, the generating function of higher spin charges of a highest weight representation is of the form
\begin{equation}
\psi(u) = \frac{(u-x_1)(u-x_2)(u-x_3)}{(u-x_1+h_3)(u-x_2+h_3)(u-x_3+h_3)}
\end{equation}
(see \ref{psihw}). From this and the formulas of appendix \ref{primaryquadraticformulas} we can determine the $w$-charges
\begin{eqnarray}\nonumber 
w_1 & = & -x_1-x_2-x_3 + \sqrt{6} \\ \nonumber
w_2 & = & \frac{1}{3} \left( x_1^2 + x_2^2 + x_3^2 - x_1 x_2 - x_1 x_3 - x_2 x_3 \right) - \frac{1}{6} \\
w_3 & = & \frac{1}{27} \left(x_1+x_2-2x_3\right) \left(x_1-2x_2+x_3\right) \left(-2x_1+x_2+x_3\right)
\end{eqnarray}
(and higher primary charges vanishing since we are in $\mathcal{W}_3$). We can now impose the $Y_{110}$ truncation relation (\ref{y110constraint}) which leads to a sextic equation in zeros $x_j$,
\begin{equation}
0 = \left( (x_1-x_2)^2 - \frac{1}{6} \right) \left( (x_1-x_3)^2 - \frac{1}{6} \right) \left( (x_2-x_3)^2 - \frac{1}{6} \right)
\end{equation}
which has a solution
\begin{equation}
  x_2-x_3 = \frac{1}{\sqrt{6}} = h_3
\end{equation}
(and other five permutations of this) with the corresponding generating function
\begin{equation}
\psi(u) = \frac{(u-x_1)(u-x_2)}{\left(u-x_1+\frac{1}{\sqrt{6}}\right)\left(u-x_2+\sqrt{\frac{2}{3}}\right)} = \frac{(u-x_1)(u-x_2)}{(u-x_1+h_3)(u-x_2+h_1)}
\end{equation}
which is exactly what we expect from the algebra $Y_{1,0,1}$, i.e. $x_2$ and $x_3$ become bound together and behave as a zero of the first type while $x_1$ remains a zero of the third type.

\section{More details of $Y_{0,1,2}$}
\label{appendixy012}
Structure constants of $Y_{012}$ in the primary basis
\begin{eqnarray*}
C_{34}^3 & = & \frac{12(c+2)(c+10)^2(5c-4) C_{33}^0}{c(c+7)(2c-1)(5c+22) C_{33}^4} \\
C_{44}^0 & = & \frac{12(c+2)(c+10)^2(5c-4) (C_{33}^0)^2}{c(c+7)(2c-1)(5c+22) (C_{33}^4)^2} \\
C_{44}^4 & = & \frac{36(c+10)(5c^3+45c^2-6c-64) C_{33}^0}{c(c+7)(2c-1)(5c+22) C_{33}^4} \\
C_{44}^{(33)} & = & \frac{72(c+4)(c+10)^2 C_{33}^0}{c(c+7)(2c-1) (C_{33}^4)^2} \\
C_{45}^5 & = & \frac{30(c+10)(85c^3+1076c^2-188c-2304) C_{33}^0}{c(2c-1)(5c+22)(7c+114) C_{33}^4} \\
C_{35}^4 & = & \frac{60(c-1)(c+13)(5c+22) C_{33}^0}{c(2c-1)(7c+114) C_{34}^5} \\
C_{45}^3 & = & \frac{C_{34}^5 C_{55}^0}{C_{33}^0} \\
C_{35}^{(33)} & = & \frac{90(c+10)^2(7c+68)C_{33}^0}{c(2c-1)(7c+114) C_{33}^4 C_{34}^5} \\
C_{45}^{(34)} & = & \frac{120(c+10)(c+13)(7c+26) C_{33}^0}{c(2c-1)(7c+114) C_{33}^4 C_{34}^5} \\
C_{55}^0 & = & \frac{720(c-1)(c+2)(c+10)^2(c+13)(5c-4) (C_{33}^0)^3}{c^2(c+7)(2c-1)^2(7c+114) (C_{33}^4)^2 (C_{34}^5)^2} \\
C_{55}^4 & = & \frac{1800(c-1)(c+10)(c+13)(85c^3+1076c^2-188c-2304) (C_{33}^0)^2}{c^2(2c-1)^2(7c+114)^2 C_{33}^4 (C_{34}^5)^2} \\
C_{55}^{(33)} & = & \frac{1800 (c+10)^2 (259c^4+6979c^3+46628c^2-26404c-154512) (C_{33}^0)^2}{c^2(2c-1)^2(7c+114)^2 (C_{33}^4)^2 (C_{34}^5)^2} \\
C_{55}^{(35)} & = & \frac{300(c+10)(49c^4+1146c^3+3222c^2-32276c-145776) C_{33}^0}{c(2c-1)(7c+114)(29c^2+533c-870) C_{33}^4 C_{34}^5} \\
& & + \frac{3c(2c-1)(7c+114) C_{33}^4 C_{34}^5 C_{55}^{(33)^{\prime\prime}}}{8(c+10)(29c^2+533c-870) C_{33}^0} \\
C_{55}^{(44)} & = & \frac{600(c+7)(c+13)(35c^3+914c^2+2412c+2568) C_{33}^0}{c(2c-1)(7c+114)(29c^2+533c-870) (C_{34}^5)^2} \\
& & -\frac{c(c+7)(2c-1)(7c+114) (C_{33}^4)^2 C_{55}^{(33)^{\prime\prime}}}{2(c+10)^2(29c^2+533c-870) C_{33}^0}
\end{eqnarray*}
Spin $3$ charge in terms of $u(1)$ charges in the free field representation
\begin{eqnarray*}
w_3 & = & - \frac{h^2q_1^3}{3(2 h^2-1)^2} + \frac{(h^2-1)q_1^2q_2}{(2h^2-1)^2} + \frac{(h^2-1)q_1^2q_3}{(2h^2-1)^2} - \frac{hq_1^2}{2(2h^2-1)}\\
& & + \frac{(h^2-1)q_1q_2^2}{(2h^2-1)^2} - \frac{4(h^2-1)q_1q_2q_3}{(2h^2-1)^2} + \frac{2(h^2-1)q_1q_2}{h(2 h^2-1)} \\
& & + \frac{(h^2-1)^2(2h^2-3)q_1q_3^2}{(2h^2-1)^2} - \frac{(h^2-1)(h^2-2)q_1q_3}{h(2h^2-1)} \\
& & - \frac{h^2q_2^3}{3(2h^2-1)^2} + \frac{(h^2-1)q_2^2q_3}{(2h^2-1)^2} + \frac{(h^2-2)q_2^2}{2h(2h^2-1)} + \frac{(h^2-1)^2(2h^2-3)q_2q_3^2}{(2h^2-1)^2} \\
& & -\frac{(h^2-1)(3h^2-4)q_2q_3}{h(2h^2-1)} + \frac{(5h^2-6)q_2}{6h^2} - \frac{2(h^2-1)^3(2h^2-3)q_3^3}{3(2h^2-1)^2} \\
& & +\frac{(h^2-1)^2(2h^2-3)q_3^2}{h(2h^2-1)}-\frac{(h^2-1)(2h^2-3)q_3}{3h^2}-\frac{q_1}{6}
\end{eqnarray*}

\bibliography{Gluing2}

\providecommand{\href}[2]{#2}\begingroup\raggedright\begin{thebibliography}{10}

\bibitem{Gaiotto:2017euk}
D.~Gaiotto and M.~Rap\v{c}\'{a}k, {\it {Vertex Algebras at the Corner}},
  \href{http://arxiv.org/abs/1703.00982}{{\tt arXiv:1703.00982}}.

\bibitem{Nekrasov:2010aa}
N.~Nekrasov and E.~Witten, {\it {The Omega Deformation, Branes, Integrability,
  and Liouville Theory}},  \href{http://arxiv.org/abs/1002.0888}{{\tt
  arXiv:1002.0888}}.

\bibitem{Gaiotto:2011nm}
D.~Gaiotto and E.~Witten, {\it {Knot Invariants from Four-Dimensional Gauge
  Theory}},  {\em Adv. Theor. Math. Phys.} {\bf 16} (2012), no.~3 935--1086,
  [\href{http://arxiv.org/abs/1106.4789}{{\tt arXiv:1106.4789}}].

\bibitem{Creutzig:2017uxh}
T.~Creutzig and D.~Gaiotto, {\it {Vertex Algebras for S-duality}},
  \href{http://arxiv.org/abs/1708.00875}{{\tt arXiv:1708.00875}}.

\bibitem{Prochazka:2017qum}
T.~Proch\'{a}zka and M.~Rap\v{c}\'{a}k, {\it {Webs of W-algebras}},
  \href{http://arxiv.org/abs/1711.06888}{{\tt arXiv:1711.06888}}.

\bibitem{Kapustin:aa}
A.~Kapustin and E.~Witten, {\it Electric-magnetic duality and the geometric
  langlands program},  \href{http://arxiv.org/abs/hep-th/0604151}{{\tt
  hep-th/0604151}}.

\bibitem{Witten:2010aa}
E.~Witten, {\it A new look at the path integral of quantum mechanics},
  \href{http://arxiv.org/abs/1009.6032}{{\tt arXiv:1009.6032}}.

\bibitem{Witten:2011aa}
E.~Witten, {\it Fivebranes and knots},
  \href{http://arxiv.org/abs/1101.3216}{{\tt arXiv:1101.3216}}.

\bibitem{Mikhaylov:2014aa}
V.~Mikhaylov and E.~Witten, {\it Branes and supergroups},
  \href{http://arxiv.org/abs/1410.1175}{{\tt arXiv:1410.1175}}.

\bibitem{Aharony:aa}
O.~Aharony, A.~Hanany, and B.~Kol, {\it {Webs of $(p,q)$ 5-branes, Five
  Dimensional Field Theories and Grid Diagrams}},
  \href{http://arxiv.org/abs/hep-th/9710116}{{\tt hep-th/9710116}}.

\bibitem{Gaiotto:2008aa}
D.~Gaiotto and E.~Witten, {\it {S-Duality of Boundary Conditions In N=4 Super
  Yang-Mills Theory}},  \href{http://arxiv.org/abs/0807.3720}{{\tt
  arXiv:0807.3720}}.

\bibitem{Gaiotto:2008ab}
D.~Gaiotto and E.~Witten, {\it {Janus Configurations, Chern-Simons Couplings,
  And The Theta-Angle in N=4 Super Yang-Mills Theory}},
  \href{http://arxiv.org/abs/0804.2907}{{\tt arXiv:0804.2907}}.

\bibitem{Gaiotto:2008ac}
D.~Gaiotto and E.~Witten, {\it {Supersymmetric Boundary Conditions in N=4 Super
  Yang-Mills Theory}},  \href{http://arxiv.org/abs/0804.2902}{{\tt
  arXiv:0804.2902}}.

\bibitem{Pope:1989sr}
C.~N. Pope, L.~J. Romans, and X.~Shen, {\it {$W$(infinity) and the Racah-wigner
  Algebra}},  {\em Nucl. Phys.} {\bf B339} (1990) 191--221.

\bibitem{Pope:1989ew}
C.~N. Pope, L.~J. Romans, and X.~Shen, {\it {The Complete Structure of
  W(Infinity)}},  {\em Phys. Lett.} {\bf B236} (1990) 173--178.

\bibitem{Pope:1990kc}
C.~N. Pope, L.~J. Romans, and X.~Shen, {\it {A New Higher Spin Algebra and the
  Lone Star Product}},  {\em Phys. Lett.} {\bf B242} (1990) 401--406.

\bibitem{Kac:1995sk}
V.~Kac and A.~Radul, {\it {Representation theory of the vertex algebra
  W(1+infinity)}},  \href{http://arxiv.org/abs/hep-th/9512150}{{\tt
  hep-th/9512150}}.

\bibitem{Yu:1991bk}
F.~Yu and Y.-S. Wu, {\it {Nonlinearly deformed W(infinity) algebra and second
  Hamiltonian structure of KP hierarchy}},  {\em Nucl. Phys.} {\bf B373} (1992)
  713--734.

\bibitem{deBoer:1993gd}
J.~de~Boer, L.~Feher, and A.~Honecker, {\it {A Class of W algebras with
  infinitely generated classical limit}},  {\em Nucl. Phys.} {\bf B420} (1994)
  409--446, [\href{http://arxiv.org/abs/hep-th/9312049}{{\tt hep-th/9312049}}].
  [,409(1993)].

\bibitem{Khesin:1994ey}
B.~Khesin and F.~Malikov, {\it {Universal Drinfeld-Sokolov reduction and
  matrices of complex size}},  {\em Commun. Math. Phys.} {\bf 175} (1996)
  113--134, [\href{http://arxiv.org/abs/hep-th/9405116}{{\tt hep-th/9405116}}].

\bibitem{Hornfeck:1994is}
K.~Hornfeck, {\it {W algebras of negative rank}},  {\em Phys. Lett.} {\bf B343}
  (1995) 94--102, [\href{http://arxiv.org/abs/hep-th/9410013}{{\tt
  hep-th/9410013}}].

\bibitem{Blumenhagen:1994wg}
R.~Blumenhagen, W.~Eholzer, A.~Honecker, K.~Hornfeck, and R.~Hubel, {\it {Coset
  realization of unifying W algebras}},  {\em Int. J. Mod. Phys.} {\bf A10}
  (1995) 2367--2430, [\href{http://arxiv.org/abs/hep-th/9406203}{{\tt
  hep-th/9406203}}].

\bibitem{Gaberdiel:2012aa}
M.~R. Gaberdiel and R.~Gopakumar, {\it Triality in minimal model holography},
  \href{http://arxiv.org/abs/1205.2472}{{\tt arXiv:1205.2472}}.

\bibitem{Prochazka:2014aa}
T.~Proch\'azka, {\it Exploring $\mathcal{W}_{\infty}$ in the quadratic basis},
  \href{http://arxiv.org/abs/1411.7697}{{\tt arXiv:1411.7697}}.

\bibitem{Linshaw:2017tvv}
A.~R. Linshaw, {\it {Universal two-parameter $\mathcal{W}_{\infty}$-algebra and
  vertex algebras of type $\mathcal{W}(2,3,\dots, N)$}},
  \href{http://arxiv.org/abs/1710.02275}{{\tt arXiv:1710.02275}}.

\bibitem{Schiffmann:2012gf}
O.~Schiffmann and E.~Vasserot, {\it {Cherednik algebras, W algebras and the
  equivariant cohomology of the moduli space of instantons on A2}},
  \href{http://arxiv.org/abs/1202.2756}{{\tt arXiv:1202.2756}}.

\bibitem{Maulik:2012rm}
D.~Maulik and A.~Okounkov, {\it Quantum groups and quantum cohomology},
  \href{http://arxiv.org/abs/1211.1287}{{\tt arXiv:1211.1287}}.

\bibitem{Braverman:2014ys}
A.~Braverman, M.~Finkelberg, and H.~Nakajima, {\it Instanton moduli spaces and
  w-algebras},  \href{http://arxiv.org/abs/1406.2381}{{\tt arXiv:1406.2381}}.

\bibitem{tsymbaliuk2017affine}
A.~Tsymbaliuk, {\it {The affine Yangian of $gl(1)$ revisited}},  {\em Advances
  in Mathematics} {\bf 304} (2017) 583--645.

\bibitem{Prochazka:2015aa}
T.~Proch\'azka, {\it {W-symmetry, topological vertex and affine Yangian}},
  \href{http://arxiv.org/abs/1512.07178}{{\tt arXiv:1512.07178}}.

\bibitem{Zhu:2015nha}
R.-D. Zhu and Y.~Matsuo, {\it {Yangian associated with 2D $\mathcal{N} = 1$
  SCFT}},  {\em PTEP} {\bf 2015} (2015), no.~9 093A01,
  [\href{http://arxiv.org/abs/1504.04150}{{\tt arXiv:1504.04150}}].

\bibitem{Gaberdiel:2017dbk}
M.~R. Gaberdiel, R.~Gopakumar, W.~Li, and C.~Peng, {\it {Higher Spins and
  Yangian Symmetries}},  {\em JHEP} {\bf 04} (2017) 152,
  [\href{http://arxiv.org/abs/1702.05100}{{\tt arXiv:1702.05100}}].

\bibitem{Gukov:2006jk}
S.~Gukov and E.~Witten, {\it {Gauge Theory, Ramification, And The Geometric
  Langlands Program}},  \href{http://arxiv.org/abs/hep-th/0612073}{{\tt
  hep-th/0612073}}.

\bibitem{bershtein}
M.~Bershtein, B.~L. Feigin, and G.~Merzon, {\it Plane partitions with a "pit":
  generating functions and representation theory},
  \href{http://arxiv.org/abs/1512.08779}{{\tt arXiv:1512.08779}}.

\bibitem{Litvinov:2016mgi}
A.~Litvinov and L.~Spodyneiko, {\it {On W algebras commuting with a set of
  screenings}},  {\em JHEP} {\bf 11} (2016) 138,
  [\href{http://arxiv.org/abs/1609.06271}{{\tt arXiv:1609.06271}}].

\bibitem{Fateev:1987zh}
V.~A. Fateev and S.~L. Lukyanov, {\it {The Models of Two-Dimensional Conformal
  Quantum Field Theory with Z(n) Symmetry}},  {\em Int. J. Mod. Phys.} {\bf A3}
  (1988) 507. [507(1987)].

\bibitem{Wang:1998bt}
W.-q. Wang, {\it {Classification of irreducible modules of $\mathcal{W}_3$
  algebra with c=-2}},  {\em Commun. Math. Phys.} {\bf 195} (1998) 113--128.

\bibitem{1207.3909}
T.~Arakawa, C.~H. Lam, and H.~Yamada, {\it Zhu's algebra, $c_2$-algebra and
  $c_2$-cofiniteness of parafermion vertex operator algebras},  2012.

\bibitem{Dotsenko:1984nm}
V.~S. Dotsenko and V.~A. Fateev, {\it {Conformal Algebra and Multipoint
  Correlation Functions in Two-Dimensional Statistical Models}},  {\em Nucl.
  Phys.} {\bf B240} (1984) 312. [,653(1984)].

\bibitem{Felder:1988zp}
G.~Felder, {\it {BRST Approach to Minimal Models}},  {\em Nucl. Phys.} {\bf
  B317} (1989) 215. [Erratum: Nucl. Phys.B324,548(1989)].

\bibitem{1409.8413}
V.~Futorny, D.~Grantcharov, and L.~E. Ramirez, {\it Irreducible generic
  gelfand-tsetlin modules of $\mathfrak{gl}(n)$},
  \href{http://arxiv.org/abs/arXiv:1409.8413}{{\tt arXiv:1409.8413}}.

\bibitem{luk1988quantization}
S.~Lukyanov, {\it {Quantization of the Gel'fand-Dikii brackets}},  {\em
  Functional Analysis and Its Applications} {\bf 22} (1988), no.~4 255--262.

\bibitem{DiFrancesco:1990qr}
P.~Di~Francesco, C.~Itzykson, and J.~B. Zuber, {\it {Classical W algebras}},
  {\em Commun. Math. Phys.} {\bf 140} (1991) 543--568.

\bibitem{Next3}
M.~Rap\v{c}\'{a}k, Y.~Soibelman, Y.~Yang, and G.~Zhao, {\it Cohomological hall
  algebras, vertex algebras and instantons: in preparation}, .

\bibitem{Fukuda:2015ura}
M.~Fukuda, S.~Nakamura, Y.~Matsuo, and R.-D. Zhu, {\it {SH$^{c}$ realization of
  minimal model CFT: triality, poset and Burge condition}},  {\em JHEP} {\bf
  11} (2015) 168, [\href{http://arxiv.org/abs/1509.01000}{{\tt
  arXiv:1509.01000}}].

\bibitem{Bouwknegt:1992wg}
P.~Bouwknegt and K.~Schoutens, {\it {W symmetry in conformal field theory}},
  {\em Phys. Rept.} {\bf 223} (1993) 183--276,
  [\href{http://arxiv.org/abs/hep-th/9210010}{{\tt hep-th/9210010}}].

\bibitem{miwa2000solitons}
T.~Miwa, M.~Jimbo, and E.~Date, {\em Solitons: Differential equations,
  symmetries and infinite dimensional algebras}, vol.~135.
\newblock Cambridge University Press, 2000.

\bibitem{Khesin:1993ru}
B.~Khesin and I.~Zakharevich, {\it {Poisson - Lie group of pseudodifferential
  symbols}},  {\em Commun. Math. Phys.} {\bf 171} (1995) 475--530,
  [\href{http://arxiv.org/abs/hep-th/9312088}{{\tt hep-th/9312088}}].

\bibitem{Brungs:1998ij}
D.~Brungs and W.~Nahm, {\it {The Associative algebras of conformal field
  theory}},  {\em Lett. Math. Phys.} {\bf 47} (1999) 379--383,
  [\href{http://arxiv.org/abs/hep-th/9811239}{{\tt hep-th/9811239}}].

\bibitem{Zhu1995ModularIO}
Y.~Zhu, {\it Modular invariance of characters of vertex operator algebras},
  1995.

\bibitem{francesco2012conformal}
P.~Francesco, P.~Mathieu, and D.~S{\'e}n{\'e}chal, {\em Conformal field
  theory}.
\newblock Springer, 2012.

\bibitem{linshaw2008invariant}
{Linshaw, Andrew R}, {\it {Invariant theory and the $W_{1+\infty}$ algebra with
  negative integral central charge}},  {\em {arXiv preprint arXiv:0811.4067}}
  (2008).

\bibitem{Creutzig:2014lsa}
T.~Creutzig and A.~R. Linshaw, {\it {Cosets of affine vertex algebras inside
  larger structures}},  \href{http://arxiv.org/abs/1407.8512}{{\tt
  arXiv:1407.8512}}.

\bibitem{Linshaw:2015bqv}
A.~R. Linshaw, {\it {The Structure of the Kac-–Wang-–Yan Algebra}},  {\em
  Commun. Math. Phys.} {\bf 345} (2016), no.~2 545--585.

\bibitem{Arakawa:2017rrn}
T.~Arakawa, T.~Creutzig, K.~Kawasetsu, and A.~R. Linshaw, {\it {Orbifolds and
  Cosets of Minimal W-Algebras}},  {\em Commun. Math. Phys.} {\bf 355} (2017),
  no.~1 339--372, [\href{http://arxiv.org/abs/1610.09348}{{\tt
  arXiv:1610.09348}}].

\bibitem{Kac:1993zg}
V.~Kac and A.~Radul, {\it {Quasifinite highest weight modules over the Lie
  algebra of differential operators on the circle}},  {\em Commun. Math. Phys.}
  {\bf 157} (1993) 429--457, [\href{http://arxiv.org/abs/hep-th/9308153}{{\tt
  hep-th/9308153}}].

\bibitem{Hornfeck:1992tm}
K.~Hornfeck, {\it {W algebras with set of primary fields of dimensions
  $(3,4,5)$ and $(3,4,5,6)$}},  {\em Nucl. Phys.} {\bf B407} (1993) 237--246,
  [\href{http://arxiv.org/abs/hep-th/9212104}{{\tt hep-th/9212104}}].

\bibitem{dong2009w}
C.~Dong, C.~H. Lam, and H.~Yamada, {\it W-algebras related to parafermion
  algebras},  {\em Journal of Algebra} {\bf 322} (2009), no.~7 2366--2403.

\bibitem{Thielemans:1991uw}
K.~Thielemans, {\it {A Mathematica package for computing operator product
  expansions}},  {\em Int. J. Mod. Phys.} {\bf C2} (1991) 787--798.

\bibitem{Mizoguchi:1988vk}
S.~Mizoguchi, {\it {Determinant Formula and Unitarity for the $W_3$ Algebra}},
  {\em Phys. Lett.} {\bf B222} (1989) 226--230.

\bibitem{mimachi1995singular}
K.~Mimachi and Y.~Yamada, {\it Singular vectors of the virasoro algebra in
  terms of jack symmetric polynomials},  {\em Communications in mathematical
  physics} {\bf 174} (1995), no.~2 447--455.

\bibitem{Awata:1995np}
H.~Awata, Y.~Matsuo, S.~Odake, and J.~Shiraishi, {\it {Excited states of
  Calogero-Sutherland model and singular vectors of the W(N) algebra}},  {\em
  Nucl. Phys.} {\bf B449} (1995) 347--374,
  [\href{http://arxiv.org/abs/hep-th/9503043}{{\tt hep-th/9503043}}].

\bibitem{Feigin:2018bkf}
B.~Feigin and S.~Gukov, {\it {VOA[$M_4$]}},
  \href{http://arxiv.org/abs/1806.02470}{{\tt arXiv:1806.02470}}.

\bibitem{Gaberdiel:2017hcn}
M.~R. Gaberdiel, W.~Li, C.~Peng, and H.~Zhang, {\it {The supersymmetric affine
  Yangian}},  {\em JHEP} {\bf 05} (2018) 200,
  [\href{http://arxiv.org/abs/1711.07449}{{\tt arXiv:1711.07449}}].

\bibitem{Wakimoto:1986gf}
M.~Wakimoto, {\it {Fock representations of the affine lie algebra A1(1)}},
  {\em Commun. Math. Phys.} {\bf 104} (1986) 605--609.

\bibitem{Feigin:1990jc}
B.~L. Feigin and E.~V. Frenkel, {\it {Affine Kac-Moody algebras and
  semiinfinite flag manifolds}},  {\em Commun. Math. Phys.} {\bf 128} (1990)
  161--189.

\bibitem{Gaiotto:2013rk}
D.~Gaiotto and J.~Lamy-Poirier, {\it {Irregular Singularities in the $H_3^+$
  WZW Model}},  \href{http://arxiv.org/abs/1301.5342}{{\tt arXiv:1301.5342}}.

\bibitem{Nekrasov:2015wsu}
N.~Nekrasov, {\it {BPS/CFT correspondence: non-perturbative Dyson-Schwinger
  equations and qq-characters}},  {\em JHEP} {\bf 03} (2016) 181,
  [\href{http://arxiv.org/abs/1512.05388}{{\tt arXiv:1512.05388}}].

\bibitem{Nekrasov:2016qym}
N.~Nekrasov, {\it {BPS/CFT correspondence II: Instantons at crossroads, moduli
  and compactness theorem}},  {\em Adv. Theor. Math. Phys.} {\bf 21} (2017)
  503--583, [\href{http://arxiv.org/abs/1608.07272}{{\tt arXiv:1608.07272}}].

\bibitem{Nekrasov:2016gud}
N.~Nekrasov and N.~S. Prabhakar, {\it {Spiked Instantons from Intersecting
  D-branes}},  {\em Nucl. Phys.} {\bf B914} (2017) 257--300,
  [\href{http://arxiv.org/abs/1611.03478}{{\tt arXiv:1611.03478}}].

\bibitem{Nekrasov:2016ydq}
N.~Nekrasov, {\it {BPS/CFT Correspondence III: Gauge Origami partition function
  and qq-characters}},  {\em Commun. Math. Phys.} {\bf 358} (2018), no.~3
  863--894, [\href{http://arxiv.org/abs/1701.00189}{{\tt arXiv:1701.00189}}].

\bibitem{Nekrasov:2017rqy}
N.~Nekrasov, {\it {BPS/CFT correspondence IV: sigma models and defects in gauge
  theory}},  \href{http://arxiv.org/abs/1711.11011}{{\tt arXiv:1711.11011}}.

\bibitem{Nekrasov:2017gzb}
N.~Nekrasov, {\it {BPS/CFT correspondence V: BPZ and KZ equations from
  qq-characters}},  \href{http://arxiv.org/abs/1711.11582}{{\tt
  arXiv:1711.11582}}.

\bibitem{Nekrasov:2017cih}
N.~Nekrasov, {\it {Magnificent Four}},
  \href{http://arxiv.org/abs/1712.08128}{{\tt arXiv:1712.08128}}.

\bibitem{Leung:1997tw}
N.~C. Leung and C.~Vafa, {\it {Branes and toric geometry}},  {\em Adv. Theor.
  Math. Phys.} {\bf 2} (1998) 91--118,
  [\href{http://arxiv.org/abs/hep-th/9711013}{{\tt hep-th/9711013}}].

\bibitem{Frenkel:2018dej}
E.~Frenkel and D.~Gaiotto, {\it {Quantum Langlands dualities of boundary
  conditions, D-modules, and conformal blocks}},
  \href{http://arxiv.org/abs/1805.00203}{{\tt arXiv:1805.00203}}.

\bibitem{Dedushenko:2017tdw}
M.~Dedushenko, S.~Gukov, and P.~Putrov, {\it {Vertex algebras and 4-manifold
  invariants}},  \href{http://arxiv.org/abs/1705.01645}{{\tt
  arXiv:1705.01645}}.

\bibitem{Gadde:2013sca}
A.~Gadde, S.~Gukov, and P.~Putrov, {\it {Fivebranes and 4-manifolds}},
  \href{http://arxiv.org/abs/1306.4320}{{\tt arXiv:1306.4320}}.

\bibitem{Gaiotto:2012sf}
D.~Gaiotto and J.~Teschner, {\it {Irregular singularities in Liouville theory
  and Argyres-Douglas type gauge theories, I}},  {\em JHEP} {\bf 12} (2012)
  050, [\href{http://arxiv.org/abs/1203.1052}{{\tt arXiv:1203.1052}}].

\bibitem{Witten:2007td}
E.~Witten, {\it {Gauge theory and wild ramification}},
  \href{http://arxiv.org/abs/0710.0631}{{\tt arXiv:0710.0631}}.

\bibitem{Ding:1996mq}
J.-t. Ding and K.~Iohara, {\it {Generalization and deformation of Drinfeld
  quantum affine algebras}},  {\em Lett. Math. Phys.} {\bf 41} (1997) 181--193.

\bibitem{Miki}
K.~{Miki}, {\it {A (q,{$\gamma$}) analog of the W$_{1+∞}$ algebra}},  {\em
  Journal of Mathematical Physics} {\bf 48} (Dec., 2007) 123520--123520.

\bibitem{Awata:2011dc}
H.~Awata, B.~Feigin, A.~Hoshino, M.~Kanai, J.~Shiraishi, and S.~Yanagida, {\it
  {Notes on Ding--Iohara algebra and AGT conjecture}},
  \href{http://arxiv.org/abs/1106.4088}{{\tt arXiv:1106.4088}}.

\bibitem{Awata:2011ce}
H.~Awata, B.~Feigin, and J.~Shiraishi, {\it {Quantum Algebraic Approach to
  Refined Topological Vertex}},  {\em JHEP} {\bf 03} (2012) 041,
  [\href{http://arxiv.org/abs/1112.6074}{{\tt arXiv:1112.6074}}].

\bibitem{Mironov:2016yue}
A.~Mironov, A.~Morozov, and Y.~Zenkevich, {\it {Ding–-Iohara-–Miki symmetry of
  network matrix models}},  {\em Phys. Lett.} {\bf B762} (2016) 196--208,
  [\href{http://arxiv.org/abs/1603.05467}{{\tt arXiv:1603.05467}}].

\bibitem{Awata:2016riz}
H.~Awata, H.~Kanno, T.~Matsumoto, A.~Mironov, A.~Morozov, A.~Morozov,
  Y.~Ohkubo, and Y.~Zenkevich, {\it {Explicit examples of DIM constraints for
  network matrix models}},  {\em JHEP} {\bf 07} (2016) 103,
  [\href{http://arxiv.org/abs/1604.08366}{{\tt arXiv:1604.08366}}].

\bibitem{Awata:2016mxc}
H.~Awata, H.~Kanno, A.~Mironov, A.~Morozov, A.~Morozov, Y.~Ohkubo, and
  Y.~Zenkevich, {\it {Toric Calabi-Yau threefolds as quantum integrable
  systems. $ \mathrm{\mathcal{R}} $ -matrix and $
  \mathrm{\mathcal{R}}\mathcal{T}\mathcal{T} $ relations}},  {\em JHEP} {\bf
  10} (2016) 047, [\href{http://arxiv.org/abs/1608.05351}{{\tt
  arXiv:1608.05351}}].

\bibitem{Awata:2016bdm}
H.~Awata, H.~Kanno, A.~Mironov, A.~Morozov, A.~Morozov, Y.~Ohkubo, and
  Y.~Zenkevich, {\it {Anomaly in RTT relation for DIM algebra and network
  matrix models}},  {\em Nucl. Phys.} {\bf B918} (2017) 358--385,
  [\href{http://arxiv.org/abs/1611.07304}{{\tt arXiv:1611.07304}}].

\bibitem{Bourgine:2016vsq}
J.-E. Bourgine, M.~Fukuda, Y.~Matsuo, H.~Zhang, and R.-D. Zhu, {\it {Coherent
  states in quantum $\mathcal{W}_{1+\infty}$ algebra and qq-character for 5d
  Super Yang-Mills}},  {\em PTEP} {\bf 2016} (2016), no.~12 123B05,
  [\href{http://arxiv.org/abs/1606.08020}{{\tt arXiv:1606.08020}}].

\bibitem{Bourgine:2017jsi}
J.-E. Bourgine, M.~Fukuda, K.~Harada, Y.~Matsuo, and R.-D. Zhu, {\it
  {$(p,q)$-webs of DIM representations, 5d $\mathcal{N}=1$ instanton partition
  functions and qq-characters}},  {\em JHEP} {\bf 11} (2017) 034,
  [\href{http://arxiv.org/abs/1703.10759}{{\tt arXiv:1703.10759}}].

\bibitem{Smirnov:2013hh}
A.~Smirnov, {\it {On the Instanton R-matrix}},  {\em Commun. Math. Phys.} {\bf
  345} (2016), no.~3 703--740, [\href{http://arxiv.org/abs/1302.0799}{{\tt
  arXiv:1302.0799}}].

\end{thebibliography}\endgroup
\bibliographystyle{JHEP}

\end{document}